\pdfoutput=1
\documentclass{article}

\usepackage[utf8]{inputenc}
\usepackage{comment}

\usepackage{geometry}  %
\usepackage{setspace}  %
\usepackage{fancyhdr}  %
\usepackage{xcolor}
\usepackage{svg}

\usepackage{graphicx}
\usepackage{caption}
\usepackage{subcaption}
\usepackage{longtable}
\usepackage{tabularray}
\usepackage{booktabs}
\usepackage{multirow}
\usepackage{float}
\usepackage{array}
\usepackage{tabularx}
\usepackage{rotating}

\usepackage{amsmath, amssymb, amsfonts}
\usepackage{amsthm}
\usepackage{mathrsfs}
\usepackage{dsfont}

\usepackage{algorithm}
\usepackage{algorithmicx}
\usepackage{algpseudocode}
\usepackage{etoolbox}

\usepackage{xcolor}
\usepackage{textcomp}
\usepackage{titlesec}
\usepackage{manyfoot}
\usepackage{appendix}
\usepackage{listings}
\usepackage{url}

\usepackage{authblk}

\usepackage{tikz}
\usetikzlibrary{shapes.geometric, arrows, positioning}

\usepackage[
    style=apa,
    sorting=ynt,
    sortcites=true,
    backend=biber,
    uniquename=false
]{biblatex}
\pretocmd{\appendix}{
  \titleformat{\section}
    {\centering\bfseries}
    {Appendix \thesection:}{1em}{}
  
  \titleformat{\subsection}[hang]
    {\bfseries}
    {\thesubsection}{1em}{}

  \titleformat{\subsubsection}[runin]
    {\bfseries\itshape}
    {\thesubsubsection}{1em}{}
}{}{}

\title{Synthesis and Perceptual Scaling of High Resolution \textcolor{black}{Naturalistic} Images Using Stable Diffusion}

\author[1,2,3,4]{Leonardo Pettini}
\author[2]{Carsten Bogler}
\author[3,4,5]{Christian Doeller}
\author[1,2,4,6,7,8,9]{John-Dylan Haynes}

\affil[1]{Department of Psychology, Humboldt-Universität zu Berlin, Berlin, Germany}
\affil[2]{Bernstein Center for Computational Neuroscience Berlin and Berlin Center for Advanced Neuroimaging, Charité Universitätsmedizin Berlin, corporate member of the Freie Universität Berlin, Humboldt-Universität zu Berlin, and Berlin Institute of Health, Berlin, Germany}
\affil[3]{Max Planck Institute for Human Cognitive and Brain Sciences, Leipzig, Germany}
\affil[4]{Max Planck School of Cognition, Leipzig, Germany}
\affil[5]{Kavli Institute for Systems Neuroscience, Centre for Neural Computation, The Egil and Pauline Braathen and Fred Kavli Centre for Cortical Microcircuits, Jebsen Centre for Alzheimer’s Disease, Norwegian University of Science and Technology, Trondheim, Norway}
\affil[6]{Clinic of Neurology, Charité-Universitätsmedizin Berlin, Berlin, Germany}
\affil[7]{Berlin Center for Advanced Neuroimaging, Charité-Universitätsmedizin Berlin, Berlin, Germany}
\affil[8]{Research Cluster of Excellence “Science of Intelligence”, Technische Universität Berlin, Berlin, Germany}
\affil[9]{Berlin School of Mind and Brain, Humboldt-Universität zu Berlin, Berlin, Germany}

\date{}

\usepackage{titlesec}

\titleformat{\section}{\centering\bfseries}{\hspace{-1em}}{0em}{} 

\titleformat{\subsection}[hang]{\bfseries}{}{0pt}{}  
\titlespacing{\subsection}{0pt}{12pt plus 2pt minus 2pt}{6pt plus 2pt minus 2pt}

\titleformat{\subsubsection}[runin]{\bfseries\itshape}{}{0em}{} 

\titleformat{\paragraph}[runin]{\bfseries}{}{0em}{} 
\titleformat{\subparagraph}[runin]{\bfseries\itshape}{}{0em}{}

\renewenvironment{abstract}
{\begin{center}\textbf{Abstract}\end{center} \noindent}
{\par\vspace{1em}}

\usepackage[skip=2pt]{caption} 
\begin{document}

\begin{titlepage}
    \centering
    \maketitle
    \thispagestyle{fancy} 
    \begin{flushleft}
\textbf{Author Note}

Leonardo Pettini: https://orcid.org/0009-0006-4855-5947 \\
Correspondence concerning this article should be addressed to Leonardo Pettini, Email: \texttt{leonardo.pettini@gmail.com}. \\Phone number: +4930209398517. \\ Address: Bernstein Center for Computational Neuroscience, Philippstraße 13/Haus 6, 10115 Berlin  \\
No conflicts of interest to declare.
\end{flushleft}

\end{titlepage}

\begin{abstract}
\textcolor{black}{Naturalistic} scenes are of key interest for visual perception, but controlling their perceptual and semantic properties is challenging. Previous work on \textcolor{black}{naturalistic} scenes has frequently focused on collections of discrete images with considerable physical differences between stimuli. However, it is often desirable to assess representations of \textcolor{black}{naturalistic} images that vary along a continuum. Traditionally, perceptually continuous variations of \textcolor{black}{naturalistic} stimuli have been obtained by morphing a source image into a target image. This produces transitions driven mainly by low-level physical features and can result in semantically ambiguous outcomes. More recently, generative adversarial networks (GANs) have been used to generate continuous perceptual variations within a stimulus category. Here we extend and generalize this approach using a different machine learning approach, a text-to-image diffusion model (Stable Diffusion XL), to generate a freely customizable stimulus set of photorealistic images that are characterized by gradual transitions, with each image representing a unique exemplar within a prompted category. We demonstrate the approach by generating a set of 108 object scenes from 6 categories. For each object scene, we generate 10 variants that are ordered along a perceptual continuum. This ordering was first estimated using a machine learning model of perceptual similarity (LPIPS) and then subsequently validated with a large online sample of human participants. In a subsequent experiment we show that this ordering is also predictive of confusability of stimuli in a working memory experiment. Our image set is suited for studies investigating the graded encoding of \textcolor{black}{naturalistic} stimuli in visual perception, attention, and memory. 

\textit{Keywords}: perceptual scaling, \textcolor{black}{naturalistic} images, stable diffusion, working memory, long-term memory
\end{abstract}

\newpage

\section{Introduction}\label{sec1}

Naturalistic stimuli are important for understanding object recognition and memory in ecologically valid settings \parencite{henderson1999high, bar2004visual, Oliva2007, Epstein2019}, but they present several challenges. They can vary widely in their semantic and perceptual dimensions, which makes them harder to select and to control experimentally in comparison to low-dimensional stimuli~\parencite{goetschalckx_generative_2021}. In contrast to traditional stimulus sets, which have relied on the manual selection of a limited set of cropped images~\parencite{Snodgrass1980, Brodeur2010}, a process that is both time-consuming and susceptible to subjectivity biases, recent attempts have strived to systematically collect and use naturalistic stimuli. These approaches are usually ``bottom-up'', involving the collection and categorization of large numbers of images from the internet (``scraping''). These can be found in large-scale image datasets used in the machine learning community for computer vision tasks, such as ImageNet~\parencite{imagenet} or the Microsoft Common Objects in Context (COCO) dataset~\parencite{lin_microsoft_2014}. For example, the Natural Scenes Dataset (NSD) initiative~\parencite{allen_massive_2022} collected a rich amount of behavioural and neuroimaging data while participants viewed scenes from the COCO dataset. Such stimuli, however, are not always ideally suited for cognitive tasks and often require manual pre-selection. To address this issue, the THINGS initiative~\parencite{hebart_things_2019} has developed a procedure to sample and evaluate a vast variety of ``object concepts'' from the web, specifically tailored for cognitive scientific tasks. Despite these advancements, these methods remain fundamentally dependent on existing images and their \textit{post hoc} categorisation. 

An alternative approach involves synthesizing stimuli using modern machine learning technology. Generative models such as Variational Autoencoders (VAEs)~\parencite{kingma2022autoencodingvariationalbayes}, Generative Adversarial Networks (GANs)~\parencite{goodfellow2014generativeadversarialnetworks} and Diffusion Models (DMs)~\parencite{ho_denoising_2020} have already provided novel methodological approaches to many areas of neuroscience, including image reconstruction from neuroimaging data~\parencite{Shen2019, Ozcelik2023, Liu2024}, analysis of neural population dynamics~\parencite{Pandarinath2018, Bashivan2019}, and clinical imaging~\parencite{Yi2019, pinaya2023generativeaimedicalimaging}. They are also promising for the synthetic generation of \textcolor{black}{naturalistic} stimuli for experimental tasks~\parencite{goetschalckx_generative_2021, son_scene_2022, cooper_standardised_2023}. Compared to ``bottom-up'' approaches that rely on scraping pre-existing images from the web, generative models allow for the synthesis of virtually unlimited high-resolution images, removing the dependency on existing sources. This is particularly appealing for object recognition and memory research, as it enables the creation of a variety of objects from a wide range of categories. 

One particularly interesting feature of artificially generated images is that they can potentially help tackle the trade-off between ecological validity and parametric experimental control. For example, low-level physical properties (such as orientation or luminance) can easily be gradually varied and they are thus suitable for psychophysical studies that involve assessment of quantitative properties ~\parencite{Schurgin2020}. In contrast, sets of discrete \textcolor{black}{naturalistic} images do not exhibit the same kind of gradual local ordering that would be necessary for such quantitative assessments. Any measurement that requires gradual variation of an image property is very difficult to realize with sets of pre-existing, discrete naturalistic stimuli.

Controlling perceptual and semantic features of high-dimensional naturalistic stimuli is inherently difficult. However, machine-learning-based approaches provide a solution. One class of machine learning methods, Generative Adversarial Networks (GANs)~\parencite{goodfellow2014generativeadversarialnetworks}, have recently gained attention in cognitive science~\parencite{goetschalckx_generative_2021}. The deep generative representations that GANs learn have been shown to be structured semantically~\parencite{yang_semantic_2020}. They allow obtaining fine-grained and relatively selective variations of the images along continuous dimensions, both for perceptual (e.g. the lightness of the scene) and for semantic features (e.g. facial attributes)~\parencite{shocher_semantic_2020, yang_l2m-gan_2021}. A recent study, ~\textcites{son_scene_2022}, used a GAN to create “scene wheels” of naturalistic indoor scenes with varying levels of similarity, which they employed in a visual working memory task, typically used for simpler features like color or orientation~\parencite{zhang_discrete_2008}. This reflects a growing interest towards a more ecologically valid assessment of cognitive function and provides the possibility to extend working memory models to naturalistic stimuli~\parencite{bates_scaling_2024}. However, despite their utility, GANs face significant limitations. Pre-trained GAN models, while available~\parencite{Karras2020}, often require fine-tuning to be suitable for specific image generation tasks. Some of these models can generate a variety of images but are restricted to certain categories and their resolution can be rather low.  

Another class of generative neural networks specialised in image synthesis is diffusion models~\parencite{ho_denoising_2020, nichol_improved_2021, song_denoising_2022}, which so far have received less attention in cognitive science. Stable Diffusion~\parencite{rombach2022high}, in particular, is an open-source model that can generate images from text prompts. Text-to-image approaches for stimulus generation are more flexible~\parencite{dhariwal_diffusion_2021} because they allow a high flexibility in the choice of visual scenes without requiring scraping of sets of dedicated training samples.  

In this study, we utilised the flexibility of diffusion-based text-to-image models to generate a novel set of naturalistic stimuli. \textcolor{black}{The terms “natural” and “naturalistic” are often used interchangeably in cognitive science~\parencites{hebart_things_2019, hebart2020revealing, gong2023largescale}. In order to avoid confusion with the machine learning terminology, which may consider the term “natural” as an antonym of “synthetic”, that is the “ground truth” images that generative models aim to replicate~\parencite{goodfellow2014generativeadversarialnetworks, dzanic2020fourier}, here we solely use the term “naturalistic” to describe synthetically generated, realistic-appearing images. We continue to use the term “natural” as opposed to “artificial” to denote the natural categories of stimuli (animals, plants, landscape elements).} \textcolor{black}{We generated sets of “object-scenes” that have a central, prominent object situated in a coherent scene, in line with previous work~\parencite{hebart_things_2019}. To generate these stimuli, we used Stable Diffusion XL~\parencite{podell_sdxl_2023}, a diffusion model that excels in the synthesis of high resolution images (1024x1024 pixels).} We also ensured and assessed the psychometric continuity in a three-step procedure. First, we ordered images using a machine-learning-based psychophysical similarity metric, Learned Perceptual Image Patch Similarity (LPIPS)~\parencite{zhang_unreasonable_2018}. Then, in a second step, we fine-tuned the ordering of stimuli using an online similarity judgement with a sample of 1113 human participants. In a third step, we assessed whether perceptual similarity generalizes to other cognitive functions, specifically to working memory representations. For this we used our novel stimulus set in a visual working memory task. Specifically, this experiment assessed visual working memory performance for stimuli at varying levels of distance in our continuous stimulus sets.  

The results confirmed that our stimuli effectively captured perceptual variations, making them a useful resource for studying memory and perception under controlled yet ecologically valid conditions. By providing the stimulus set publicly (see Data Availability section), we hope to contribute a valuable tool for the research community, bridging the gap between ecological validity and experimental control in visual cognition studies.

\section{Methods}\label{sec2}

The study consisted of three main parts: In Stage 1 ("Stimulus set generation"), we used a diffusion model~\parencite{ho_denoising_2020, nichol_improved_2021, song_denoising_2022}, specifically Stable Diffusion XL~\parencite{podell_sdxl_2023}, to generate large sets of exemplars of object scenes, each described by a text prompt (e.g. for "fish", we used \textit{"award-winning marine photo of a colorful fish in a coral reef, centered in the scene, vibrant underwater scene, high detail"}). We then estimated the perceived similarity of the different exemplars based on a computational model (Learned Perceptual Image Patch Similarity, LPIPS)~\parencite{zhang_unreasonable_2018} and ordered the scenes accordingly. In Stage 2 ("Perceptual Similarity Judgement Experiment") we validated and subsequently fine-tuned the perceptual ordering by performing a similarity judgement task with an online sample of 1113 participants. In Stage 3 ("Memory Validation Experiment") we additionally validated that the proximity along our perceptual continua predicted performance in a separate working memory task.

\subsection{Stage 1: Image Generation using the Diffusion Model}

The image generation proceeded in five steps (shown in Figure~\ref{fig:stage_1}; for full details see the Appendix): \textbf{(S1)} We used Stable Diffusion XL in combination with text prompts to generate naturalistic images ("object scenes") with a clear, central object situated in a coherent scene, rich in detail (see Appendix~\ref{appendix:generative_model} for full specifications). Since text-to-image models~\parencite{zhang2023texttoimagediffusionmodelsgenerative} can generate multiple images from a textual prompt, there are potentially infinite images that can be synthesized. We therefore narrowed down the prompt space by defining \textbf{6 different categories} from which our objects were taken. These categories were split into natural (animals, plants, landscape elements) and artificial (vehicles, items, buildings). For each category, we identified \textbf{18 unique object scenes} that were distinct yet representative of the category (e.g. "fish" and "beaver" were two object scenes of the category "animals"). Each of the 108 object scenes had an associated \textbf{unique text prompt}. See Appendix~\ref{appendix:generation_pipeline} for details about the prompt selection. \textbf{(S2)} Based on these text prompts, we used Stable Diffusion XL~\parencite{podell_sdxl_2023} to generate \textbf{60 exemplar images} per each of the 108 object scenes. These were different realisations of the same object scene (e.g. 60 different realisations of the object scene "fish"; see Figure~\ref{fig:stage_1} and Appendix~\ref{appendix:generation_pipeline}). \textbf{(S3)} For each pair of these 60 exemplar images from one object scene, we computed a \textbf{coarse-grained model-based perceptual similarity score} using an established metric based on neural network activation patches, the Learned Perceptual Image Patch Similarity (LPIPS) metric~\parencite{zhang_unreasonable_2018}. From this, we selected an \textbf{``anchor image''} and a \textbf{``guide image''}. These two images were chosen such that they were representative of the set as well as perceptually similar to one another (for details see Appendix~\ref{appendix:selection-anchors}). \textbf{(S4)} We then used \textbf{spherical interpolation} to yield 200 further images per object scene on the continuum between the anchor image and the guide image (this two-step procedure was done in order to go from coarse-grained to fine-grained steps). This interpolation was done at the level of noise latents rather than at the semantic level of the text prompts in order to ensure that interpolated images varied perceptually but not in terms of meaning (see Appendix~\ref{appendix:interpolation_algorithm}). \textbf{(S5)} We used the artificial neural network model (LPIPS) once more to compute the \textbf{fine-grained model-based perceptual similarity} between these 200 interpolations. Based on this, we selected a total of ten images that were dissimilar from the anchor image in an approximately linear way (see Appendix~\ref{appendix:selection_interpol_images}).

\subsection{Stage 2: Psychophysical Similarity Judgement Task}

\subsubsection{Participants}

We recruited 1,285 participants on the online platform Prolific for an experiment involving a  perceptual similarity judgement task. This sample was larger than the target minimum (at least 20 participants judging an object, for a total of 1,080) because we anticipated excluding up to 20\% of the data, as reported previously~\parencite{Gagne2023Jan, uittenhove_effectiveness_2023}. The final sample included 1,113 participants (age 18--40, $M = 28.9$, $SD = 5.7$; 760 male, 521 female, and 5 "prefer not to say").
 Participants received remuneration of 8.53 GBP (approximately 10 EUR) per hour. They were selected from a standard sample (aged 18--40 years, fluent in English) to maximize data quality. We excluded participants who did not engage properly with the task or experienced technical problems (see exclusion criteria in the Appendix~\ref{appendix:control_quality_exclusion_criteria}).
All participants provided informed consent before taking part in the study. The consent process included detailed information about the purpose and duration of the experiment, voluntariness of participation, and data protection policies. The study was approved by the Ethics Committee of the Institute of Psychology of the Humboldt University of Berlin, Germany.

\subsubsection{Experimental Procedures and Design}

Participants performed a similarity judgement task based on triplet comparisons, or “method of triads”~\parencite{torgerson_multidimensional_1952}, which is used both in psychophysics and machine learning~\parencite{aguilar_comparing_2017, demiralp2014learning, haghiri_estimation_2020, kunstle_estimating_2022, li_extracting_2016, wichmann_methods_2017}. Given a set of stimuli \( S = \{s_1, s_2, \ldots, s_n\} \), where \( n \) is the total number of stimuli, participants are asked to judge the similarity of three stimuli at a time. In particular, given a triplet of stimuli \( (s_i, s_j, s_k) \), they are asked which between two probe stimuli \( s_j \) and \( s_k \) is most similar to a reference stimulus \( s_i \). 
After providing informed consent, participants received detailed instructions on the task procedure and had to pass an attention check to make sure they understood the instructions. Before starting the main task, they underwent a training session where they performed 12 trials using an independent set of stimuli. The main task consisted of 288 trials divided in 6 blocks of 48 trials each. In between blocks, participants were prompted to take a short break (maximum 2-3 minutes) to maintain attention and accuracy. Each trial began with a fixation target~\parencite{thaler_what_2013} presented at the centre of the screen for a jittered duration between 500 and 1000 ms, which was randomized in 100 ms steps. Then, participants were shown three images: one reference image on top and two probe images below (see Figure~\ref{fig:online_task_design}). Their task was to judge which of the two probe images was more similar to the reference image. Full details of the experimental procedures can be found in the supplementary material.

\subsubsection{Analyses}
The main goal of the similarity judgement analysis was to estimate a perceptual scale for each object and its variations, given a set of perceptual judgements. To analyse the triplet judgements, we used three algorithms: Maximum Likelihood Difference Scaling (MLDS)~\parencite{maloney_maximum_2003}, which is a well-established scaling method in psychophysics, and two ordinal embedding algorithms, Soft Ordinal Embedding (SOE)~\parencite{terada_local_2014} and t-Distributed Stochastic Triplet Embedding (t-STE)~\parencite{van_der_maaten_stochastic_2012}. MLDS can be used only in one-dimensional cases, whereas t-STE and SOE have been proposed to find ordinal embeddings in higher dimensional spaces~\parencite{haghiri_estimation_2020}. Embeddings, in this context, are Euclidean representations that preserve the ordinal relationships among data points based on a set of perceptual triplet judgements~\parencite{agarwal2007generalized, jamieson2011low}. We performed a comparative analysis between the three algorithms, assessing both their performance and the stability of the embedding estimates. To assess their performance, we used the cross-validated triplet error \(E_t\)~\parencite{haghiri_estimation_2020}.
For each object and category, we calculated the mean cross-validated triplet error of each algorithm by averaging the triplet errors from all cross-validation steps.  The object-wise and category-wise error estimated how well the model performed for individual objects and for categories. We also calculated the overall mean triplet error for each algorithm, which provided a general measure of how they performed across objects. As a final step, we reordered the images according to the embedding results and compared this order to the one defined by the LPIPS metric.

\subsection{Stage 3: Memory Validation Task}

\subsubsection{Participants}

For the second experiment we recruited \textcolor{black}{338} participants from the online platform Prolific using the same criteria as in the Psychophysical Similarity Judgement Task. The final sample included 240 participants (age 18--40, \textcolor{black}{$M = 28.42$, $SD = 5.86$;  138 male, 101 female, and 1 "prefer not to say"}). \textcolor{black}{We excluded participants who did not engage properly with the task, who showed behavioural performance below chance and who reported technical problems (for detailed criteria see Appendix~\ref{appendix:memory_data_quality_exclusion}). The study was approved by the Ethics Committee of the Institute
of Psychology of the Humboldt University of Berlin, Germany.}

\subsubsection{Stimuli}

From the 10-image set for each of the 108 object scenes, we selected a target image and three foils with increasing perceptual dissimilarity. Since our final image sets contained ten images per object, we implemented an algorithm to subsample four images using their embedding values and avoiding abrupt perceptual changes as much as possible (see Appendix~\ref{appendix:target_foil_selection}). 

\subsubsection{Experimental Procedures and Design}
Participants began by reviewing and signing an online consent form, after which they received detailed instructions about the experimental procedure. They then completed a brief training session within their browser to familiarise themselves with the task. At the end of the experiment, they were asked to fill in a short debriefing questionnaire. In line with previous work~\parencite{Daniel2016}, we used a delayed match-to-sample paradigm with repeated and non-repeated target images (Figure \ref{fig:stage_3}). The experiment was divided into six blocks of 18 trials each, for a total of 108 trials. At the beginning of each trial, a target image, which participants were instructed to memorise, was shown in the centre of the screen for two seconds, followed by an eight-second delay with a blank screen. Then, participants were tested on whether they could recognise the same target image among a perceptually similar foil chosen from our novel stimulus set. We selected three foils with increasing levels of dissimilarity (see below for details) to manipulate task difficulty across three levels: Easy, Medium, and Hard. The target and foil were sequentially presented for one second each, separated by a 500-ms gap, and their order was counterbalanced so that sometimes the target was shown first and sometimes second. Once both images had appeared, a new response mapping screen was presented with two options displayed in a monospaced font to the left and right of a question mark, respectively, with their positions on the screen counterbalanced (see below). Participants used this screen to indicate which image was the target by selecting either “one” or “two,” corresponding to pressing the ‘f’ key if the chosen option appeared on the left, and the ‘j’ key if it appeared on the right. The response window had a fixed duration of two seconds, regardless of when they pressed a key. Upon key press, participants were given visual feedback indicating the selected option (but not whether the response was correct). An inter-trial interval was randomly selected from 1000 ms, 1500 ms, and 2000 ms to introduce temporal variability and reduce predictability. Participants were instructed to maintain stable fixation throughout the task by fixating on a central target (a bull's-eye with crosshairs) that remained visible at the centre of the screen, even during stimulus presentation~\parencite{thaler_what_2013}. The fixation target was also presented at the beginning and end of each block. The total trial duration was about 19 seconds.
Participants were assigned 9 repeated targets and 54 non-repeated targets from the stimulus set including 108 objects (see above). Each experimental block contained 9 repeated targets and 9 non-repeated targets. We ensured that repeated and non-repeated targets were systematically counterbalanced across participants. Each object appeared as a repeated target for an equal number of participants, while the distribution of non-repeated targets varied across the stimulus set, though the final count was similar. Within each block, difficulty levels were distributed evenly across repeated and non-repeated targets (i.e., three Easy, three Medium, and three Hard trials for each category), so that the average difficulty of each block is constant. For each repeated target, we showed all difficulty levels in a random order before repeating them, which means that the overall object-specific difficulty was fully counterbalanced every 3 blocks. The presentation order of target and foil, as well as the left/right placement of the two response choices (“one” and “two”), was evenly split and randomised throughout the experiment so that the correct response occurred equally often on each button press. This constraint was designed to limit any systematic response bias.

\subsubsection{Analyses}
We calculated participants' average accuracies over blocks for each target image condition (repeated vs non-repeated) and difficulty level (Easy, Medium, Hard). In order to additionally quantify learning effects over time for each experimental condition, we adopted a Bayesian framework. Bayesian methods offer several advantages for this type of analysis, including the ability to compute estimates and credible intervals for predicted quantities directly \parencite{Wagenmakers2018}. Moreover, they facilitate the use of hierarchical structures to model individual differences in learning trajectories. We used a Bayesian hierarchical logistic regression model, which was implemented using the \texttt{brms} package \parencite{Burkner2017} in R \parencite{r_core_team_r_2024}. The Bayesian model was estimated using Markov chain Monte Carlo (MCMC) sampling with 4 chains, each consisting of 10,000 iterations, including 2,000 “warm up” iterations. This resulted in a total of 32,000 post-warmup draws. The model was fitted to a total of 25,920 trials. Convergence was assessed using the Potential Scale Reduction Factor (PSRF), denoted as $\hat{R}$. The resolution of the chains was evaluated using the effective sample size (ESS) \parencite{Kruschke2021}. The full specification of the model is provided in the Appendix~\ref{appendix:memory_experiment_model_spec}.

\begin{figure}[htbp]
    \centering
        \includegraphics[width=\linewidth]{figures/stage_1_v3_dpi-300.jpg}
        \vspace{3pt}
        \caption{
        Overview of the pipeline to generate the stimulus set (Stage 1). We defined six global categories (three natural and three artificial), each including eighteen object scenes. To generate varied exemplar images for each object (shown here for ``fish''), we used a single prompt, and variability was induced by using sixty starting noise latents (see Methods). From each set of exemplar images, we selected an anchor and a guide image for the interpolation, which was performed between their respective noise latents. Out of 200 interpolated images, we pre-selected 10 (the anchor image and nine interpolations) using a computational model-based metric (LPIPS, \parencite{zhang_unreasonable_2018}) as an initial approximation to quantify their perceptual similarity.
        }
        \label{fig:stage_1}
\end{figure}

\section{Results}\label{results}

\subsection{Stage 1: Stimulus set generation}

\begin{figure}[htbp]
    \centering
    \caption{Table showing the categories and objects in the stimulus set. The categories in this table are classified as \textit{artificial}. For each object, the \textit{anchor image} is shown.}
    \includegraphics[width=\textwidth]{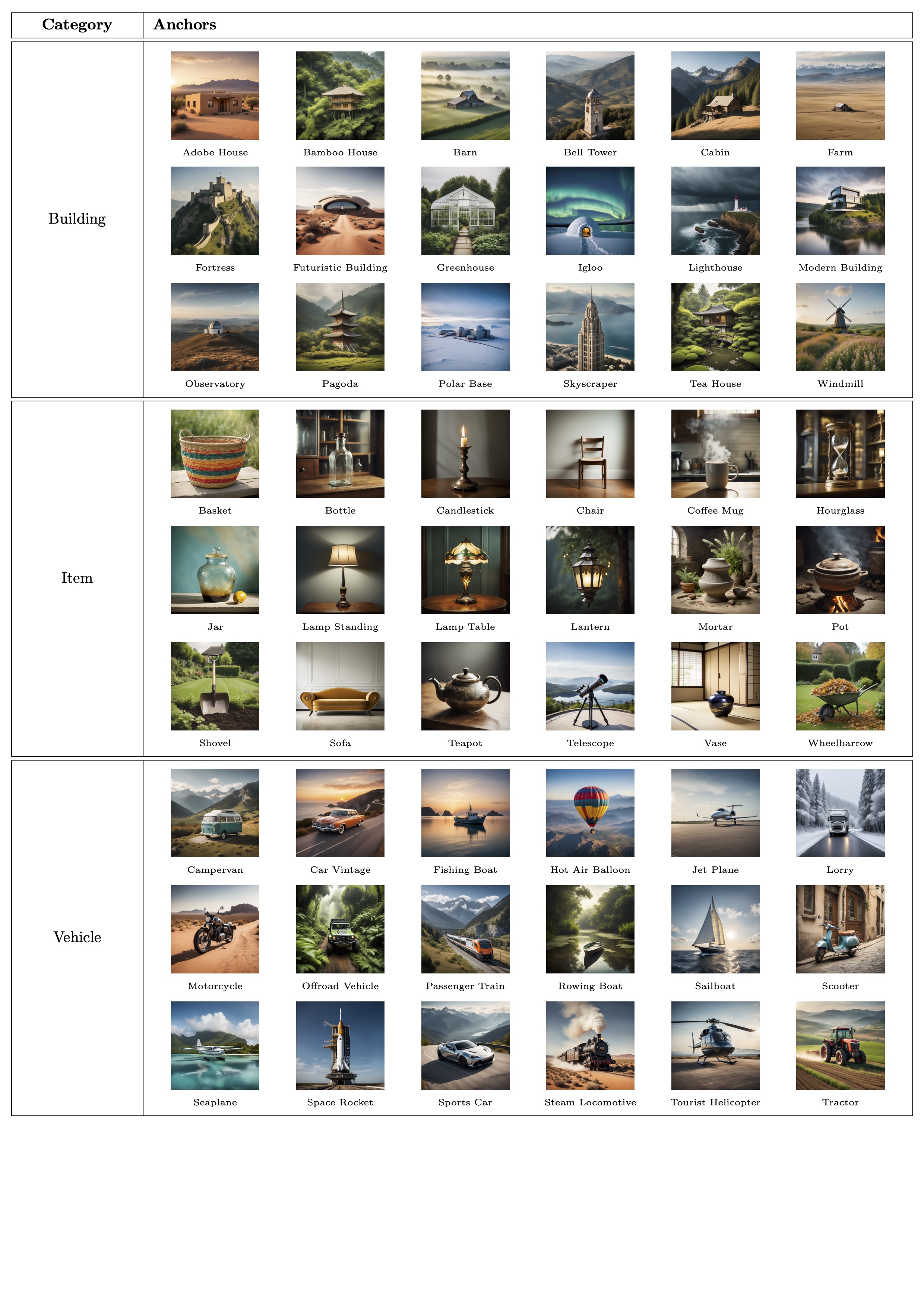}
    \label{fig:table-anchor-images-2}
\end{figure}

In the first steps (S1-S2), our procedure generated 60 object scenes for 18 objects taken out of six categories (shown in Figure~\ref{fig:stage_1} for the example object scene "fish"). After the next steps (S3-S5) our procedure generated an ordered set of 10 images for each object scene based on the LPIPS, a machine learning model of perceptual similarity~\parencite{zhang_unreasonable_2018}. Specifically, each set comprised 10 slightly different realisations of the same object (see Figure~\ref{fig:stage_1}, bottom left for "fish"). Within each object set the images formed a perceptual continuum. Interpolation at the level of ``noise latents'' with a fixed text prompt successfully generated image sets where each image represented a distinct realisation of the same semantic content (see Appendix~\ref{appendix:image-generation}). Figure~\ref{fig:table-anchor-images-1} shows examples for each of the 18 objects in the \textcolor{black}{three natural categories}, and Figure~\ref{fig:table-anchor-images-2} shows examples for each of the 18 objects in the three artificial categories. Out of the 60 exemplar images created for each object we selected an anchor and a guide image (see Appendix~\ref{appendix:selection-anchors}). Figures ~\ref{fig:table-anchor-images-1} and ~\ref{fig:table-anchor-images-2} show the anchor images that were used as a starting point for the subsequent interpolation. 

As in previous studies~\parencite{cao2024synartifactclassifyingalleviatingartifacts}, we implemented a systematic process to exclude images with artefacts, resulting in a 3.3\% exclusion rate (Figure~\ref{fig:exclusion_fig_exemplars}). Generating smooth transitions was challenging, likely due to the non-linear nature of the high-dimensional space in which the interpolations were performed. Even though increasing the interpolation steps did make the transitions smoother, we occasionally observed perceptual “jumps” up to 1000 interpolation steps. Visual inspection revealed that most interpolated images were of high quality. A small proportion (12.4\%) of images had to be replaced because it exhibited artefacts  (see Appendix~\ref{appendix:image-generation}; Figure~\ref{fig:swapped_interpols}; Table~\ref{tab:exclusion-rates-exemplars}). We also checked whether there were any overall differences in model-based similarity (LPIPS) scores across the different categories. We found that only \textit{buildings} and \textit{items} deviated in their scores (for full details, see Appendix~\ref{appendix:pair-selection-lpips}). Furthermore, we assessed physical properties of the anchor images and found they were  similar across categories (Figures~\ref{fig:img_physical_characteristics} and \ref{fig:img_physical_characteristics_by_category} in Appendix~\ref{appendix:physical-properties}). 

\begin{figure}[htbp]
    \centering
    \caption{Table showing the categories and objects in the stimulus set. The categories in this table are classified as \textit{natural}. For each object, the \textit{anchor image} is shown.}
    \includegraphics[width=\textwidth]{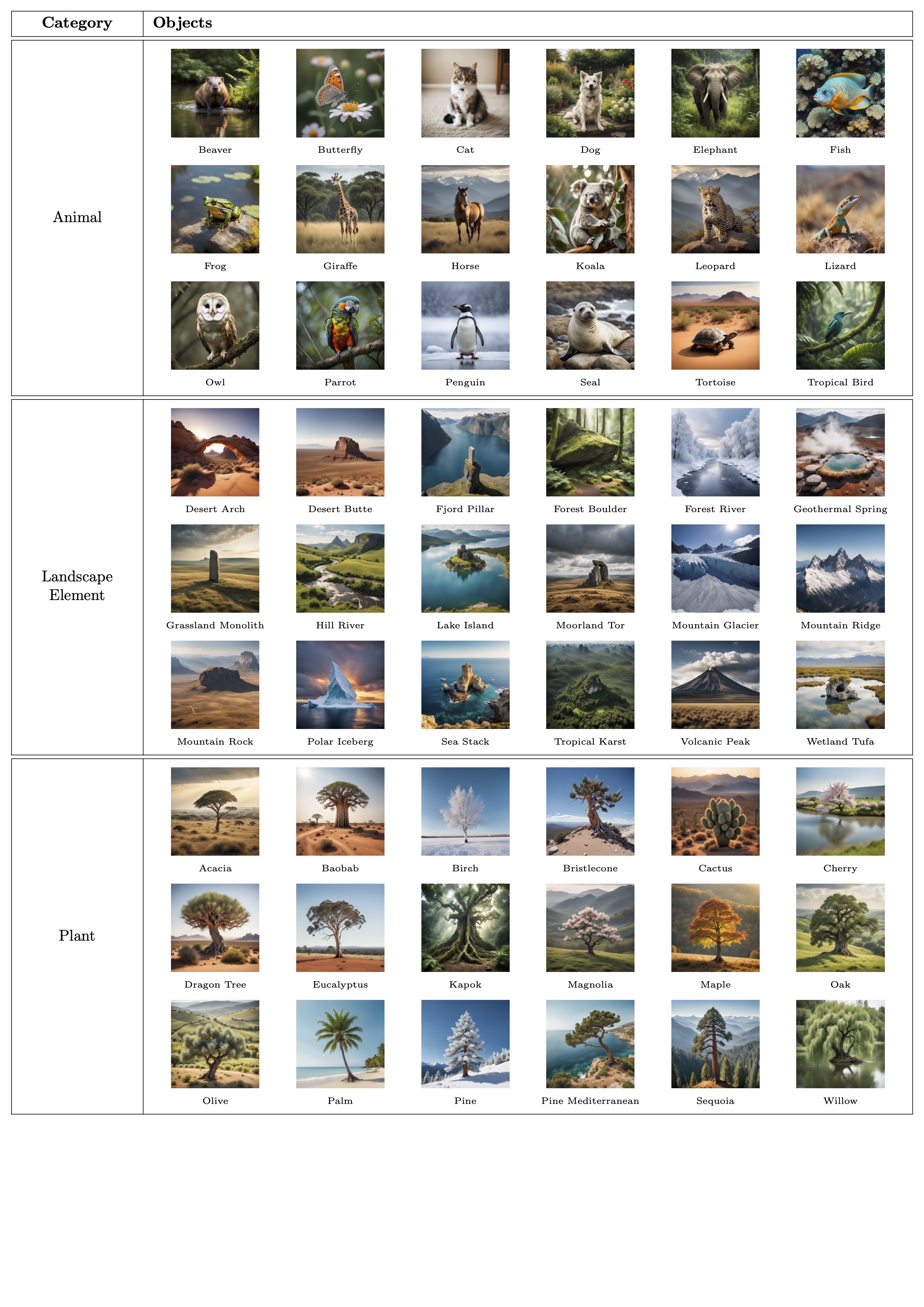}
    \label{fig:table-anchor-images-1}
\end{figure}

\subsection{Stage 2: Similarity judgement task}

\begin{figure}[htbp]
\centering
\includegraphics[width=0.9\textwidth]{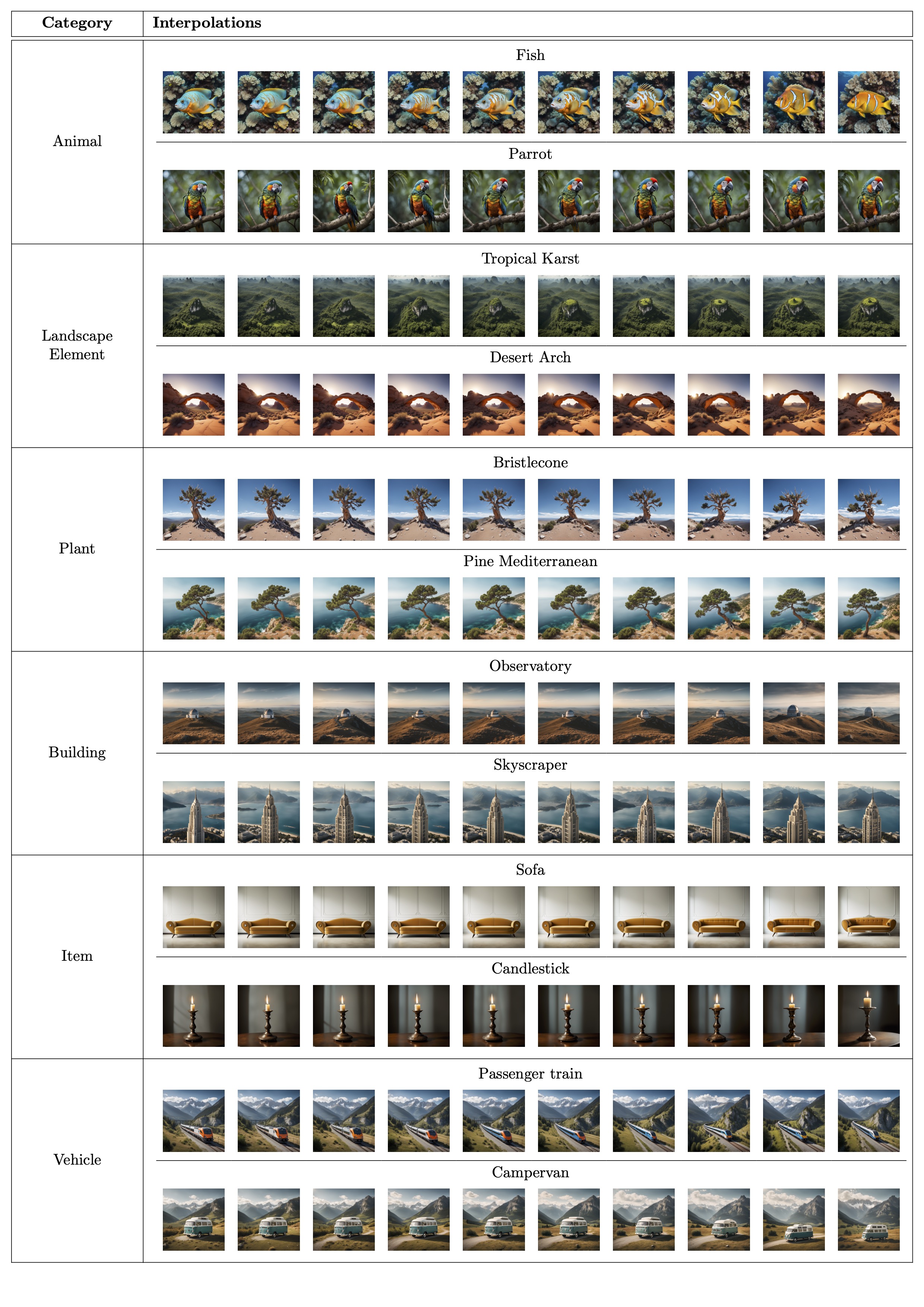}
\caption{
    Examples of the final ordered image sets based on the psychophysical crowd-sourcing task. The figure shows the series for 2 out of 18 exemplars of each category. The anchor image is on the left. 
}
\label{fig:final_stimulus_set}
\end{figure}

To validate the LPIPS-based ordering of each object set, we ran an online crowdsourced psychophysical similarity task using triplet comparisons (Figure~\ref{fig:stage_2}a, for details about the task cf. Appendix~\ref{appendix:similarity_ratings}). These judgements were used to estimate a one-dimensional embedding representing the perceived similarity scale within each object set. We assessed to which degree the ordering derived from human judgements aligned with the order based on the LPIPS score. The matrix shown in Figure~\ref{fig:stage_2}b reports how often images occupied the same rank position within an object set across both methods. Overall, we observed strong alignment between the two orderings, with a Spearman rank correlation of $\rho = 0.73$. Figure~\ref{fig:stage_2}c shows the output of this stage for the example object scene "fish". Figure~\ref{fig:final_stimulus_set} displays the final image sets for two example objects from each of the 18 categories, ordered according to psychophysical judgements. Additional comparisons between the original LPIPS-based orderings and those adjusted using behavioural responses are shown in Figure~\ref{fig:example_swaps} (Appendix~\ref{appendix:reordering}).

\begin{figure}[htb]
    \centering
    \includegraphics[width=\linewidth, keepaspectratio]{figures/stage_2_v-2.jpg}
    \setlength{\abovecaptionskip}{10pt plus 3pt minus 2pt}
    \caption{Psychophysical validation (Stage 2). (a) We conducted an online similarity judgement task using triplet comparisons to confirm and potentially fine-tune the order for each object set. The x-axis shows the ten images, and the y-axis shows the psychophysical embedding score. Relative distances on this embedding score correspond to perceptual differences. (b) Alignment between the LPIPS-based order and the embedding-based order derived from human similarity judgements. Each cell indicates the number of images assigned to the corresponding rank combination by both metrics. Higher counts along the diagonal indicate stronger agreement between LPIPS distances and perceived similarity. (c) Final image set ordered by embedding values. For the object "fish", LPIPS distances fully matched human similarity judgements.}
    \label{fig:stage_2}
\end{figure}

\subsection{Stage 3: Memory task}

The mean accuracy for 240 participants across six blocks for Repeated and Non-repeated conditions at three difficulty levels (Easy, Medium, and Hard)  is shown in Figure \ref{fig:stage_3}. Accuracy systematically varied with task difficulty, showing a clear graded pattern: performance was highest in the Easy condition, intermediate in the Medium condition, and lowest in the Hard condition. Across all difficulty levels, mean accuracy for Repeated targets started at the same level of Non-repeated targets, to then outperform it in later blocks. The graded accuracy distinctions were preserved throughout learning. We fitted a Bayesian hierarchical model to quantify these graded effects of task difficulty and stimulus repetition across blocks. All parameters and derived quantities showed robust convergence ($\hat{R}$ close to $1.00$; bulk-ESS$> 8000$; tail-ESS$> 15,000$; see diagnostic plots in Figure~\ref{fig:diagnostic-panel}, Appendix~\ref{appendix:memory_experiment_results}). Posterior parameter estimates are provided in Table~\ref{tab:bayesian_model_results} and visualised in Figure~\ref{fig:parameter-estimates} in Appendix~\ref{appendix:memory_experiment_results}.
A more detailed description of the modelling result is provided in Appendix~\ref{appendix:memory_experiment_results}.

\begin{figure}[htbp]
    \centering
    \includegraphics[width=\linewidth]{figures/stage_3_v-2.jpg}
    \setlength{\abovecaptionskip}{10pt plus 3pt minus 2pt} 
    \caption{Results of the Memory Validation Task (Stage 3). (a) Example image sets showing a target alongside three foils for four distinct object scenes (from top to bottom: Mediterranean pine, lake island, observatory, chair). (b) Schematic of a trial in the delayed match-to-sample paradigm. (c) Mean accuracy for repeated and non-repeated targets. (d) Mean accuracy across difficulty levels, grouped by block and target condition. Error bars indicate standard error of the mean (SEM).}
    \label{fig:stage_3}
\end{figure}

\section{Discussion}\label{discussion}

\subsection{Perceptual scaling and similarity judgement task}

The interpolation procedure successfully generated sets of images with a gradient of perceptual similarity and without impacting the semantic meaning across image variations excessively. The use of text prompts and text-to-image model allowed for very high flexibility in the choice of categories. Keeping the text prompt constant while interpolating between the starting noise latents proved to be an effective and controlled approach for manipulating the  perceptual characteristics of a scene while maintaining its semantic meaning. 

Our procedure synthesised a set of high resolution naturalistic object stimuli from different categories that are systematically ordered along a psychophysically validated perceptual continuum. They could prove useful for many cognitive tasks that involve measurements on perceptual continua, such as change detection or working memory tasks ~\parencite{zhang_discrete_2008}. In comparison to scraping images from the internet~\parencite{hebart_things_2019}, or using pre-trained GANs~\parencite{son_scene_2022}, using a text-to-image model in combination with psychophysical ordering also showed extreme versatility in semantic contents and high image quality. This method could be used to generate new images and categories, tailoring them to specific experimental needs.

\textcolor{black}{Previous image sequences and data sets have typically used considerably different exemplars~\parencite{greene2009recognition, krizhevsky_imagenet_2012, lin_microsoft_2014, hebart_things_2019}. In contrast, our stimulus set is primarily intended for paradigms that aim for local confusability and threshold-level variations, as exemplified in our memory validation study. For each object-scene we generated a set of image variations that (i) can be placed on a monotonic perceptual continuum, (ii) preserve the same semantics (e.g., “picture of a cat on a carpet”), and (iii) differ only in low-level perceptual details. Please note that our approach in its current form does not aim to control the specific dimensions along which an image is varied. For synthetically generated images as here this can be achieved by using either extensions of diffusion models~\parencite{gandikotaConceptSlidersLoRA2024} or GANs~\parencite{goodfellow2014generativeadversarialnetworks}.} \textcolor{black}{One possible future application could be the augmentation of existing datasets. For example, one could use the semantic categorisation of the THINGS dataset~\parencite{hebart_things_2019}, which was implemented using a data-driven method, to generate a higher number of exemplars for each category. Another possible application is the extension of work on natural scene properties~\parencite{greene2009recognition}, for example by implementing the possibility of varying specific feature dimensions~\parencite{gandikotaConceptSlidersLoRA2024}.}

\textcolor{black}{Our object-centred layout deliberately placed a single, prominent object at the centre of the image. This design choice preserves clear semantics and limits contextual clutter, in line with object-centric datasets such as THINGS~\parencite{hebart_things_2019} and neuroimaging work using ImageNet~\parencite{chang2019bold5000, gong2023largescale}. While central placement exploits the well-documented oculomotor ‘centre bias’ in scene viewing~\parencite{tatler2007central}, it also limits the stimulus set’s applicability. Our stimulus set is in fact designed for research on object-based attention, perception and memory, rather than for complex multi-object scenes~\parencite{greene2009recognition}.  Future work could extend the pipeline to generate scene variants to study cognitive functions such as naturalistic gaze behaviour for complex scenes.}

The LPIPS metric~\parencite{zhang_unreasonable_2018} played a central role as a quantifiable prediction of perceptual similarity and enabled us to efficiently manage and screen a large set of images before fine-tuning the ordering based on psychophysics. In general, LPIPS showed a good alignment with the similarity judgements from the online task and proved to be a valid first approximation. One limitation of our approach is the absence of a comparable baseline similarity across object scenes. Although we found the score to be reliable and aligned with human perception within a single object scene (e.g. indicating a graded increase in perceptual dissimilarity), the metric magnitude across object scenes was less interpretable. Future work could address this limitation by accounting for similarity scores not only within but also across object scenes.

\textcolor{black}{One limitation of the present pipeline is that both the high‑level categories (natural vs artificial) as well as the specific objects (e.g., “butterfly,” “sailboat”) were chosen \textit{a priori} by the experimenters, leaving room for selection bias. To address the subjectivity inherent in prompt selection, future work could combine generative image synthesis with data‑driven concept‑selection frameworks to represent semantic spaces more uniformly. Another question is whether similar text prompts could also be used to identify and retrieve images from web search engines. Compared to such web-based retrieval of existing images, we believe the use of image synthesis offers several key advantages: First, it allows to flexibly generate arbitrary stimuli that might not be available on the internet~\parencite{podell_sdxl_2023}. Second, it provides a fully parametrized framework that allows images to be generated with variable number and similarity, beyond what is available in image databases~\parencite{hebart_things_2019, krizhevsky_imagenet_2012, lin_microsoft_2014}.}

As is common for AI-generated images~\parencite{cao2024synartifactclassifyingalleviatingartifacts}, we also identified artefacts, which exhibited high variability across categories. Although the exclusion rates were generally low, some objects were more prone to the presence of artefacts than others. This variability can be attributed to several factors. A first possibility is that there were differences in the availability of training data for different objects~\parencite{Yang2023}. Another possibility is the challenge of finding effective prompts~\parencite{liu2023designguidelinespromptengineering}, given the virtually infinite number of possibilities. Finally, the levels of familiarity vary across images, which may make artefacts of more familiar objects more noticeable. The category with the highest incidence of artefacts was \textit{animals}, followed by \textit{items} and \textit{vehicles}. For example, \textit{giraffe} images had the highest exclusion rate, often due to inaccuracies such as an incorrect number of legs, which is a very salient feature~\parencite{kamali2025characterizingphotorealismartifactsdiffusion}. Also, there might be tighter normative constraints of features for certain objects, for example, the number of legs for a giraffe is fixed, whereas the number of leaves on a plant might be more variable. These are all known problems in generative AI, although models are quickly improving~\parencite{cao2024synartifactclassifyingalleviatingartifacts}. \textcolor{black}{In contrast, artefacts were less frequently identified in images of plants, landscape elements, and buildings (which are less normatively constrained). Although we did not investigate the cognitive implications of these category-specific differences empirically, such variability may be informative for future studies investigating models of human and AI cognition~\parencite{lu2023seeingbelievingbenchmarkinghuman, huang2024analysishumanperceptiondistinguishing, kamali2025characterizingphotorealismartifactsdiffusion}. In particular, some categories might have very rigid norms for specific features (e.g., the fixed number of extremities in animal anatomy) and pose greater challenges for generative models than less constrained categories (e.g., trees, where the number of branches is not so constrained). Future work could examine the alignment between generative models and humans in how visual features are represented and integrated across semantic categories. From a practical perspective, category-dependent differences should be explicitly considered when selecting target images in line with experimental goals. Object scenes with smoother interpolations may be preferable for tasks focused on graded perceptual continua or memory precision, whereas object scenes with structurally constrained features may be more suited for studies investigating feature-binding processes or anomaly detection.} Another limitation of our approach is that the exclusion criteria were rather subjective and future work could use crowd-sourcing platforms or artefact classifiers~\parencite{cao2024synartifactclassifyingalleviatingartifacts} to tackle the problem systematically.

\subsection{Memory validation task}

We validated our stimulus set in an additional behavioural experiment, investigating how the precision of scene encoding in visual working memory (vWM) varies as a function of proximity along our modelled continua. We observed a clear gradient with more similar images being memorized better. Despite the fact that performance increased with repeated experience with the objects (see Figure~\ref{fig:stage_3}, bottom right), the similarity-related gradient was largely maintained. VWM performance is known to benefit from contributions of LTM, where aspects such as meaning, familiarity, and previous exposure strengthen memory representations~\parencite{Xie2018, Schurgin2020, Brady2022}. In paradigms rooted in Hebbian principles~\parencite{Hebb1961}, repeated exposure reinforces stimulus-specific traces in both vWM and LTM, enabling sequential recall improvements through synaptic reinforcement~\parencite{Johnson2019, Mizrak2022, Souza2022}. Continuous feature-report tasks show that, under repeated exposure, LTM can support high-fidelity representations with precision comparable to that of vWM~\parencite{Brady2013, Miner2020}. \textcolor{black}{Note that performance improved across all difficulty levels even if object scenes were not repeated. We interpret the overall pattern as the joint expression of (i) a stimulus-specific component (repetition-based strengthening of long-term memory traces for individual scenes) and (ii) a stimulus-nonspecific component (general adaptation to the task)~\parencite{sagiPerceptualLearningVision2011}. Future studies could further minimise the nonspecific effects by expanding the practice phase until accuracy stabilises across a broad subset of object-scenes.}

Our results align with previous findings linking latent space distances to working memory performance~\parencite{son_scene_2022, bates_scaling_2024}, but extend the investigation to a broader range of object categories. The systematic manipulation of perceptual similarity allowed us to directly assess how prior exposure strengthens LTM representations and facilitates recognition, particularly in high-interference conditions where targets and foils were most confusable. This level of experimental control was essential for isolating the contribution of LTM to WM, a distinction typically challenging to capture with naturalistic images due to their inherent variability and complexity.

\subsection{Concluding remarks}
 
The use of generative models for stimulus synthesis has the potential to set a new standard in cognitive research by enabling the creation of customised, naturalistic stimuli and psychometrically ordered stimuli tailored to the needs of cognitive experiments. These models can be of great help in tackling the trade-off between ecological validity and experimental control. Here, we introduced a novel approach using diffusion-based generative models, specifically Stable Diffusion XL, in combination with flexible text prompts to synthesise a stimulus set of high resolution, photorealistic scenes with graded perceptual differences and consistent semantic content. 

Our three-step validation procedure demonstrated that our stimuli effectively captured a perceptual continuum, first quantified objectively using the LPIPS metric~\parencite{zhang_unreasonable_2018} and subsequently confirmed through human psychophysical judgements collected in an online task. Moreover, this perceptual scaling was further validated in a visual working memory task, where recognition accuracy decreased systematically as perceptual similarity between targets and foils increased. Performance was highest in the Easy condition and progressively lower in the Medium and Hard conditions, where foils more closely resembled the target. This pattern validates that proximity along the perceptual continua affects working memory performance in a graded way. 

By providing the stimulus set publicly (see Data Availability section), we hope to contribute a valuable tool for the research community, bridging the gap between ecological validity and experimental control in visual cognition studies. These stimuli are particularly well-suited for addressing key challenges in visual cognitive research, including change detection, memory, and object recognition.

\section*{Declarations}

\subsection*{Funding}
LP was supported by the Max Planck Society and by the German Federal Ministry of Education and Research (Bundesministerium für Bildung und Forschung). JDH was supported by the Deutsche Forschungsgemeinschaft (DFG, Exzellenzcluster Science of Intelligence; SFB 940 “Volition and Cognitive Control”; and SFB-TRR 295 “Retuning dynamic motor network disorders using neuromodulation”).

\subsection*{Conflicts of interest/Competing interests}
The authors declare no competing interests.

\subsection*{Ethics approval}
The studies were approved by the Ethics Committee of the Institute of Psychology of the Humboldt University Berlin, Germany.

\subsection*{Consent to participate}
All participants provided informed written consent prior to participation in the study, in accordance with institutional guidelines.

\subsection*{Consent for publication}
Participants were informed that anonymised data may be published in scientific journals and provided consent for publication.

\subsection*{Availability of data and materials}
The code, stimulus set (including images and metadata), and behavioural data from the online task will be made available on OSF at the following link: \url{https://osf.io/pf4tv}. 

\subsection*{Code availability}
Custom code will be made available via GitHub upon publication.

\subsection*{Authors' contributions}
Leonardo Pettini, Carsten Bogler, and John-Dylan Haynes conceived the study and designed the experimental paradigms. Leonardo Pettini collected and analysed the data, and wrote the first draft of the manuscript. Carsten Bogler, John-Dylan Haynes, and Christian Doeller provided supervision, conceptual input, and revisions. All authors reviewed and approved the final version of the manuscript.

\subsection*{Acknowledgements}
We thank Joram Soch for helpful comments on the manuscript.

\newpage

\begin{refcontext}[sorting=nyt]
\printbibliography
\end{refcontext}

\clearpage

\appendix

\setcounter{secnumdepth}{3}
\section{Extended Methods}\label{appendix:methods}

\subsection{Image generation}\label{appendix:image-generation_methods}

In the next section, we provide detailed information about the model and its parameters. We will then describe the criteria used to generate exemplar images, followed by an explanation of how fine-grained image variations were obtained through interpolation between initial noise latents. A schematic overview of the procedure is shown in Figure~\ref{fig:stage_1} (main text).

\subsubsection{Model}\label{appendix:generative_model}
Text-to-image diffusion models such as Stable Diffusion~\parencite{rombach2022high} have gained considerable interest in the past years. They rely on two fundamental components to generate images: a starting noisy latent and a textual prompt~\parencite{Yang2023}. The noisy latent provides an initial random state, which the model progressively refines through a series of denoising steps, outputting an image. The textual prompt guides the model, conditioning the denoising process to generate images that align with the described content. Several text-to-image models are available on the market. We opted for Stable Diffusion XL (SDXL)~\parencite{podell_sdxl_2023} for two main reasons: output quality and customizability. In comparison to previous versions, SDXL relies on a larger UNet backbone and excels at generating high-quality images, especially at high resolutions (1024x1024 pixels). Moreover, the SDXL architecture incorporates two text encoders instead of one (OpenCLIP ViT-bigG and the original CLIP ViT-L text encoder), creating a richer textual conditioning for the diffusion process. Stable Diffusion XL benefits from a two-step generation process involving a base model and a refiner model, known as the ensemble of expert denoisers~\parencite{balaji_ediff-i_2023}: the base model generates a latent image, which is then fed into the refiner model. The refiner model completes the denoising process, enhancing the details of the image. In SDXL, the refiner model can be used to improve output quality of the fully-refined image. The base and refiner model pipelines were instantiated from pretrained model weights, which were downloaded from the Hugging Face repository\footnote{The base model was downloaded from \url{huggingface.co/stabilityai/stable-diffusion-xl-base-1.0}. The refiner model was downloaded from \url{huggingface.co/stabilityai/stable-diffusion-xl-refiner-1.0}.}. Images were generated using 50 inference steps for the base model to fully denoise the image and an additional 15 steps for the refiner model to improve detail quality. The default invisible watermark was added to each image to ensure it is marked as AI-generated. The ensemble of expert denoisers was integrated using a pre-trained Low-Rank Adaptation (LoRA)~\parencite{hu_lora_2021} model\footnote{The LoRA model was downloaded from: \url{https://huggingface.co/ffxvs/lora-effects-xl/blob/main/xl_more_art-full_v1.safetensors}.} to further improve the quality and detail of the generated images. Unlike the refiner model, which improves detail quality through additional inference steps, LoRA modifies the base model weights by adding a low-rank update matrix, allowing for efficient fine-tuning.

We used the Hugging Face diffusers 0.26.02 Python library, alongside standard libraries such as PyTorch 2.0.1~\parencite{ansel_pytorch_2024}, which provides high usability and customizability. We ran the model on the GPU cluster at the Bernstein Center for Computational Neuroscience, Charité-Universitätsmedizin Berlin, Germany. The cluster included four NVIDIA GeForce RTX 3090 GPUs (24GB each). To ensure efficient utilisation of the GPU resources, we distributed the image generation tasks across the available GPUs using Pytorch Multiprocessing. Each GPU processed a single image independently.

\subsubsection{Generation pipeline: design and diffusion of exemplar images}\label{appendix:generation_pipeline}

Generating images from textual prompts has the unprecedented advantage of letting users create any possible composition using natural language, but it also presents the problem of how to narrow down the prompt space to generate a finite and usable number of stimuli. We established the following criteria to generate our stimuli: first, they should have a clear subject, situated in the centre and foreground (e.g. a picture of a cat sitting on the carpet); second, they should be situated in a coherent and realistic scene (e.g. an animal in its natural habitat); finally, there should be no human figures in the scenes. 

Moreover, we divided the main subjects into six broad categories, three natural (“animals”, “plants”, “landscape elements”) and three artificial (“items”, “buildings”, “vehicles”). For each category, we defined a set of 18 objects (e.g. “cat”, “olive tree”, “sailboat”, “sofa” etc.), which we chose with the aim of creating a relatively diverse sample (e.g. animals from diverse biospheres, different kinds of landscape etc.). All categories and corresponding objects are listed in Table~\ref{tab:categories_object_overview}. For each object, we generated 60 exemplars using the same prompt, but varying the starting noise tensor. 

The structure of textual prompts was designed to balance two main aspects. On the one hand, the resulting images should vary across categories and be rich in detail. On the other hand, the whole stimulus set should have some consistency in terms of image layout. Therefore, we used relatively specific prompts, but with a shared main structure across the sample. With few exceptions, prompts started with a declaring the main subject (e.g. “film solitary bristlecone pine” or “photo of a barn owl” ), followed by information about its position (e.g. “in the foreground”, “in the centre”) and about the background or scene it is situated in (e.g. “rocky terrain and blue sky in the background”), to finally describe the image style (“high resolution photography”) and some additional attributes (e.g. “cinematic”, “bright light”, “beautiful”). In order to avoid the undesired addition of human figures, we also used a negative prompt (“people, person, human figures, human body parts, humans”), which was constant across all generations. We checked that this worked as intended by visual inspection. Using this procedure, for each object we generated a set of 60 various but semantically homogenous exemplars, maintaining its layout and general composition fairly consistent. 

We initialised the starting noise latents of each exemplar using unique seed values (for 60 exemplars and 108 objects, 6480 seeds in total). Seeds were used to create a PyTorch Gaussian noise tensor with dimensions (1, 4, 128, 128), where 128 is determined by dividing the target resolution (1024) by the Variational Autoencoder (VAE) scale factor of 8 used in Stable Diffusion XL. Storing the seeds ensured the reproducible initialization of the noise latents, making it possible to generate the same exact image at later stages of the pipeline, as well as to interpolate between noise latents.

After generation, we visually inspected all images and excluded those with prominent artefacts (such as an animal with too many legs), especially in central positions. The reason behind this is that they can be distracting and therefore make the stimulus less suitable for a cognitive scientific experiment. However, we can provide these images upon reasonable request to researchers who may be interested in using them for further studies, such as those focusing on human perception of AI-generated artefacts.

\renewcommand\tabularxcolumn[1]{m{#1}} 
\begin{table}[H]
\centering
\begin{tabularx}{\textwidth}{>{\raggedright\arraybackslash}p{3cm} X}
\toprule
\textbf{Category} & \textbf{Objects} \\
\midrule
animal & beaver, butterfly, cat, dog, elephant, fish, frog, giraffe, horse, koala, leopard, lizard, owl, parrot, penguin, seal, tortoise, tropical\_bird \\
\midrule
building & adobe\_house, bamboo\_house, barn, bell\_tower, cabin, farm, fortress, futuristic\_building, greenhouse, igloo, lighthouse, modern\_building, observatory, pagoda, polar\_base, skyscraper, tea\_house, windmill \\
\midrule
item & basket, bottle, candlestick, chair, coffee\_mug, hourglass, jar, lamp\_standing, lamp\_table, lantern, mortar, pot, shovel, sofa, teapot, telescope, vase, wheelbarrow \\
\midrule
landscape\_element & desert\_arch, desert\_butte, fjord\_pillar, forest\_boulder, forest\_river, geothermal\_spring, grassland\_monolith, hill\_river, lake\_island, moorland\_tor, mountain\_glacier, mountain\_ridge, mountain\_rock, polar\_iceberg, sea\_stack, tropical\_karst, volcanic\_peak, wetland\_tufa \\
\midrule
plant & acacia, baobab, birch, bristlecone, cactus, cherry, dragon\_tree, eucalyptus, kapok, magnolia, maple, oak, olive, palm, pine, pine\_med, sequoia, willow \\
\midrule
vehicle & campervan, car\_vintage, fishing\_boat, hot\_air\_balloon, jet\_plane, lorry, motorcycle, offroad\_vehicle, passenger\_train, rowing\_boat, sailboat, scooter, seaplane, space\_rocket, sports\_car, steam\_locomotive, tourist\_helicopter, tractor \\
\bottomrule
\end{tabularx}
\caption{Overview of the six categories and their corresponding objects: three natural (animals, plants, landscape elements) and three artificial (items, buildings, vehicles), each with 18 diverse objects.}
\label{tab:categories_object_overview}
\end{table}

\subsubsection{Selection of the anchor images}\label{appendix:selection-anchors}
In order to create variations of increasing dissimilarity from the exemplar image, we set up a procedure\footnote{We did not use the Hugging Face \textit{StableDiffusionImg2ImgPipeline}, which allows a reference image to influence the generation at various denoising steps, resulting in variations that incorporate elements of the reference image with adjustable strength. We decided against this option, because we observed that the control over the degree of perceptual similarity between the output images and the reference image can be rather inconsistent. 
} relying on the interpolation of the noise latents corresponding to a selected image pair. Each image pair included two anchor images, an \textit{anchor image} and a \textit{guide image} for the interpolation. These images were selected according to the following criteria:
\begin{enumerate}
    \item The chosen images should be representative for the generated batch (e.g. no outliers).
    \item They should be relatively similar to one another in terms of composition and perceptual features, avoiding changes in orientation or scene layout as much as possible.
\end{enumerate}

Such criteria sought to optimise the interpolation smoothness (in the next step) while minimising our manual interference with the process. In order to satisfy both conditions, for each object we identified two images based on their relative perceptual similarity within the generated set. To quantify their perceptual similarity, we used the Learned Perceptual Image Patch Similarity (LPIPS) metric~\parencite{zhang_unreasonable_2018}. The LPIPS metric quantifies the perceptual differences between images by comparing deep features extracted from a pre-trained neural network and it has been shown to correlate well with human judgement~\parencite{zhang_unreasonable_2018}. For a few test images, we compared metrics using features from the SqueezeNet~\parencite{iandola_squeezenet_2016}, AlexNet~\parencite{krizhevsky_imagenet_2012}, and VGG~\parencite{simonyan_very_2015} architectures. The results across these models were comparable. Therefore, we opted to use SqueezeNet due to its computational efficiency and speed. We first computed the pairwise similarity of each image within the set. Then, for each image, we computed its mean LPIPS score against all other images in the set to assess its representativeness. The assumption behind this approach is that an image with a lower mean LPIPS score is, on average, more perceptually similar to the other images in the set. This higher similarity makes it a better representative of the overall set's perceptual characteristics. The first image of the pair (the \textit{anchor image}) was the image with the highest overall similarity to the rest of the set, that is, with the lowest mean LPIPS score. To select the \textit{guide image}, we used the following procedure: first, we ranked images based on their mean LPIPS scores, and considered the images with the five lowest scores (excluding the score of the \textit{anchor image}), which amounted to the top 8\%-10\% of the image set, depending on how many images had been excluded. We then ranked the images within this subset by their similarity with the \textit{anchor image}, selected the most similar as the \textit{guide image} and other ones as backup images. We then visually inspected the selected pairs. If the \textit{guide image} had a significant visual disparity from the \textit{anchor image} (e.g., opposite orientation), we selected the next most similar image as a backup. This process continued, if necessary, to the third most similar image and so forth, to ensure the pair maintained a high degree of perceptual coherence while maintaining the selection as objective as possible. In this way, we made sure that the interpolation would be computed between two images that are both representative of the set and similar to each other.

\subsubsection{Interpolation algorithm}\label{appendix:interpolation_algorithm}
\begin{figure}[H]
    \centering
    \begin{subfigure}[b]{0.48\linewidth}
        \caption{}
        \centering
        \includegraphics[height=5.5cm]{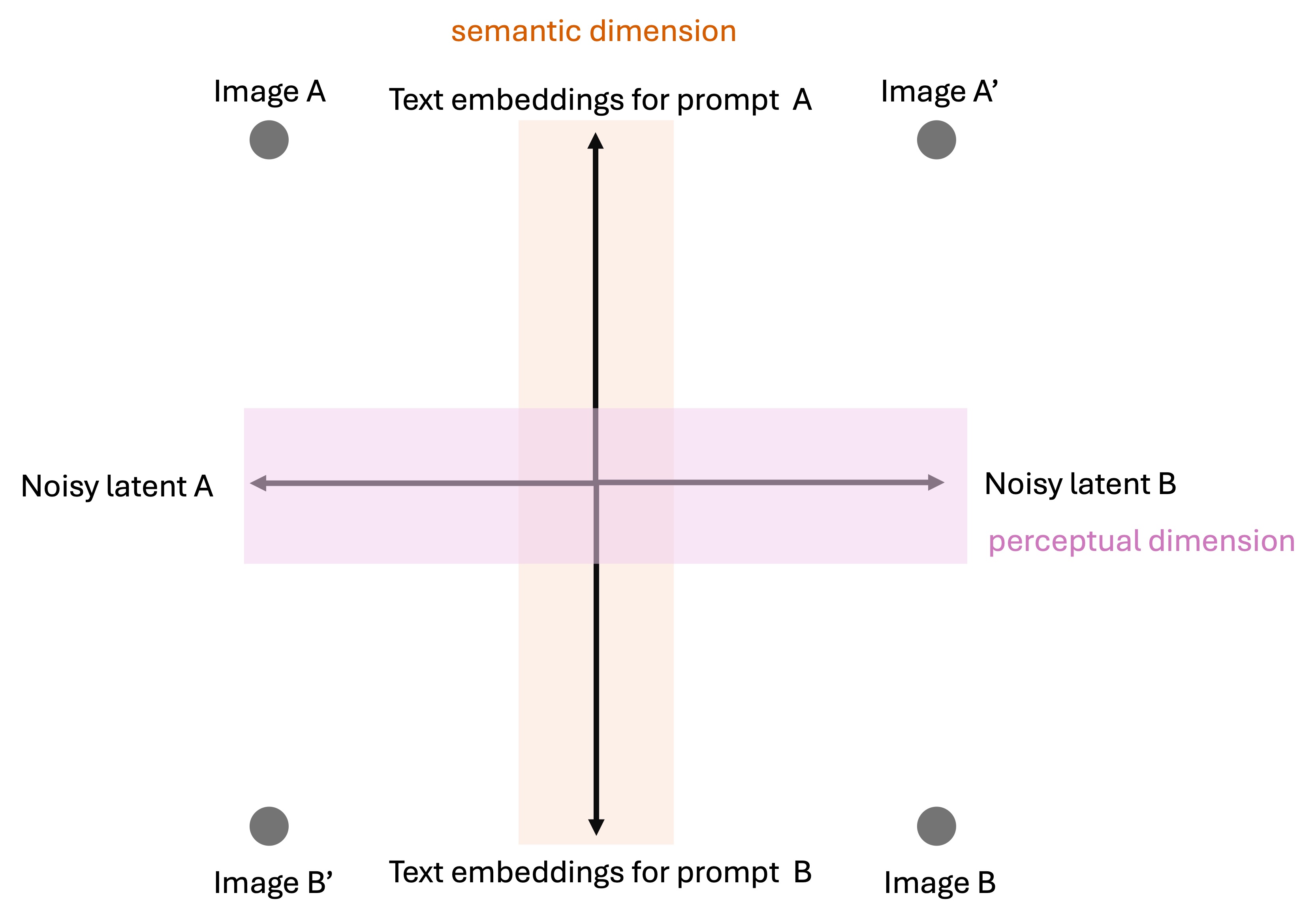}
    \end{subfigure}%
    \begin{subfigure}[b]{0.48\linewidth}
        \caption{}
        \centering
        \includegraphics[height=5.5cm]{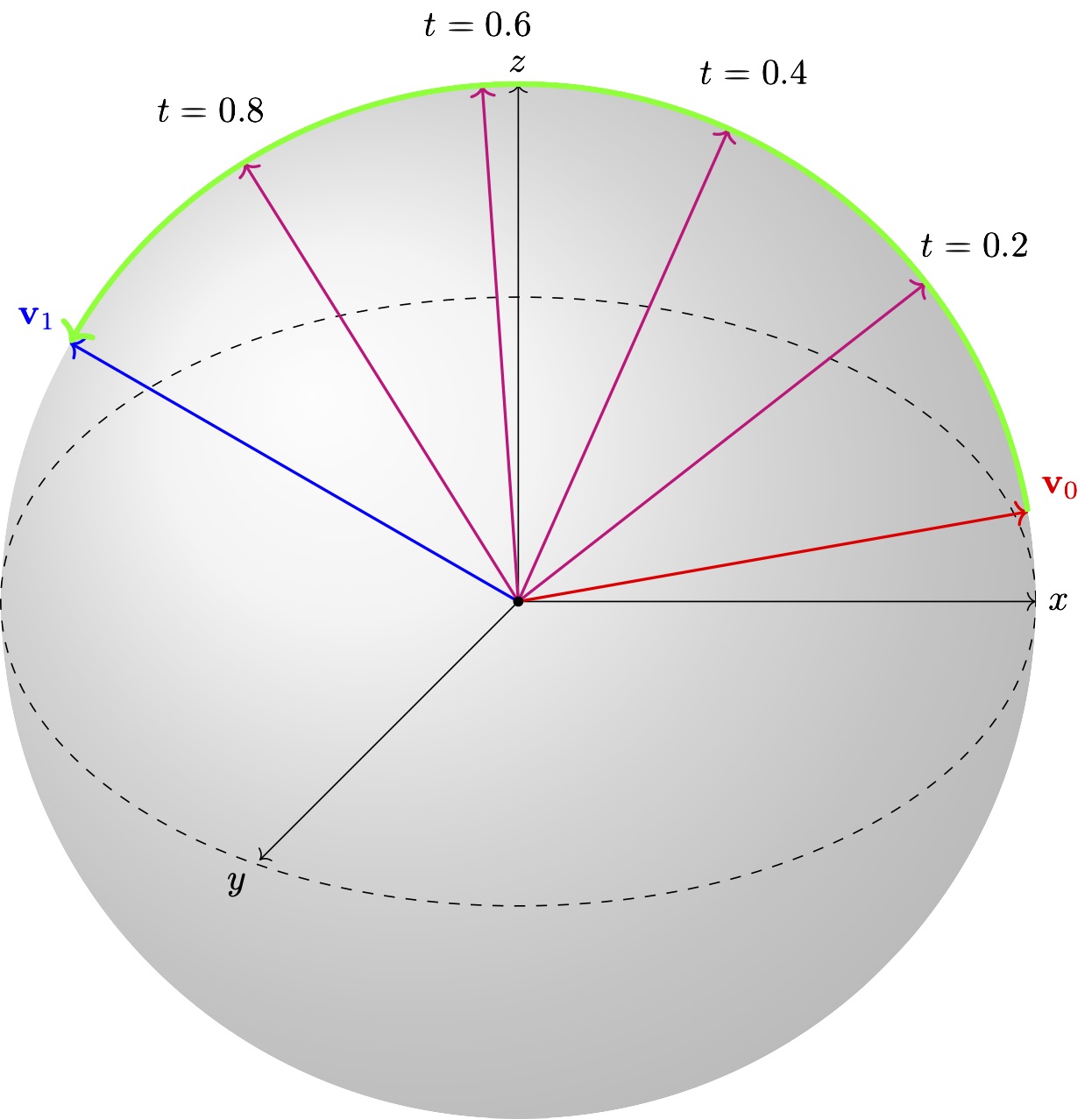}
    \end{subfigure}
    \vspace{3pt}
    \caption{a) Schematic representation of the two possible spaces in which the interpolation can be operated. It is possible to interpolate either between the text embeddings (encoded by the CLIP models) or the initial noisy latents. Interpolation between text embeddings will result in significant semantic variations (e.g. between Image A and Image B, or B'), whereas interpolations between noisy latents will mostly involve perceptual variations, keeping the meaning of the image fairly constant (e.g. between Image A and A' or Image B and B'). It should be noted that this is a simplification because semantic variations cause perceptual variations and vice versa. The aim here is to highlight the main dimension of variation. Moreover, to have fine-grained control of the scene layout, methods relying on models like ControlNet should be used. b) A three-dimensional simplification of the spherical linear interpolation (slerp) between two vectors $v_0$ and $v_1$. In this study, the interpolation is applied between $128 \times 128 \times 4 \times 1$ tensors.}
    \label{fig:interpolation_space}
\end{figure}

Once the two anchor images were selected, we generated the interpolations between the \textit{anchor} and the \textit{guide image} using their starting noise latents. In contrast to GANs, where the interpolation between two latent vectors corresponds to smooth image transitions, direct interpolation in diffusion is less straightforward. The interpolation between images can be operated either in the prompt space (i.e., between text embeddings), between the starting noise latents or both (Figure~\ref{fig:interpolation_space}). Since the goal of this work was to maintain the semantic aspects of the images fairly constant, while changing their perceptual features, we interpolated between noise latents while maintaining the text embeddings constant\footnote{Interpolating in prompt space results in a semantic interpolation between two images (e.g. interpolating between ``picture of a cat'' and ``picture of a dog'' will result in intermediate pictures morphing from a cat into a dog). Interpolating between noise tensors while keeping the prompt constant (e.g. ``picture of a cat sitting on the carpet'') results in the interpolation between two exemplars of the same semantic meaning (a cat sitting on a carpet). Depending on how specific the prompt is and on model parameters such as the guidance scale, which determines how much the diffusion process will adhere to the textual prompt, other minor semantic variations might occur. In order to control for this, it is important to engineer the textual prompts carefully. However, for relatively specific textual prompts, the main semantic meaning of the image will remain fairly constant.}. Spherical linear interpolation (slerp) is particularly suitable for interpolating high-dimensional data as it preserves the length of the vectors and ensures smooth transitions~\parencite{shoemake_animating_1985}. Given two vectors \(\mathbf{v}_0\) and \(\mathbf{v}_1\) and an interpolation parameter \(t\) where \(0 \le t \le 1\), slerp can be defined as:

\[
\text{Slerp}(\mathbf{v}_0, \mathbf{v}_1; t) = \frac{\sin[(1 - t) \theta]}{\sin[\theta]} \mathbf{v}_0 + \frac{\sin[t \theta]}{\sin \theta} \mathbf{v}_1
\]

where \(\theta\) is the angle between \(\mathbf{v}_0\) and \(\mathbf{v}_1\), determined using the arccosine function on the normalized dot product between the vectors:

\[
\theta = \arccos\left(\frac{\mathbf{v}_0 \cdot \mathbf{v}_1}{\|\mathbf{v}_0\| \|\mathbf{v}_1\|}\right)
\]

We first defined 200 interpolation values \(t\) or “steps”, which are linearly spaced between 0 and 1. The number of interpolation steps determines the granularity of this transition. We chose 200 steps as a compromise to balance between computational time and smoothness of transition. Feeding the interpolation steps in the slerp function, we generated a series of intermediate noise latents that transition between the noise latents that were used to synthesize the \textit{anchor image} and the \textit{guide image}. We then used SDXL to denoise each interpolated latent and generate intermediate images. Importantly, we diffused the images iteratively in two sequential batches. In the first batch, we used the PyTorch generator initialised with the seed from the \textit{anchor image} and diffused each intermediate noise tensor up to the midpoint. In the second batch, we used the PyTorch generator initialised with the seed from the \textit{guide image} and diffused each intermediate noise tensor back to the midpoint. This dual-batch approach helped mitigate small variations or inconsistencies in the final output images, making sure that the diffusion process was equally controlled by both seeds. 

\subsubsection{Selection of the interpolated images}\label{appendix:selection_interpol_images}
For each object, we selected ten variations, the \textit{anchor image} and nine of its interpolations. To do so, we again used the LPIPS metric to compute the perceptual similarity between the \textit{anchor image} and every interpolated image. We defined target similarity scores, linearly spaced between the minimum and maximum LPIPS values observed\footnote{In most cases, but not always, this was the \textit{guide image}. We used this method to avoid large perceptual discrepancies as much as possible between the interpolations and the \textit{anchor image}}. For each target score, we selected the image whose similarity score was closest to the target score, ensuring the absence of duplicates. This involves calculating the absolute differences between the similarity scores and the target scores, and selecting the closest matches. One option could have been defining the variations in terms of interpolation steps (e.g. generating 10 interpolation steps in the previous step). Our approach, however, presents two advantages. First, the interpolation process in high-dimensional latent spaces can result in complex, non-linear changes in perceptual similarity. Consequently, simply selecting images based on fixed intervals of interpolation steps might not accurately represent a linear increase in dissimilarity from the \textit{anchor image}. Our approach thus partially compensates for the non-linear perceptual changes by focusing on the actual observed similarities, rather than the potentially misleading interpolation steps. Secondly, the image generation process is susceptible to the introduction of artefacts. Since we wanted to make sure that the final image set maintained a high quality and, at the same time, a perceptual gradient, we conducted a final visual inspection to identify and exclude images with noticeable artefacts. If artefacts were detected, we selected immediate neighbours with very similar LPIPS scores as replacements. Therefore, generating a large number of interpolated images allowed us to exclude interpolated images with significant artefacts in favour of neighbouring images, without affecting the perceptual gradient significantly and minimising the interference of our own subjective judgement.

\subsection{Perceptual Similarity Judgement Task}\label{appendix:similarity_ratings}

We used the LPIPS metric to pre-select ten images per object because it aligns reasonably well with human perceptual judgement~\parencite{zhang_unreasonable_2018}, and having participants evaluate all 108 objects with 200 images each would have been experimentally challenging. Although LPIPS provides a good approximation, we conducted a crowdsourcing task to experimentally validate the pre-selected images and adjust their ordinal positioning if needed.

\subsubsection{Stimuli}\label{appendix:smilarity_judgement_stimuli}
The final stimulus set included a total of 1,080 images, comprising 18 objects from 6 categories, with 10 variations per object (see above for the selection procedure). Images were  downscaled to 512x512 to decrease potential latencies from the client side, while maintaining high quality.

\subsubsection{Similarity judgement task using triplet comparisons}\label{appendix:smilarity_judgement_triplets}
Participants performed a similarity judgement task based on triplet comparisons, or “method of triads”~\parencite{torgerson_multidimensional_1952}, which is used both in psychophysics and machine learning~\parencite{aguilar_comparing_2017, demiralp2014learning, haghiri_estimation_2020, kunstle_estimating_2022, li_extracting_2016, wichmann_methods_2017}. Given a set of stimuli \( S = \{s_1, s_2, \ldots, s_n\} \), where \( n \) is the total number of stimuli, participants are asked to judge the similarity of three stimuli at a time. In particular, given a triplet of stimuli \( (s_i, s_j, s_k) \), they are asked which between two probe stimuli \( s_j \) and \( s_k \) is most similar to a reference stimulus \( s_i \). From the set of stimuli \( S \), the number of all possible unique triplets is determined by the binomial coefficient:

$$
\binom{n}{3} = \frac{n!}{3!(n-3)!}
$$

However, since each element of the triplet can be a reference stimulus, the total number of possible triplet questions is obtained by multiplying the binomial coefficient by three:

$$
C = 3 \cdot \binom{n}{3}
$$

Since we planned to let participants assess the similarity between each combination of variations of the same object, but not across objects, in our case \( n \) was 10 (total number of image variations for each object) and the total number of triplet combinations \( C \) was 360.

Prior to data collection, we ran a set of simulations to estimate a) the number of triplet judgements that we would need to estimate object-specific psychophysical similarity functions robustly, and b) a target sample size of participants to judge each object. Previous work has shown that for \( n \) objects and \( d \) perceptual embedding dimensions, a subset of \( 2dn \log(n) \) triplets should be sufficient to reconstruct the embeddings with small error~\parencite{haghiri_estimation_2020, jamieson2011low, jain_finite_2016}, which for \( n=10 \) and \( d=1 \) amounts to \( 2 \cdot 1 \cdot 10 \log(10) = 20 \log(10) \approx 46 \) triplets. Following the simulation results and expecting a relatively high noise for naturalistic images that are perceptually very similar, we estimated that a subset of 144 triplet judgements for one object from each single participant (40\% of the total possible triplet combinations \( C \)) and at least 20 participants for each object would be necessary. The subset of triplet combinations was randomly sampled for each participant independently. Because of experimental constraints, i.e. not making the online task too long, each participant was assigned two objects. The assignment of the objects to participants was randomised.

\subsubsection{Task procedure}\label{appendix:smilarity_judgement_procedure}
\begin{figure}[h!]
    \centering
    \includegraphics[width=0.8\textwidth]{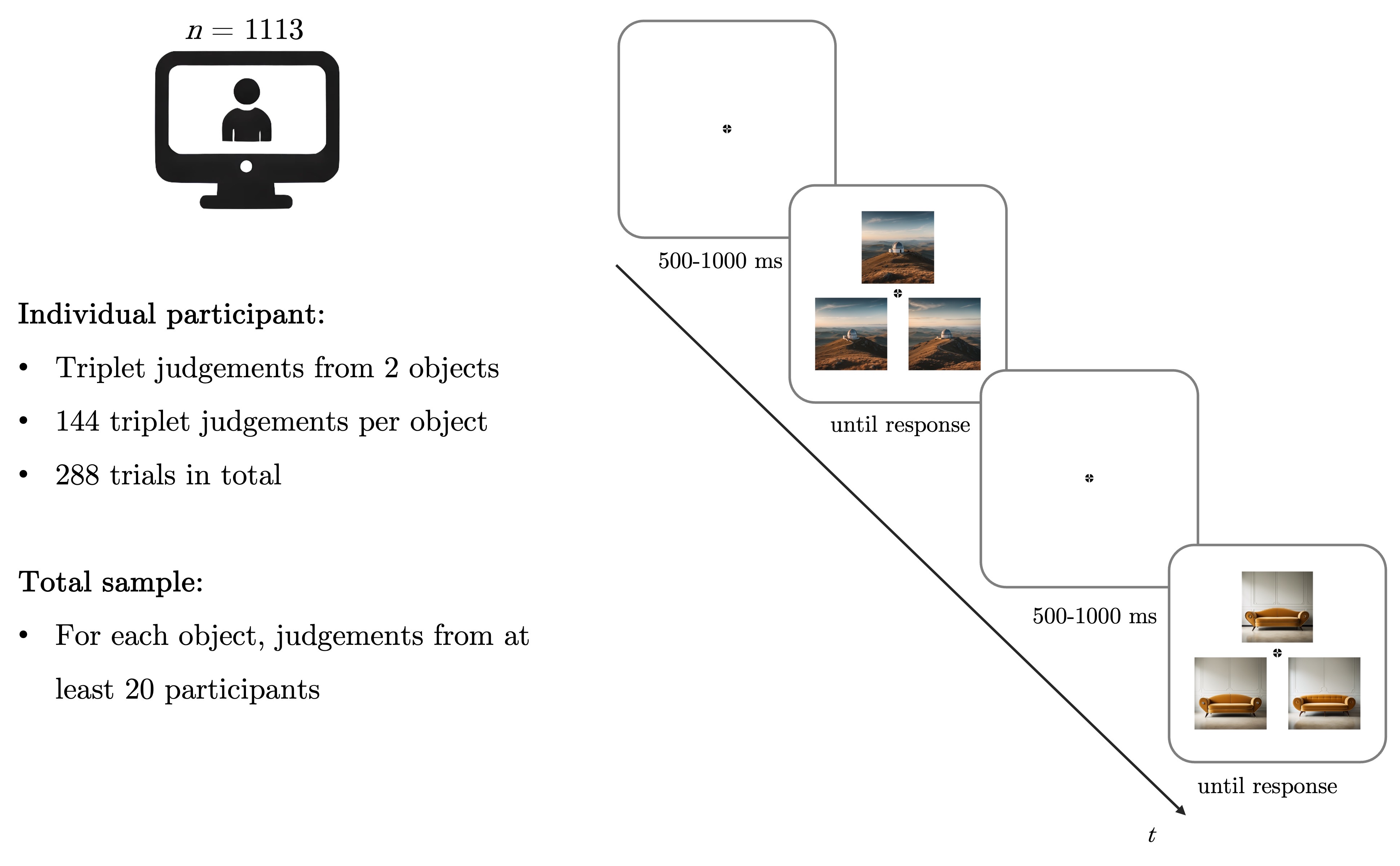}
    \caption{Online task design with two example trials of the similarity judgement task (for \textit{observatory} and \textit{sofa}). Participants were tasked to indicate which of the two bottom images is most similar to the top image. Each participant was assigned two objects and the triplet comparisons were always between variations of the same object.}
    \label{fig:online_task_design}
\end{figure}

After providing informed consent, participants received detailed instructions on the task procedure and had to pass an attention check to make sure they understood the instructions. Before starting the main task, they underwent a training session where they performed 12 trials using an independent set of stimuli. The main task consisted of 288 trials divided in 6 blocks of 48 trials each. In between blocks, participants were prompted to take a short break (maximum 2-3 minutes) to maintain attention and accuracy. Each trial began with a fixation target~\parencite{thaler_what_2013} presented at the centre of the screen for a jittered duration between 500 and 1000 ms, which was randomized in 100 ms steps. Then, participants were shown three images: one reference image on top and two probe images below (see Figure~\ref{fig:online_task_design}). Their task was to judge which of the two probe images was more similar to the reference image. Participants made their choices by pressing the “f” and “j” keys to select, respectively, the left or the right probe. They were asked to respond within 5 seconds and – if they did not – after their response they were prompted to respond more quickly. The opacity of the selected image was reduced to 0.5 for 1000 ms, giving participants visual feedback that their response had been registered. At the end of the experiment, participants were given the option to complete a debriefing questionnaire about their experience. They were asked for their overall impressions of the study, whether they encountered any technical problems, if they used any specific strategies during the experiment, and for any additional feedback they might have.

\subsubsection{Experimental control, data quality, and exclusion criteria}\label{appendix:control_quality_exclusion_criteria}
\begin{figure}[H]
    \centering
    \includegraphics[width=\linewidth]{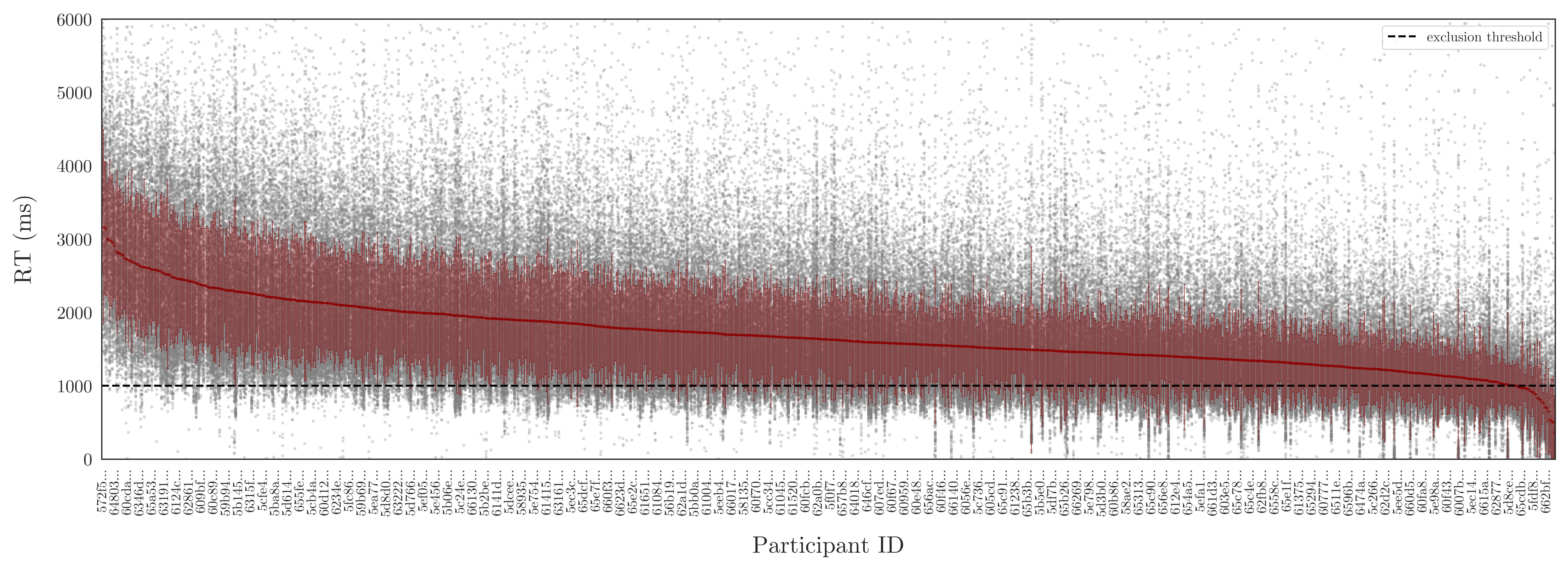}
    \caption{Distribution of reaction times for all participants, ordered by mean reaction time. The 1000 ms cutoff threshold is indicated by the dashed line.}
    \label{fig:reaction_times_all}
\end{figure}

\begin{figure}[H]
    \centering
    \includegraphics[width=0.6\linewidth]{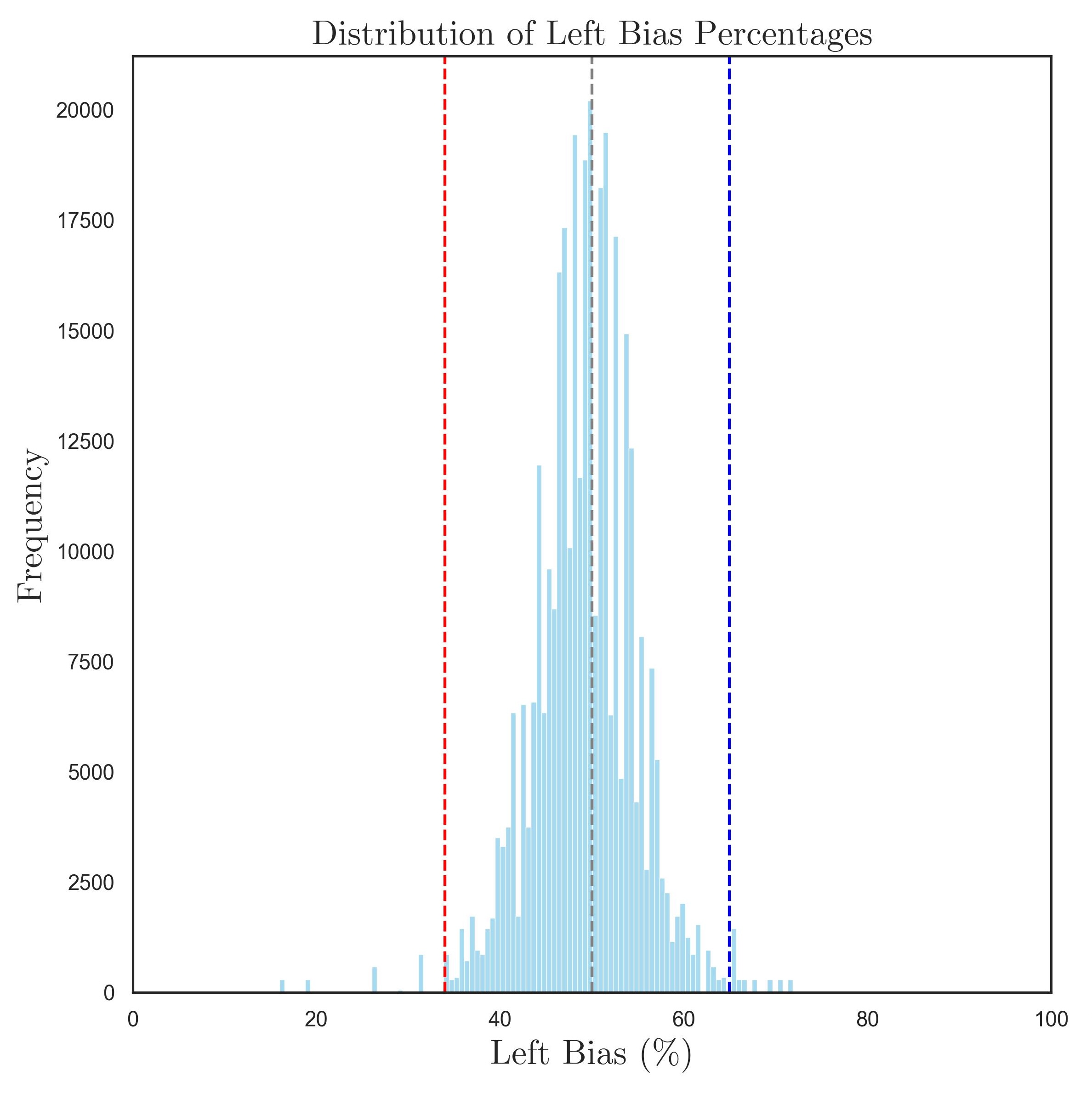}
    \caption{Distribution of left bias percentages across participants. The dashed blue and red lines indicate the thresholds for extreme biases, while the central black dashed line represents the expected 50\% bias.}
    \label{fig:response_bias_distributions}
\end{figure}

A key drawback of online studies compared to on-site experiments is the lack of experimental control, but there are several strategies to maintain data quality~\parencite{rodd_moving_2024}. We implemented both experimental measures during data collection and an analysis pipeline with sanity checks on the collected data. In order to ensure a relative consistency in the experimental environment, participants were required to use a desktop device, the Google Chrome web browser with a minimum screen resolution of 800x600 pixels and to remain in fullscreen mode throughout the experiment. These requirements were enforced through a preliminary device and browser check before the experiment began and by pausing the experiment if participants exited fullscreen mode. Every interaction with the browser (e.g. if participant exited fullscreen) was recorded, in addition to the time spent in each phase of the task (e.g. how much time they spent on each break).  Such data was crucially important to identify and potentially exclude participants who showed unreliable engagement with the experiment. Participants who did not finish the experiment were also excluded. In addition to the interaction data, we used two behavioural indicators to flag participants with outlying performance: reaction times and response biases. We first examined the distribution of RTs (Figure~\ref{fig:reaction_times_all}) and identified these outliers with very short RTs. Any participant with mean RTs below 1000 milliseconds was excluded, assuming that a mean RT below 1000 ms throughout the task would indicate that a participant might have performed the task inattentively. We also examined the distribution of response biases (Figure~\ref{fig:response_bias_distributions}), making the assumption that participants who pressed one of the two keys with high frequency were not fully engaging with the task. We first calculated the proportion of choices for ‘f’ (left probe) and ‘j' (right probe) key presses for each participant, to then exclude participants with biases falling 1.85 times outside the interquartile range bounds. Finally, we conducted an analysis of participant feedback provided at the end of the experiment to identify potential technical problems or other issues encountered during the study. 

\begin{figure}[H]
    \centering
    \includegraphics[width=\linewidth]{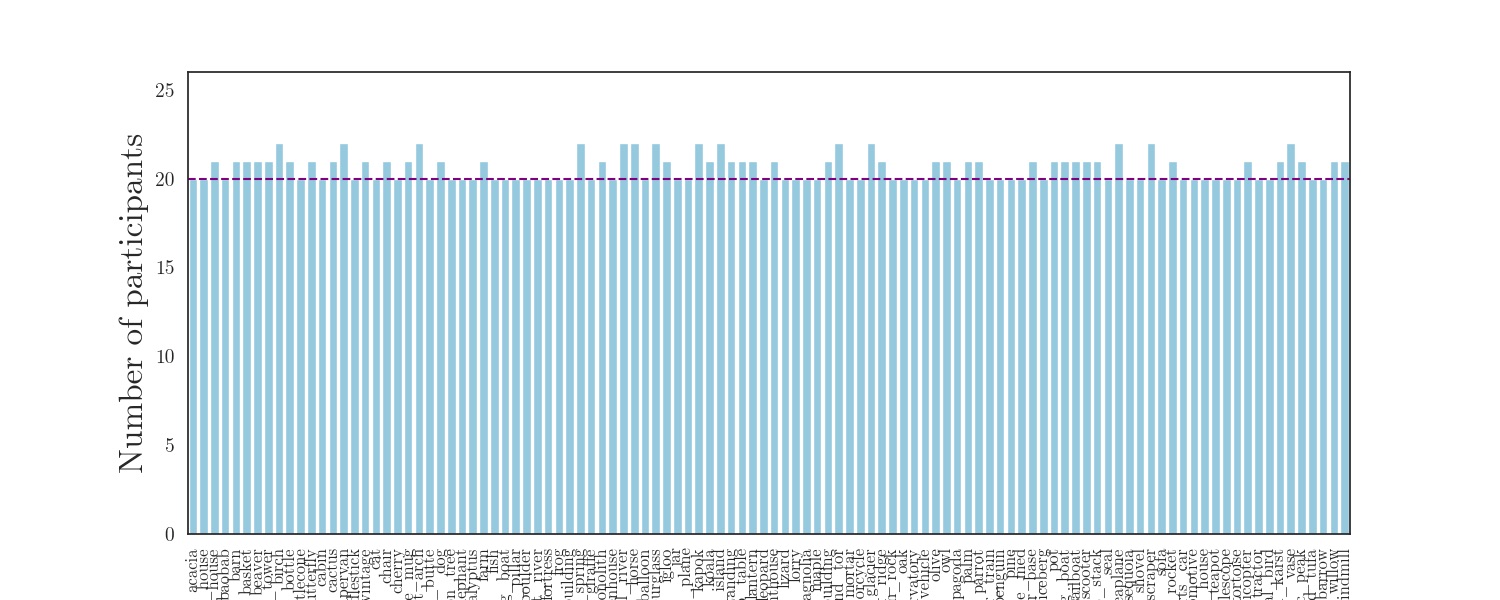}
    \caption{Participant count for each object.}
    \label{fig:participant_count_for_objects}
\end{figure}

Of 1253 participants who participated in the experiment, 140 (11\%) were excluded either because they did not complete all trials (92 participants, 7.3\%) or because they did not satisfy the inclusion criteria based on reaction times (36 participants, 2.9\%) and response biases (20 participants, 1.6\%), with 4 (0.3\%) participants satisfying more than one exclusion criterion. A final sample of 1113 participants passed the criteria and was used for the analysis. The final sample had an average age of 28.9 years (SD = 5.7). The majority of participants were male, accounting for 60\% of the sample, while 39.6\% were female. A small proportion, 0.4\%, chose not to disclose their gender. Each participant provided judgements only for two objects and each object was assigned to at least 20 participants. The majority of objects (56) were judged by 20 participants, 38 objects were judged by 21 participants, and 14 objects were judged by 22 participants (Figure~\ref{fig:participant_count_for_objects} and Table~\ref{tab:participant_count_for_objects}). Because of the very small difference, we included all data in the final analysis.

\begin{longtable}{lr}
\caption{Number of participants assigned to each object.} \label{tab:participant_count_for_objects} \\

\toprule
object name &
number of participants assigned  \\
\midrule
acacia & 20 \\
adobe\_house & 20 \\
bamboo\_house & 21 \\
baobab & 20 \\
barn & 21 \\
basket & 21 \\
beaver & 21 \\
bell\_tower & 21 \\
birch & 22 \\
bottle & 21 \\
bristlecone & 20 \\
butterfly & 21 \\
cabin & 20 \\
cactus & 21 \\
campervan & 22 \\
candlestick & 20 \\
car\_vintage & 21 \\
cat & 20 \\
chair & 21 \\
cherry & 20 \\
coffee\_mug & 21 \\
desert\_arch & 22 \\
desert\_butte & 20 \\
dog & 21 \\
dragon\_tree & 20 \\
elephant & 20 \\
eucalyptus & 20 \\
farm & 21 \\
fish & 20 \\
fishing\_boat & 20 \\
fjord\_pillar & 20 \\
forest\_boulder & 20 \\
forest\_river & 20 \\
fortress & 20 \\
frog & 20 \\
futuristic\_building & 20 \\
geothermal\_spring & 22 \\
giraffe & 20 \\
grassland\_monolith & 21 \\
greenhouse & 20 \\
hill\_river & 22 \\
horse & 22 \\
hot\_air\_balloon & 20 \\
hourglass & 22 \\
igloo & 21 \\
jar & 20 \\
jet\_plane & 20 \\
kapok & 22 \\
koala & 21 \\
lake\_island & 22 \\
lamp\_standing & 21 \\
lamp\_table & 21 \\
lantern & 21 \\
leopard & 20 \\
lighthouse & 21 \\
lizard & 20 \\
lorry & 20 \\
magnolia & 20 \\
maple & 20 \\
modern\_building & 21 \\
moorland\_tor & 22 \\
mortar & 20 \\
motorcycle & 20 \\
mountain\_glacier & 22 \\
mountain\_ridge & 21 \\
mountain\_rock & 20 \\
oak & 20 \\
observatory & 20 \\
offroad\_vehicle & 20 \\
olive & 21 \\
owl & 21 \\
pagoda & 20 \\
palm & 21 \\
parrot & 21 \\
passenger\_train & 20 \\
penguin & 20 \\
pine & 20 \\
pine\_med & 20 \\
polar\_base & 21 \\
polar\_iceberg & 20 \\
pot & 21 \\
rowing\_boat & 21 \\
sailboat & 21 \\
scooter & 21 \\
sea\_stack & 21 \\
seal & 20 \\
seaplane & 22 \\
sequoia & 20 \\
shovel & 20 \\
skyscraper & 22 \\
sofa & 20 \\
space\_rocket & 21 \\
sports\_car & 20 \\
steam\_locomotive & 20 \\
tea\_house & 20 \\
teapot & 20 \\
telescope & 20 \\
tortoise & 20 \\
tourist\_helicopter & 21 \\
tractor & 20 \\
tropical\_bird & 20 \\
tropical\_karst & 21 \\
vase & 22 \\
volcanic\_peak & 21 \\
wetland\_tufa & 20 \\
wheelbarrow & 20 \\
willow & 21 \\
windmill & 21 \\
\bottomrule
\end{longtable}

\subsubsection{Similarity judgement analysis}\label{appendix:similarity_judgement_analysis}
The main goal of the similarity judgement analysis was to estimate a perceptual scale for each object and its variations, given a set of perceptual judgements. To analyse the triplet judgements, we used three algorithms: Maximum Likelihood Difference Scaling (MLDS)~\parencite{maloney_maximum_2003}, which is well-established scaling method in psychophysics, and two ordinal embedding algorithms, Soft Ordinal Embedding (SOE)~\parencite{terada_local_2014} and t-Distributed Stochastic Triplet Embedding (t-STE)~\parencite{van_der_maaten_stochastic_2012}. MLDS can be used only in one-dimensional cases, whereas t-STE and SOE have been proposed to find ordinal embeddings in higher dimensional spaces~\parencite{haghiri_estimation_2020}. Embeddings, in this context, are Euclidean representations that preserve the ordinal relationships among data points based on a set of perceptual triplet judgements~\parencite{agarwal2007generalized, jamieson2011low}. Given a set of triplet constraints \(\mathcal{T} = \{ (i, j, k) \}\), derived from perceptual judgements where each triplet \((i, j, k)\) indicates that image \(i\) is perceived as more similar to image \(j\) than to image \(k\), an embedding is a set of points \(\{ x_i \in \mathbb{R}^d \}\), where \(d\) denotes the dimensionality of the embedding space, such that:
\[
\| x_i - x_j \| < \| x_i - x_k \|, \quad \forall\, (i, j, k) \in \mathcal{T}.
\]

Because perceptual judgements are noisy and have inconsistencies, it may not be possible to satisfy all triplet constraints at the same time. For this reason, the embedding \(\{ x_i \in \mathbb{R}^d \}\) is obtained by solving an optimization problem that minimizes a loss function over the set of triplet constraints~\parencite{agarwal2007generalized, jamieson2011low}. 

We performed a comparative analysis between the three algorithms, assessing both their performance and the stability of the embedding estimates.
To assess their performance, we used the cross-validated triplet error \(E_t\)~\parencite{haghiri_estimation_2020}. Given a set of embedding representations \( y_1, \ldots, y_n \) estimated on a triplet set \( T \), the triplet error measures how many triplets from a validation set \( T' \) are not correctly represented by those embeddings. It is defined as follows~\parencite{haghiri_estimation_2020}:

$$
E_t = \frac{1}{|T'|} \sum_{t=(i,j,k) \in T'} \mathds{1}\{R_t \cdot \text{sgn}(\|\hat{y}_i - \hat{y}_k\|^2 - \|\hat{y}_i - \hat{y}_j\|^2) = 1\},
$$

where \( \mathds{1} \) is the indicator function that equals 1 if the given triplet does not satisfy the expected distances and equals 0 if the relationship is satisfied. The triplet error thus counts the number of violations where the distance between the reference point \( i \) and point \( k \) is less than or equal to the distance between the reference point \( i \) and point \( j \). We used a Leave-One-Participant-Out cross-validation to evaluate embeddings for each object using the three different estimators. First, the triplet judgement data were grouped by object. For each object group, we divided the triplet judgements into training and test data. In each fold of the cross-validation, data from one participant were left out as the test set \( T' \), while the data from the remaining participants constituted the training set \( T \). This ensured that the model was evaluated on data from a participant it had not seen during training, providing an unbiased estimate of its performance. Moreover, it also provided information about how generalisable the embedding estimates were across participants. For each fold of the cross-validation, we ran 10 iterations of the embedding estimation procedure for the t-STE and SOE algorithms. These methods, which rely on stochastic optimisation techniques like gradient descent, are prone to converging on local optima. To mitigate this, we ran the algorithm with multiple initializations and selected the embedding solutions that best fit the data, using the triplet error (calculated on the training set only) as the metric~\parencite{haghiri_estimation_2020}. This approach increases the likelihood of identifying a more globally optimal solution by sampling various potential embeddings and reducing the impact of any single suboptimal result. Unlike t-STE and SOE, which often require multiple runs to achieve stable results, MLDS uses a maximum likelihood estimation approach that does not require multiple independent runs. Importantly, in each iteration the training data was used both to compute the embeddings and to calculate the triplet error to prevent data leakage. After running 10 iterations, we selected the embedding with the lowest triplet error as best fitting. 

When the embeddings are computed, the scale and orientation of the solutions can vary across different iterations, resulting in inconsistencies such as varying scales and mirror images. To achieve a consistent scale and orientation, the embedding values were subjected to two transformations. First, we normalized them to the [0,1] interval. Second, we standardized the direction of the embeddings by enforcing a descending order using the Theil-Sen estimator \parencite{theil_rank-invariant_1950, sen1968estimates}. The estimator fits a line by calculating the median slope from all pairwise slopes of the data points within a sample, providing a robust measure of the general trend of the data. If the slope of the fitted line was positive, indicating an ascending order, we flipped and translated the embeddings \( (1-y) \) to achieve the desired descending orientation within the [0,1] interval. Importantly, this method does not affect the triplet error calculation, since the relative distances are preserved. This method is suitable for our use case, which involves aligning the embeddings on a one-dimensional scale. For multidimensional embeddings, Generalized Procrustes Analysis or unsupervised methods \parencite{gower_generalized_1975, pmlr-v89-grave19a}.

For each object and category, we calculated the mean cross-validated triplet error of each algorithm by averaging the triplet errors from all cross-validation steps.  The object-wise and category-wise error estimated how well the model performed for individual objects and for categories. We also calculated the overall mean triplet error for each algorithm, which provided a general measure of how they performed across objects.

As a final step, we reordered the images according to the embedding results and compared this order to the one defined by the LPIPS metric. 

\subsubsection{Physical characteristics of the stimuli set}\label{appendix:physical_characteristics_stimuli}

We calculated the physical characteristics of the stimulus set\footnote{Compare with Cooper et al. \parencite{cooper_standardised_2023}}. These measures were calculated only for the anchor images. When necessary, color images were converted to grayscale using the OpenCV \textit{cvtColor} function with the \textit{COLOR\_RGB2GRAY} conversion code. This function implements a color space conversion applying the following weighted sum of the RGB channels~\parencite{opencvOpenCVColor}:

\[
\text{gray\_image} = 0.299 \cdot R + 0.587 \cdot G + 0.114 \cdot B
\]

We calculated the following measures:

\begin{itemize}
    \item \textbf{Contrast:}  Defined as the standard deviation of the pixel intensity values in the normalized grayscale image. This measure captures the variation in intensity within the image and indicates the overall variability in brightness:
    
    \[
    \text{contrast} = \sqrt{\frac{1}{N} \sum_{i=1}^{N} \left( \frac{\text{gray\_image}(i)}{255.0} - \mu \right)^2}
    \]
    
    where \( N \) is the total number of pixels, \(\text{gray\_image}(i)\) is the intensity of the \(i\)-th pixel, and \(\mu\) is the mean intensity of the normalized grayscale image:
    
    \[
    \mu = \frac{1}{N} \sum_{i=1}^{N} \frac{\text{gray\_image}(i)}{255.0}
    \]

    Normalizing by 255, which is the maximum intensity value for an 8-bit grayscale image, scales the pixel intensity values to a range between 0 and 1 and allows for consistent contrast evaluation across different images.
        
    \item \textbf{Mean intensity:} Calculated as the mean pixel intensity of the normalized grayscale image. The luminance was calculated using the formula:
    
    \[
    \text{mean\_intensity} = \frac{1}{N} \sum_{i=1}^{N} \text{gray\_image}(i)
    \]
    
    where \( N \) is the total number of pixels and \(\text{gray\_image}(i)\) is the intensity of the \(i\)-th pixel.

    \item \textbf{Relative Luminance:} Relative luminance takes into account the perception of brightness by considering the different sensitivities of the human eye to red, green, and blue light~\parencite{bt2002parameter}. Calculated by first converting gamma-compressed RGB values to linear RGB values using a 2.2 power curve:
    \[
    R_{\text{lin}} = R'^{2.2}, \quad G_{\text{lin}} = G'^{2.2}, \quad B_{\text{lin}} = B'^{2.2}
    \]
    Then applying the relative luminance formula according to the BT.709 standard \parencite{bt2002parameter}:
    \[
    Y = 0.2126 \cdot R_{\text{lin}} + 0.7152 \cdot G_{\text{lin}} + 0.0722 \cdot B_{\text{lin}}
    \]

    \item \textbf{Colorfulness:} The method for calculating image colorfulness is based on the approach by Hasler and Süsstrunk \parencite{Hasler2003}. This method uses an opponent color space to quantify colorfulness and they show it correlates with human perceptual judgements. Two channels, \(rg\) and \(yb\), represent the differences between the red and green channels, and a combination of the red, green, and blue channels, respectively. These channels are computed as follows:

    \[
    rg = R - G
    \]
    
    \[
    yb = 0.5 \cdot (R + G) - B
    \]
    
    The mean (\(\mu\)) and standard deviation (\(\sigma\)) for these channels are calculated as:
    
    \[
    \sigma_{rgyb} = \sqrt{\sigma_{rg}^2 + \sigma_{yb}^2}
    \]
    
    \[
    \mu_{rgyb} = \sqrt{\mu_{rg}^2 + \mu_{yb}^2}
    \]
    
    The colorfulness metric is then defined as:
    
    \[
    \hat{M}(3) = \sigma_{rgyb} + 0.3 \cdot \mu_{rgyb}
    \]
    
    Here, \(\sigma_{rgyb}\) combines the variability of the \(rg\) and \(yb\) channels, while \(\mu_{rgyb}\) represents their combined mean value, adjusted by a factor of 0.3.

    \item \textbf{Shannon Entropy:} Entropy measures the amount of information or randomness in an image by analyzing the distribution of pixel intensity values. Higher entropy indicates more complexity and variation in pixel intensity values and reflects richness of detail in the image. It was computed using Shannon's formula \parencite{shannon}:
    \[
    S = -\sum p_k \log(p_k),
    \]
    where \( p_k \) represents the probability of a pixel having intensity value \( k \), calculated from the frequency of each intensity value. \( S \) quantifies how unpredictable the pixel intensities are within the image. This calculation was performed using the \textit{skimage.measure.shannon\_entropy} function from the \textit{scikit-image 0.24.0} package.

    \item \textbf{Edge Density:} Edge density is defined as the proportion of edge pixels to the number of pixels in a portion of the image \parencite{4217308}. This measure indicates the amount of structural detail present, with higher edge density suggesting more edges and thus more complexity~\parencite{rosenholtz2007measuring}. Edge density was calculated using the Canny edge detection algorithm~\parencite{cannny1986} and considering all pixels within the image. The Canny algorithm identifies areas of steep intensity change through a multi-stage process and outputs a binary map where edge pixels are marked. The formula for edge density is given by:
    
    \[
    \text{edge\_density} = \frac{N_{\text{edge}}}{W \times H}
    \]
    
    \begin{itemize}
        \item \(N_{\text{edge}}\) is the number of edge pixels identified by the Canny edge detection algorithm.
        \item \(W\) is the width of the image.
        \item \(H\) is the height of the image.
    \end{itemize}

    The algorithm was implemented using the \textit{feature.canny} function from \textit{scikit-image} 0.24.0.

    \item \textbf{Blur Strength:} Blur strength estimates the amount of blurring present in an image. It was calculated using the \textit{skimage.measure.blur\_effect} function, based on the method described by Crete-Roffet et al. \parencite{creteroffet:hal-00232709}, which developed a perceptual blur metric that correlates well with human perception of blur and has been validated through subjective tests. 

    \item \textbf{Sharpness:} Sharpness measures the clarity of edges and fine details in an image. To quantify sharpness, we follow the method outlined by \parencite{903548}, which relies on a Laplacian operator detecting regions of rapid intensity change (edges) by computing the second spatial derivative of the image. The sharpness metric, \(\text{LAP\_VAR}(I)\), is defined as the variance of the Laplacian values. 
    
    First, a Laplacian mask is applied to the image to compute the Laplacian values \(L(m,n)\) at each pixel \((m,n)\). The mask is:
    
    \[
    \text{laplacian\_mask} = \begin{bmatrix}
    0 & -1 & 0 \\
    -1 & 4 & -1 \\
    0 & -1 & 0
    \end{bmatrix}
    \]
    
    This operation enhances the edges by emphasizing regions where there is a significant change in intensity.
    
    Next, the mean of these absolute Laplacian values, \(\bar{L}\), is calculated as follows:
    
    \[
    \bar{L} = \frac{1}{NM} \sum_{m=1}^{M} \sum_{n=1}^{N} |L(m,n)|
    \]
    
    where \(M\) and \(N\) are the dimensions of the image, and \(|L(m,n)|\) represents the absolute value of the Laplacian at pixel \((m,n)\).
    
    The sharpness metric, \(\text{LAP\_VAR}(I)\), is then computed as the variance of these Laplacian values around the mean:
    
    \[
    \text{LAP\_VAR}(I) = \sum_{m=1}^{M} \sum_{n=1}^{N} \left( |L(m,n)| - \bar{L} \right)^2
    \]
    
    In this formula:
    \begin{itemize}
    \item \(L(m,n)\) is the Laplacian value at the pixel \((m,n)\).
    \item \(\bar{L}\) is the mean of the absolute Laplacian values.
    \end{itemize}
    
    High variance in these values indicates a sharp image with well-defined edges, whereas low variance suggests a blurred image with less distinct edges.

    \item \textbf{Spectral energy:} Image statistics vary depending on scene content and viewing conditions~\parencite{torralba2003statistics}. Second-order statistics such as the power spectrum, which describe the relationships between pixel intensities, are strongly correlated with scene scale (the spatial extent and distance of scene elements) and scene category (e.g., natural or man-made environments). Analyzing the spectral energy of images helps in understanding low-level features of the stimuli~\parencite{cooper_standardised_2023}. Using the Natural Image Statistical Toolbox for MATLAB~\parencite{bainbridge2015toolbox}, adapted for Python, we measured the spectral characteristics of each image. We determined the dominant spatial frequencies by finding the frequency below which 80\% of the total spectral energy is contained. We also assessed high spatial frequencies by calculating the proportion of frequencies above 10 cycles per image. Such metric quantifies the amount of fine details and textures within the image.

\end{itemize}

\subsubsection{Programming tools}\label{appendix:programming_tools}

The experiment was implemented using the jsPsych library 7.3.1~\parencite{leeuw_jspsych_2023} and JATOS~\parencite{lange_just_2015}, which was installed on a server at the Bernstein Center for Computational Neuroscience (Berlin, Germany). 
Experimental condition assignments were managed via an external database\footnote{\url{https://supabase.com}} and a backend application using Python FastAPI 0.111.0, hosted on Vercel\footnote{\url{https://vercel.com}}. To run the embedding analyses, we used the Python package \textit{cblearn} \parencite{kunstle_cblearn_2024}, as well as the original MATLAB code for t-STE\footnote{\url{https://lvdmaaten.github.io/ste/Stochastic_Triplet_Embedding.html}} and the original R code for MLDS \parencite{knoblauch_mlds_2008} for replication tests. Since we could successfully replicate the results, we report only the final results obtained using the \textit{cblearn} package.

\pagebreak
\subsection{Memory experiment}\label{appendix:memory_experiment_methods}

\subsubsection{\textcolor{black}{Experimental control, data quality, and exclusion criteria}}\label{appendix:memory_data_quality_exclusion}

\textcolor{black}{To ensure high data quality, we applied similar experimental measures as in the Perceptual Similarity Judgement Task (see \ref{appendix:control_quality_exclusion_criteria}). Participants were required to use a desktop device, Google Chrome as the web browser, a minimum screen resolution of 800×600 pixels, and fullscreen mode throughout the task. A preliminary device and browser check ensured that these requirements were met before the experiment began. If participants exited fullscreen mode during the task, the experiment was automatically paused. Participants were excluded if they did not finish the experiment (12.43\%), encountered technical issues (8.88\%), or had a mean reaction time below 500 ms or an extreme response bias (1.85 times outside the interquartile range bounds)(3.85\%). Mean reaction time and response bias were used as indicators of participants who might have engaged poorly with the task (e.g. repeatedly pressing the same key without performing the task). In addition, we also excluded participants with very poor behavioural performance (17.16\%), quantified as a mean accuracy below chance level (50\%).}

\subsubsection{Selection of the targets and foils}\label{appendix:target_foil_selection}

First, we computed the absolute differences between consecutive embedding values for each object’s set of images. Next, we set a threshold at the 90th percentile of these differences. Consecutive pairs exceeding this threshold were considered discontinuities. If a discontinuity occurred at the beginning or end of a sequence, we marked the corresponding images as outliers. We then selected the target image. This was the image with the highest embedding value that was not marked as an outlier and was not immediately followed by a discontinuity. This procedure made sure that the next image, the ``Hard'' foil, was perceptually similar to the target. After choosing the target and removing outliers, we divided the remaining images into three bins. The bins corresponded to the 25th, 50th, and 75th percentile of embedding values. From each bin, we selected one foil image whose embedding value was closest to the bin's percentile.

\subsubsection{Model specification}\label{appendix:memory_experiment_model_spec}

The primary goal of our analysis was to examine how learning progressed over time, measured as changes in accuracy across experimental blocks, and to determine whether these learning trajectories differed depending on two primary experimental manipulations: target condition and difficulty. Target condition was modelled as a two-level factor (repeated versus non-repeated targets), difficulty as a three-level factor (Easy, Medium, Hard), and block as a continuous variable ranging from 1 to 6. The fixed effects accounted for the population-level effects of block, difficulty, target condition, for all two-way interactions (block × difficulty, block × target condition, difficulty × target condition), and for the three-way interaction (block × difficulty × target condition). The random effects structure accounted for participant-specific variability through random intercepts, representing individual baseline performance levels, and random slopes for block, reflecting individual differences in learning rates over time. 

For each trial $i$ of participant $j$, the binary accuracy outcome $y_{ij}$ is modeled as a Bernoulli-distributed random variable:

\begin{equation}
y_{ij} \sim \text{Bernoulli}(p_{ij})
\end{equation}

The probability of a correct response $p_{ij}$ is linked to a linear predictor via the logit function:

\begin{equation}
\begin{aligned}
\text{logit}(p_{ij}) =\; & \beta_0 + \beta_1 \cdot \text{block}_{ij} \\
& + \beta_2 \cdot \text{difficulty\_medium}_{ij} + \beta_3 \cdot \text{difficulty\_hard}_{ij} \\
& + \beta_4 \cdot \text{target\_condition}_{ij} \\
& + \beta_5 \cdot (\text{block}_{ij} \times \text{difficulty\_medium}_{ij}) + \beta_6 \cdot (\text{block}_{ij} \times \text{difficulty\_hard}_{ij}) \\
& + \beta_7 \cdot (\text{block}_{ij} \times \text{target\_condition}_{ij}) \\
& + \beta_8 \cdot (\text{difficulty\_medium}_{ij} \times \text{target\_condition}_{ij}) + \beta_9 \cdot (\text{difficulty\_hard}_{ij} \times \text{target\_condition}_{ij}) \\
& + \beta_{10} \cdot (\text{block}_{ij} \times \text{difficulty\_medium}_{ij} \times \text{target\_condition}_{ij}) \\
& + \beta_{11} \cdot (\text{block}_{ij} \times \text{difficulty\_hard}_{ij} \times \text{target\_condition}_{ij}) \\
& + u_{0j} + u_{1j} \cdot \text{block}_{ij}.
\end{aligned}
\end{equation}

where $\beta_0$ represents the baseline log-odds of a correct response under the reference conditions (e.g., initial block, Easy difficulty, non-repeated stimuli). The coefficients $\beta_1$ through $\beta_{11}$ model the fixed effects of the predictors and their interactions. To account for individual differences among participants, the model incorporates random effects for both the intercept and the slope of the block variable:

\begin{equation}
\begin{pmatrix} u_{0j} \\ u_{1j} \end{pmatrix} 
\sim \mathcal{N} \left( 
\begin{pmatrix} 0 \\ 0 \end{pmatrix}, \mathbf{\Sigma} \right)
\end{equation}

where $u_{0j}$ is the random intercept (modelling participant-specific baseline performance) and $u_{1j}$ is the random slope for the block (modelling participant-specific learning rates). 

The random intercept ($u_{0j}$) and random slope ($u_{1j}$) were allowed to correlate, with covariance matrix:

\begin{equation}
\mathbf{\Sigma} = 
\begin{bmatrix} 
\sigma_{u_0}^2 & \rho \sigma_{u_0} \sigma_{u_1} \\ 
\rho \sigma_{u_0} \sigma_{u_1} & \sigma_{u_1}^2 
\end{bmatrix}
\end{equation}

where $\sigma_{u_0}$ and $\sigma_{u_1}$ are the standard deviations of the random intercept and slope, respectively, and $\rho$ represents their correlation.

The prior distributions for the model parameters were selected to be weakly informative to help stabilise parameters estimates \parencite{Gelman2006}. Fixed effects were assigned normal priors:
\begin{equation}
\beta_k \sim \mathcal{N}(0, \sigma^2), \quad k = 0,1,\dots,11
\end{equation}
with $\sigma^2 = 4$, constraining coefficients to plausible ranges on the log-odds scale (±4 log-odds at 2SD) and centering the effect at 0. 
Random effect standard deviations ($\sigma_{u0}, \sigma_{u1}$) were assigned:
\begin{equation}
\sigma_u \sim \text{Half-Cauchy}(0, 1)
\end{equation}
priors, which ensure positive values and favor smaller variability across participants while still permitting larger deviations \parencite{Gelman1995, PolsonScott2012}. 

The prior for the correlation between random effects ($\rho$) was modeled using a Lewandowski-Kurowicka-Joe (LKJ) distribution \parencite{Lewandowski2009}:
\begin{equation}
\rho \sim \text{LKJ}(\zeta)
\end{equation}
with $\zeta = 2$ as a weakly informative prior that discourages extreme correlations, unless supported by the data \parencite{Burkner2017}.
These priors were evaluated through prior predictive checks to confirm they generated reasonable simulated data. In addition, we analysed other weakly informative priors to assess if they would produce similar results. Posterior predictive checks were conducted to evaluate model fit. 

\pagebreak

\section{Extended Results}\label{appendix:supplementary_results}
\subsection{Image Generation}\label{appendix:image-generation}
\subsubsection{Exemplar images}\label{appendix:image-generation-exemplars}
We generated 60 exemplar images for each one of the 108 objects, for a total of 6480 images. The synthesised images were visually evaluated for their quality and fidelity to the input textual prompts, which was generally high. The application of the LoRA model visibly enhanced the quality of the generated images, especially making contrast and scene lighting more realistic.

\begin{figure}[H]
    \centering
    \includegraphics[width=0.7\textwidth]{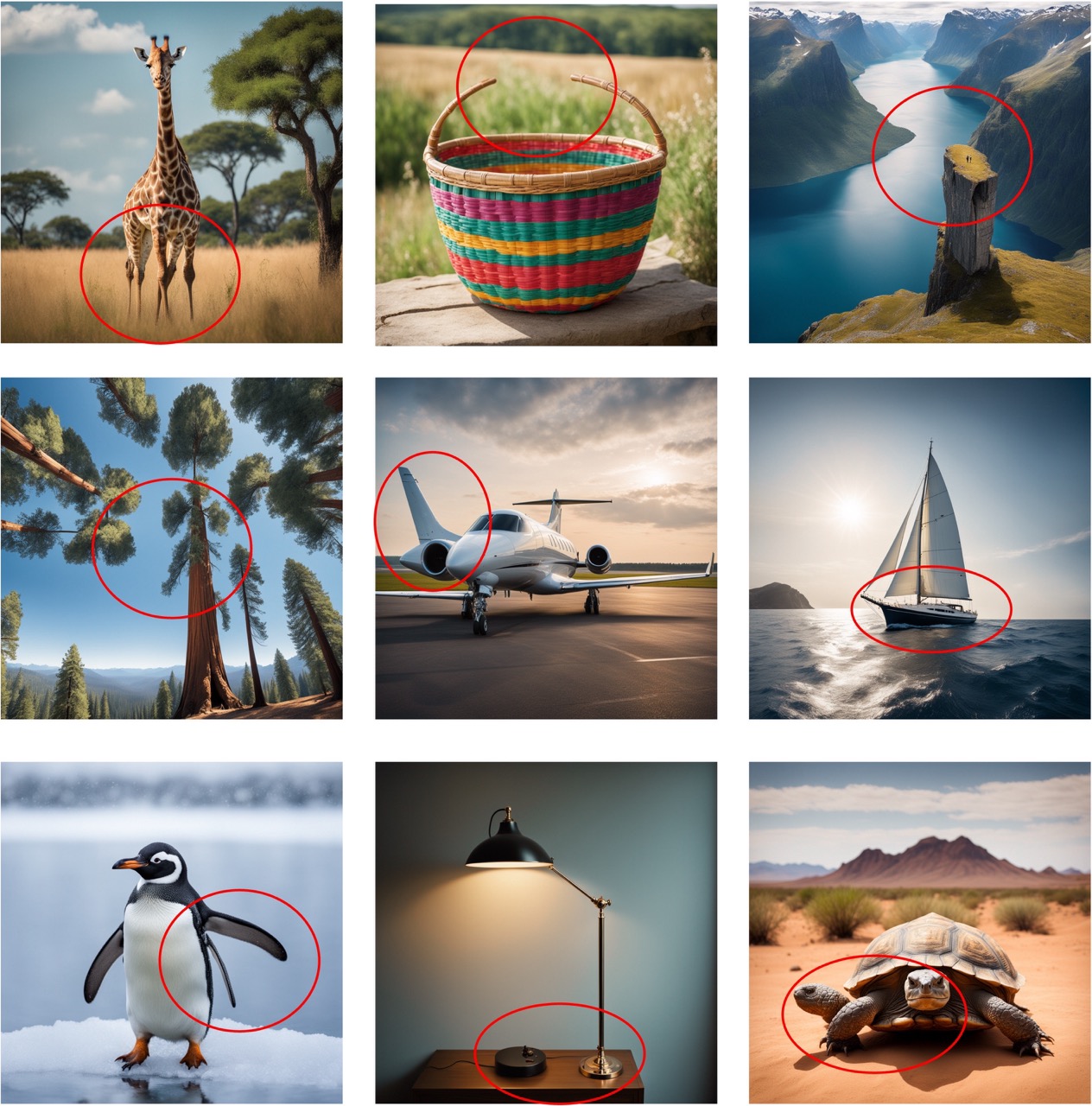}
    \caption{Some examples of excluded exemplars, highlighting the artefacts. A common artefact when generating animals was the wrong number of limbs (\textit{giraffe} and \textit{penguin}) or other body parts (the \textit{head} of the \textit{tortoise}). 
    Other artefacts depicted here are layouts that did not follow the prompt (e.g., the \textit{sequoia}), non-realistic perspectives (e.g., the horizon in \textit{sailboat}), or compositions (e.g., the handle in \textit{basket}, the reactor in \textit{jet plane}, and the lamp stand in \textit{lamp (table)}). Finally, despite the negative prompt, sometimes human figures made an appearance (e.g., in \textit{fjord pillar}). Such images were also excluded.}
    \label{fig:excluded_exemplars_examples}
\end{figure}

\begin{figure}[H]
    \centering
    \begin{subfigure}[b]{0.48\textwidth}
        \centering
        \includegraphics[width=\linewidth, height=5.5cm]{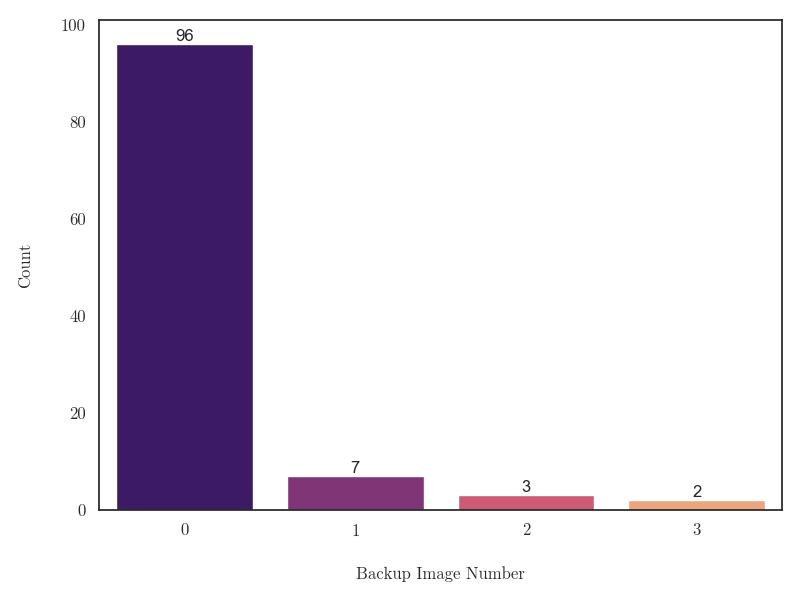} 
        \caption{Exclusion rates anchors}
        \label{fig:exclusion_rate_exemplars}
    \end{subfigure}
    \hfill
    \begin{subfigure}[b]{0.48\textwidth}
        \centering
        \includegraphics[width=\linewidth, height=5.5cm]{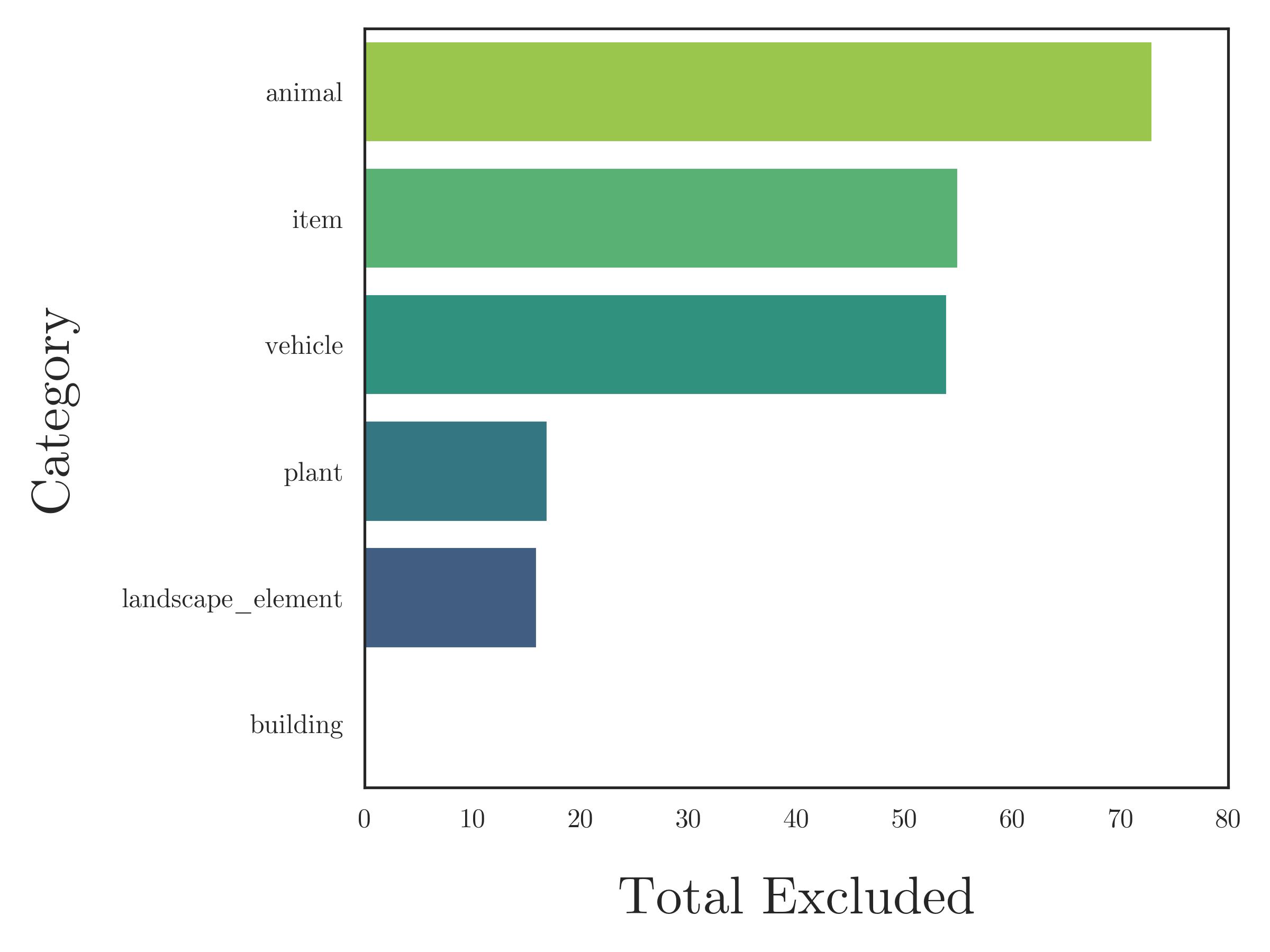}
        \caption{Excluded by category}
        \label{fig:excluded_by_category}
    \end{subfigure}
    
    \vspace{0.3cm}
    
    \begin{subfigure}[b]{0.48\textwidth}
        \centering
        \includegraphics[width=\linewidth, height=5.5cm]{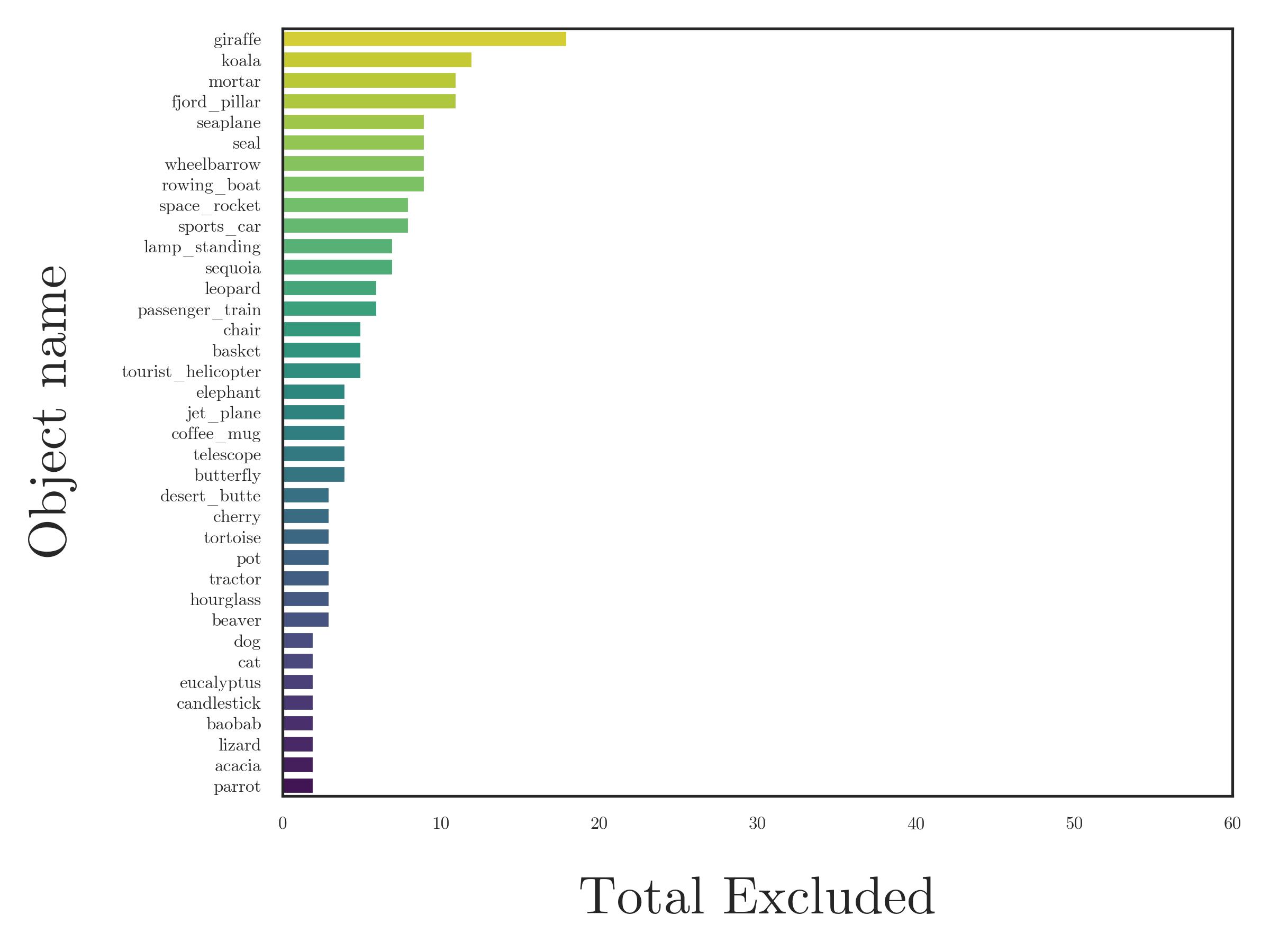}
        \caption{Excluded by object}
        \label{fig:excluded_by_object}
    \end{subfigure}
    
    \caption{The giraffe had the highest number of exclusions, with 18 images excluded, which represents 30\% of the 60 images for this object and 8.4\% of the total 215 excluded images. This was followed by the koala with 12 exclusions (20\% of its images and 5.6\% of the total exclusions), and both the fjord pillar and mortar each with 11 exclusions (18.3\% of their images and 5.1\% of the total exclusions). The seal, seaplane, rowing boat, and wheelbarrow each had 9 exclusions (15\% of their images and 4.2\% of the total exclusions). The space rocket and sports car each had 8 exclusions (13.3\% of their images and 3.7\% of the total exclusions). Other notable exclusions included the lamp standing and leopard, each with 7 exclusions (11.7\% of their images and 3.3\% of the total exclusions), and the passenger train with 6 exclusions (10\% of its images and 2.8\% of the total exclusions). Objects such as the tourist helicopter, elephant, coffee mug, butterfly, and telescope had between 3 to 5 exclusions (ranging from 5\% to 8.3\% of their images and 1.4\% to 2.3\% of the total exclusions).}
    \label{fig:exclusion_fig_exemplars}
\end{figure}

After visual inspection, 215/6480 images (3.3\%) were excluded (Figure~\ref{fig:exclusion_fig_exemplars}). We provide some examples of images that we excluded, highlighting their artefacts in Figure~\ref{fig:excluded_exemplars_examples}. The distribution of these exclusions by category showed variability in exclusion rates across different categories: the animal category had the highest number of exclusions, with 73 images excluded, accounting for approximately 34\% of the total exclusions. The item category followed with 55 excluded images, making up about 26\% of the exclusions. The vehicle category had 54 exclusions, representing 25\% of the total. The plant category saw 17 images excluded, which is roughly 8\% of the total. The landscape element category had 16 exclusions, constituting about 7\% of the total. Notably, the building category had no exclusions. An object-specific breakdown (Table~\ref{tab:exclusion-rates-exemplars}) shows that while some objects faced high exclusion rates (e.g., 30\% of the giraffe images presented visible artefacts), 79/108 (73.1\%) objects had fewer than 3 exclusions, of which 57/108 objects (52.8\%) had no exclusions at all.

\begin{longtable}{llrrr}
\caption{Exclusion rates for all exemplar images.} \label{tab:exclusion-rates-exemplars} \\
\toprule
 & object & total\_excluded & total\_images & exclusion\_rate \\
\midrule
\endfirsthead

\multicolumn{5}{c}{{\tablename\ \thetable{} -- continued from previous page}} \\
\toprule
 & object & total\_excluded & total\_images & exclusion\_rate \\
\midrule
\endhead

\midrule
\multicolumn{5}{r}{{Continued on next page}} \\
\endfoot

\bottomrule
\endlastfoot

0 & windmill & 0 & 60 & 0.000000 \\
1 & polar\_iceberg & 0 & 60 & 0.000000 \\
2 & grassland\_monolith & 0 & 60 & 0.000000 \\
3 & greenhouse & 0 & 60 & 0.000000 \\
4 & hill\_river & 0 & 60 & 0.000000 \\
5 & hot\_air\_balloon & 0 & 60 & 0.000000 \\
6 & polar\_base & 0 & 60 & 0.000000 \\
7 & igloo & 0 & 60 & 0.000000 \\
8 & pine\_med & 0 & 60 & 0.000000 \\
9 & kapok & 0 & 60 & 0.000000 \\
10 & pine & 0 & 60 & 0.000000 \\
11 & lake\_island & 0 & 60 & 0.000000 \\
12 & lamp\_table & 0 & 60 & 0.000000 \\
13 & lantern & 0 & 60 & 0.000000 \\
14 & willow & 0 & 60 & 0.000000 \\
15 & lighthouse & 0 & 60 & 0.000000 \\
16 & lorry & 0 & 60 & 0.000000 \\
17 & magnolia & 0 & 60 & 0.000000 \\
18 & maple & 0 & 60 & 0.000000 \\
19 & modern\_building & 0 & 60 & 0.000000 \\
20 & moorland\_tor & 0 & 60 & 0.000000 \\
21 & palm & 0 & 60 & 0.000000 \\
22 & motorcycle & 0 & 60 & 0.000000 \\
23 & mountain\_glacier & 0 & 60 & 0.000000 \\
24 & mountain\_ridge & 0 & 60 & 0.000000 \\
25 & mountain\_rock & 0 & 60 & 0.000000 \\
26 & oak & 0 & 60 & 0.000000 \\
27 & observatory & 0 & 60 & 0.000000 \\
28 & offroad\_vehicle & 0 & 60 & 0.000000 \\
29 & olive & 0 & 60 & 0.000000 \\
30 & geothermal\_spring & 0 & 60 & 0.000000 \\
31 & fortress & 0 & 60 & 0.000000 \\
32 & futuristic\_building & 0 & 60 & 0.000000 \\
33 & forest\_boulder & 0 & 60 & 0.000000 \\
34 & adobe\_house & 0 & 60 & 0.000000 \\
35 & bamboo\_house & 0 & 60 & 0.000000 \\
36 & wetland\_tufa & 0 & 60 & 0.000000 \\
37 & barn & 0 & 60 & 0.000000 \\
38 & volcanic\_peak & 0 & 60 & 0.000000 \\
39 & vase & 0 & 60 & 0.000000 \\
40 & bell\_tower & 0 & 60 & 0.000000 \\
41 & birch & 0 & 60 & 0.000000 \\
42 & bottle & 0 & 60 & 0.000000 \\
43 & bristlecone & 0 & 60 & 0.000000 \\
44 & cabin & 0 & 60 & 0.000000 \\
45 & car\_vintage & 0 & 60 & 0.000000 \\
46 & forest\_river & 0 & 60 & 0.000000 \\
47 & tea\_house & 0 & 60 & 0.000000 \\
48 & steam\_locomotive & 0 & 60 & 0.000000 \\
49 & teapot & 0 & 60 & 0.000000 \\
50 & dragon\_tree & 0 & 60 & 0.000000 \\
51 & fishing\_boat & 0 & 60 & 0.000000 \\
52 & farm & 0 & 60 & 0.000000 \\
53 & scooter & 0 & 60 & 0.000000 \\
54 & sea\_stack & 0 & 60 & 0.000000 \\
55 & sofa & 0 & 60 & 0.000000 \\
56 & pagoda & 0 & 60 & 0.000000 \\
57 & skyscraper & 0 & 60 & 0.000000 \\
58 & cactus & 1 & 60 & 0.016667 \\
59 & shovel & 1 & 60 & 0.016667 \\
60 & tropical\_karst & 1 & 60 & 0.016667 \\
61 & tropical\_bird & 1 & 60 & 0.016667 \\
62 & horse & 1 & 60 & 0.016667 \\
63 & jar & 1 & 60 & 0.016667 \\
64 & fish & 1 & 60 & 0.016667 \\
65 & desert\_arch & 1 & 60 & 0.016667 \\
66 & sailboat & 1 & 60 & 0.016667 \\
67 & owl & 1 & 60 & 0.016667 \\
68 & frog & 1 & 60 & 0.016667 \\
69 & campervan & 1 & 60 & 0.016667 \\
70 & penguin & 1 & 60 & 0.016667 \\
71 & parrot & 2 & 60 & 0.033333 \\
72 & acacia & 2 & 60 & 0.033333 \\
73 & lizard & 2 & 60 & 0.033333 \\
74 & baobab & 2 & 60 & 0.033333 \\
75 & candlestick & 2 & 60 & 0.033333 \\
76 & eucalyptus & 2 & 60 & 0.033333 \\
77 & cat & 2 & 60 & 0.033333 \\
78 & dog & 2 & 60 & 0.033333 \\
79 & beaver & 3 & 60 & 0.050000 \\
80 & hourglass & 3 & 60 & 0.050000 \\
81 & tractor & 3 & 60 & 0.050000 \\
82 & pot & 3 & 60 & 0.050000 \\
83 & tortoise & 3 & 60 & 0.050000 \\
84 & cherry & 3 & 60 & 0.050000 \\
85 & desert\_butte & 3 & 60 & 0.050000 \\
86 & butterfly & 4 & 60 & 0.066667 \\
87 & telescope & 4 & 60 & 0.066667 \\
88 & coffee\_mug & 4 & 60 & 0.066667 \\
89 & jet\_plane & 4 & 60 & 0.066667 \\
90 & elephant & 4 & 60 & 0.066667 \\
91 & tourist\_helicopter & 5 & 60 & 0.083333 \\
92 & basket & 5 & 60 & 0.083333 \\
93 & chair & 5 & 60 & 0.083333 \\
94 & passenger\_train & 6 & 60 & 0.100000 \\
95 & leopard & 6 & 60 & 0.100000 \\
96 & sequoia & 7 & 60 & 0.116667 \\
97 & lamp\_standing & 7 & 60 & 0.116667 \\
98 & sports\_car & 8 & 60 & 0.133333 \\
99 & space\_rocket & 8 & 60 & 0.133333 \\
100 & rowing\_boat & 9 & 60 & 0.150000 \\
101 & wheelbarrow & 9 & 60 & 0.150000 \\
102 & seal & 9 & 60 & 0.150000 \\
103 & seaplane & 9 & 60 & 0.150000 \\
104 & fjord\_pillar & 11 & 60 & 0.183333 \\
105 & mortar & 11 & 60 & 0.183333 \\
106 & koala & 12 & 60 & 0.200000 \\
107 & giraffe & 18 & 60 & 0.300000 \\
\end{longtable}

\subsubsection{Pair selection using the LPIPS metric}\label{appendix:pair-selection-lpips}
\begin{figure}[htbp]
    \centering
    \begin{subfigure}[b]{\textwidth}
        \centering
        \caption{}
        \includegraphics[width=0.7\linewidth]{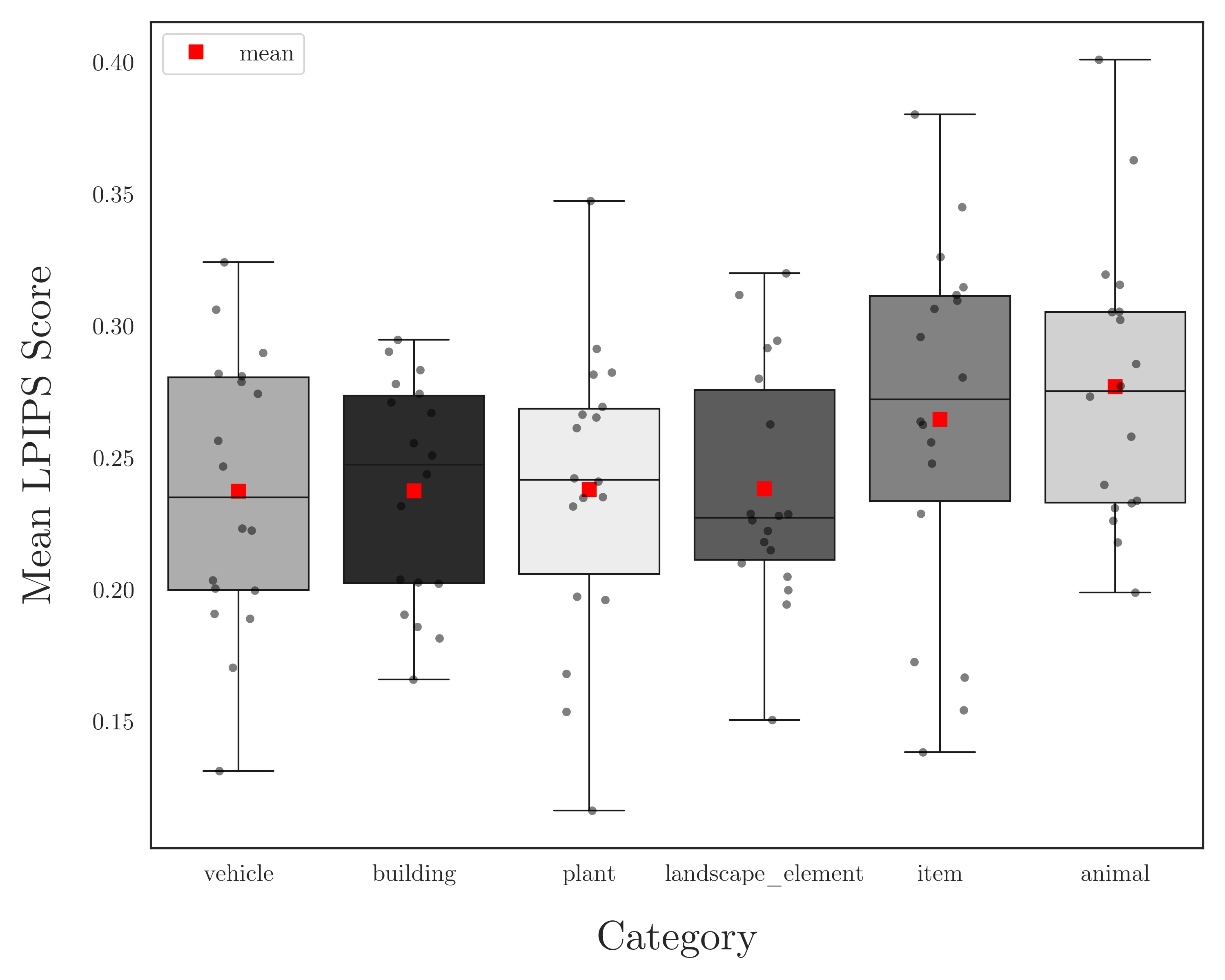}
        \label{fig:exemplar_mean_lpips_cat}
    \end{subfigure}
    
    \vspace{0.3cm}
    
    \begin{subfigure}[b]{\textwidth}
        \centering
        \caption{}
        \includegraphics[width=\linewidth]{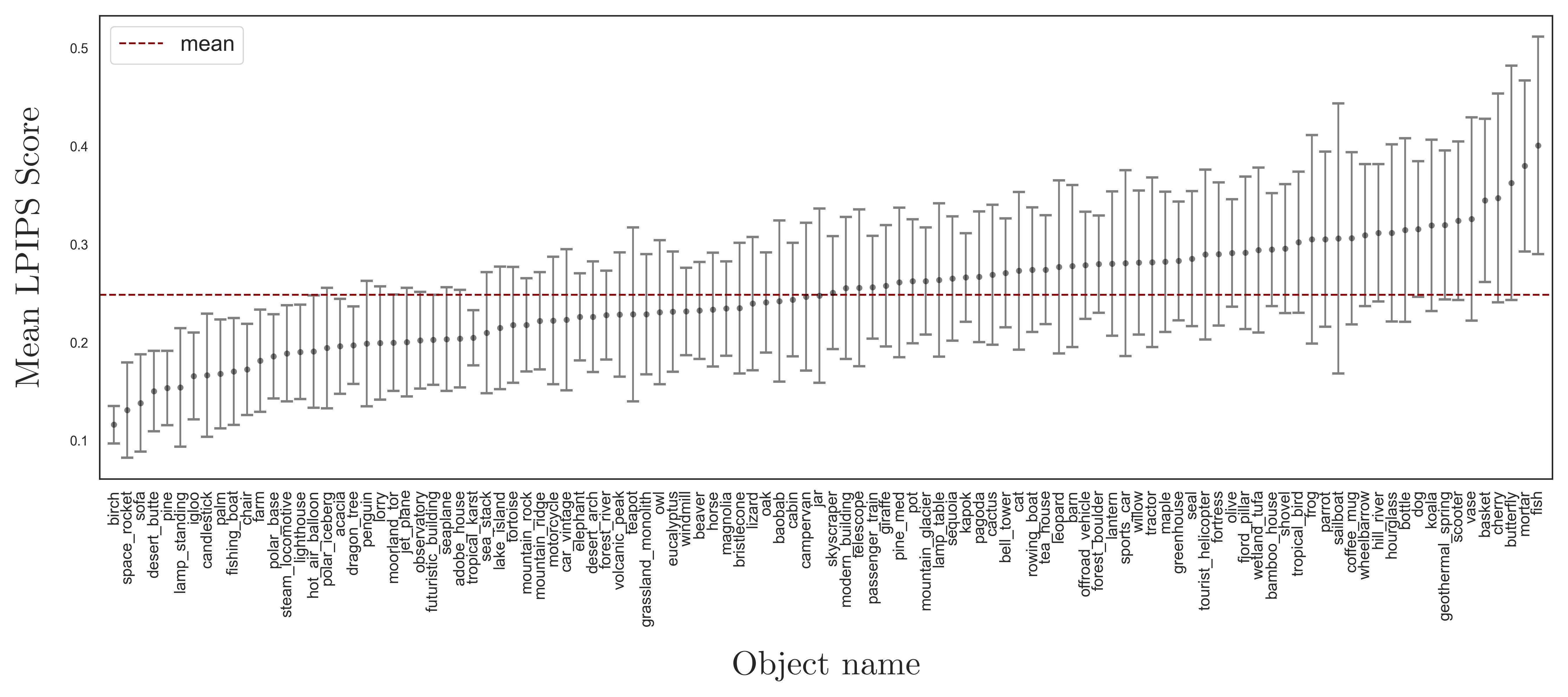}
        \label{fig:exemplar_mean_lpips_by_obj}
    \end{subfigure}
    \caption{Mean LPIPS scores between exemplar images, grouped by a) category and b) object.}
    \label{fig:lpips_exemplars}
\end{figure}

For each object, we calculated the LPIPS scores between all pairs of the exemplar images that had not been excluded (Figure~\ref{fig:lpips_exemplars}). Overall, the mean LPIPS score across all objects was $0.453$ ($SD = 0.054$). The object with the highest mean LPIPS score was fish ($M = 0.589$, $SD = 0.039$) and the one with the lowest was birch ($M = 0.286$, $SD = 0.344$). We then examined the distribution of mean LPIPS scores for each category. The category items had the highest mean LPIPS score ($M = 0.483$, $SD = 0.055$), followed by animal ($M = 0.469$, $SD = 0.068$), vehicle ($M = 0.450$, $SD = 0.028$), landscape element ($M = 0.446$, $SD = 0.054$), plant ($M = 0.442$, $SD = 0.055$), and building had the lowest mean score ($M = 0.430$, $SD = 0.043$). A one-way Analysis of Variance (ANOVA) revealed a statistically significant difference in mean LPIPS scores by category ($F(5, 24) = 2.54$, $p = 0.033$). To investigate the pairwise differences more in detail, we conducted a post-hoc Tukey HSD test \parencite{tukey_comparing_1949}, which indicated a significant difference in mean LPIPS scores between the categories building and item (mean difference $= 0.0538$, $p = 0.030$). No other pairwise comparisons showed statistically significant differences in mean LPIPS scores ($p > 0.05$). All test results are reported in Table~\ref{tab:exemplars-tukey-hsd}.

\begin{center}
\begin{table}
    \centering
    \begin{tabular}{ccccccc}
        \toprule
        \textbf{group1}   &  \textbf{group2}   & \textbf{meandiff} & \textbf{p-adj} & \textbf{lower} & \textbf{upper} & \textbf{reject}  \\
        \midrule
          animal       &      building      &      -0.0399      &     0.2052     &    -0.0904     &     0.0106     &      False       \\
          animal       &        item        &       0.0139      &     0.9671     &    -0.0366     &     0.0644     &      False       \\
          animal       & landscape\_element &      -0.0238      &     0.7459     &    -0.0743     &     0.0267     &      False       \\
          animal       &       plant        &      -0.0274      &     0.6156     &    -0.0779     &     0.0231     &      False       \\
          animal       &      vehicle       &      -0.0192      &     0.8779     &    -0.0697     &     0.0313     &      False       \\
         building      &        item        &       0.0538      &     0.0297     &     0.0033     &     0.1043     &       True       \\
         building      & landscape\_element &       0.0161      &     0.9383     &    -0.0344     &     0.0666     &      False       \\
         building      &       plant        &       0.0125      &     0.9792     &     -0.038     &     0.063      &      False       \\
         building      &      vehicle       &       0.0207      &     0.8406     &    -0.0298     &     0.0712     &      False       \\
           item        & landscape\_element &      -0.0377      &     0.2624     &    -0.0882     &     0.0128     &      False       \\
           item        &       plant        &      -0.0413      &     0.1747     &    -0.0918     &     0.0092     &      False       \\
           item        &      vehicle       &      -0.0331      &     0.4052     &    -0.0836     &     0.0174     &      False       \\
    landscape\_element &       plant        &      -0.0036      &     0.9999     &    -0.0541     &     0.0469     &      False       \\
    landscape\_element &      vehicle       &       0.0046      &     0.9998     &    -0.0459     &     0.055      &      False       \\
          plant        &      vehicle       &       0.0082      &     0.997      &    -0.0423     &     0.0587     &      False       \\
            \bottomrule
    \end{tabular}
    \caption{Multiple Comparison of Means - Tukey HSD, Family-Wise Error Rate (FWER) = 0.05, between LPIPS scores from all non-excluded anchor images. The table reports pairwise comparisons among different categories with the mean difference, adjusted p-values, and confidence intervals. Significant differences at the 0.05 level are indicated in the reject column.}
    \label{tab:exemplars-tukey-hsd}
\end{table}
\end{center}

\begin{figure}[htbp]
    \centering
    \begin{subfigure}[b]{\textwidth}
    \centering
    \caption{}
    \includegraphics[width=0.7\linewidth]{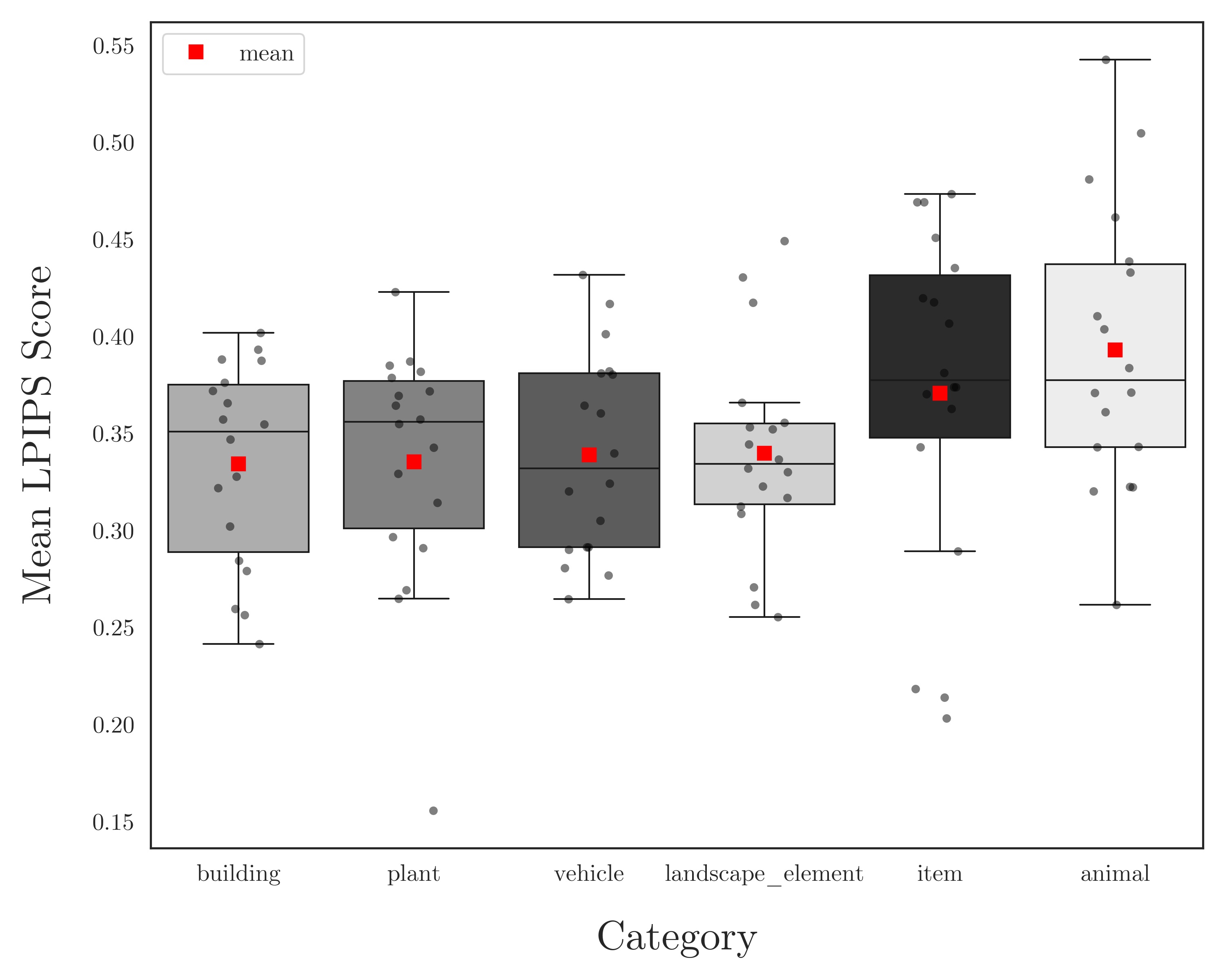}
    \label{fig:lpips_scores_pairs_by_category}
    \end{subfigure}
    
    \vspace{0.3cm}
    
    \begin{subfigure}[b]{\textwidth}
    \centering
    \caption{}
    \includegraphics[width=\linewidth]{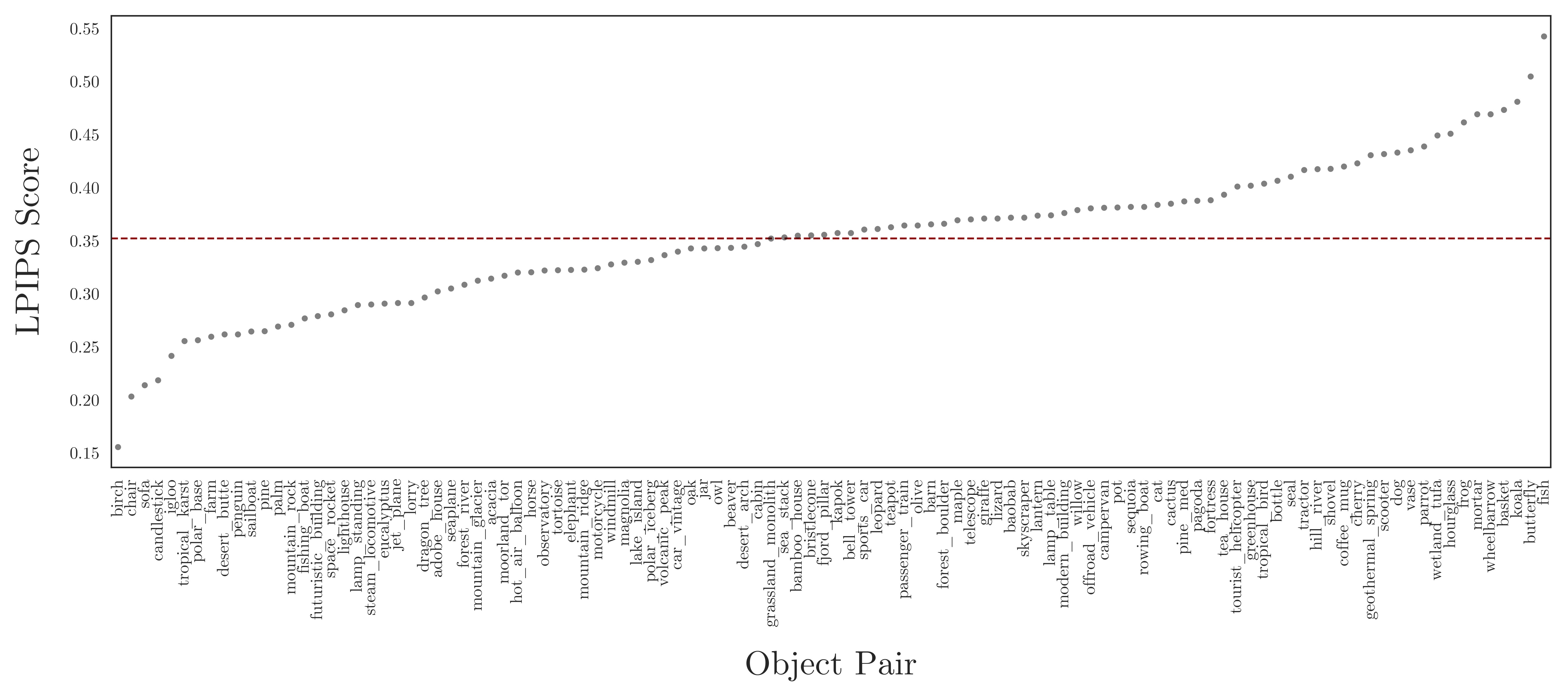}
    \label{fig:lpips_scores_pairs_by_object}
    \end{subfigure}
    \caption{Mean LPIPS scores for selected pairs, grouped by object and category. a) Distribution of mean LPIPS scores by category. Each data point is the mean LPIPS score for an object belonging to that category (18 data points in total). The category means are shown in red. b) Mean LPIPS scores by object, ordered by magnitude. Lower scores indicate more homogeneity (i.e. highest similarity) between the exemplar images. The red dotted line is the overall mean across all similarity scores.}
    \label{fig:lpips_pairs}
\end{figure}

Following the procedure illustrated in the Methods (see Appendix~\ref{appendix:selection-anchors} and ~\ref{appendix:interpolation_algorithm}), for each object we selected a pair of images (an \textit{anchor image} and a \textit{guide image}) to interpolate. All anchor images are shown in Figure~\ref{fig:table-anchor-images-1} and ~\ref{fig:table-anchor-images-2}. The performance of this procedure was assessed through visual inspection. In the vast majority of cases ($89.81\%$ of image pairs), our procedure successfully selected a suitable pair of images on the first attempt. However, in 11 cases ($10.18\%$), the initially selected second image did not meet our specified conditions (as detailed in the Methods section), necessitating the use of backup images. Specifically, we resorted to the second choice for 7 object pairs, the third choice for 3 pairs, and the fourth choice for 2 pairs. We assessed the statistical distribution of LPIPS scores between the selected pairs, that is the score between the \textit{anchor} and the \textit{guide image} (Figure~\ref{fig:lpips_pairs}). The mean LPIPS score was $0.35$ ($SD = 0.067$). Pairs from the animal category had a mean score of $0.393$ ($SD = 0.073$), from buildings had a mean score of $0.334$ ($SD = 0.052$), from items had a mean score of $0.370$ ($SD = 0.087$), from landscape elements had a mean score of $0.340$ ($SD = 0.053$), from plants had a mean score of $0.335$ ($SD = 0.063$), and from vehicles had a mean score of $0.339$ ($SD = 0.052$). A one-way ANOVA test indicated a statistically significant difference in mean LPIPS scores across categories ($F(5, 120) = 2.525$, $p = 0.034$). However, subsequent pairwise comparisons using the Tukey HSD test did not reveal significant differences between any specific pairs of categories (all $p > 0.05$), see Table~\ref{tab:pairs-tukey-hsd}. This suggests that while pairs from different categories differed in mean LPIPS scores, these differences were relatively subtle.

\begin{table}[ht]
    \centering
    \begin{tabular}{ccccccc}
        \toprule
        \textbf{group1} & \textbf{group2} & \textbf{meandiff} & \textbf{p-adj} & \textbf{lower} & \textbf{upper} & \textbf{reject} \\
        \midrule
        animal & building & -0.0589 & 0.0791 & -0.1216 & 0.0039 & False \\
        animal & item & -0.0225 & 0.9033 & -0.0852 & 0.0403 & False \\
        animal & landscape\_element & -0.0533 & 0.1434 & -0.1161 & 0.0094 & False \\
        animal & plant & -0.0577 & 0.0902 & -0.1204 & 0.0051 & False \\
        animal & vehicle & -0.0541 & 0.1325 & -0.1168 & 0.0086 & False \\
        building & item & 0.0364 & 0.5448 & -0.0263 & 0.0991 & False \\
        building & landscape\_element & 0.0055 & 0.9998 & -0.0572 & 0.0683 & False \\
        building & plant & 0.0012 & 1.0 & -0.0616 & 0.0639 & False \\
        building & vehicle & 0.0048 & 0.9999 & -0.0580 & 0.0675 & False \\
        item & landscape\_element & -0.0309 & 0.7096 & -0.0936 & 0.0319 & False \\
        item & plant & -0.0352 & 0.5803 & -0.0980 & 0.0275 & False \\
        item & vehicle & -0.0316 & 0.6873 & -0.0944 & 0.0311 & False \\
        landscape\_element & plant & -0.0044 & 1.0 & -0.0671 & 0.0584 & False \\
        landscape\_element & vehicle & -0.0008 & 1.0 & -0.0635 & 0.0620 & False \\
        plant & vehicle & 0.0036 & 1.0 & -0.0591 & 0.0663 & False \\
        \bottomrule
    \end{tabular}
    \caption{Multiple Comparison of Means - Tukey HSD, FWER=0.05, between LPIPS scores from selected pairs of \textit{anchor} and \textit{guide} images in the interpolations. The table presents mean differences, adjusted p-values, and confidence intervals for each pairwise comparison.}
        \label{tab:pairs-tukey-hsd}
\end{table}

\subsubsection{Interpolations}
For each of the 108 object pairs, we generated 200 interpolated images (including the \textit{anchor} and the \textit{guide} images). Following the criteria explained in the Methods (Appendix~\ref{appendix:selection_interpol_images}), we selected 10 images for each object, including the \textit{anchor image} and 9 of its interpolations. To achieve this, we first computed the LPIPS score between the \textit{anchor image} and each interpolated image. 

\begin{figure}[htbp]
    \centering
    \begin{subfigure}[b]{\textwidth}
        \centering
        \caption{}
        \includegraphics[width=0.75\linewidth]{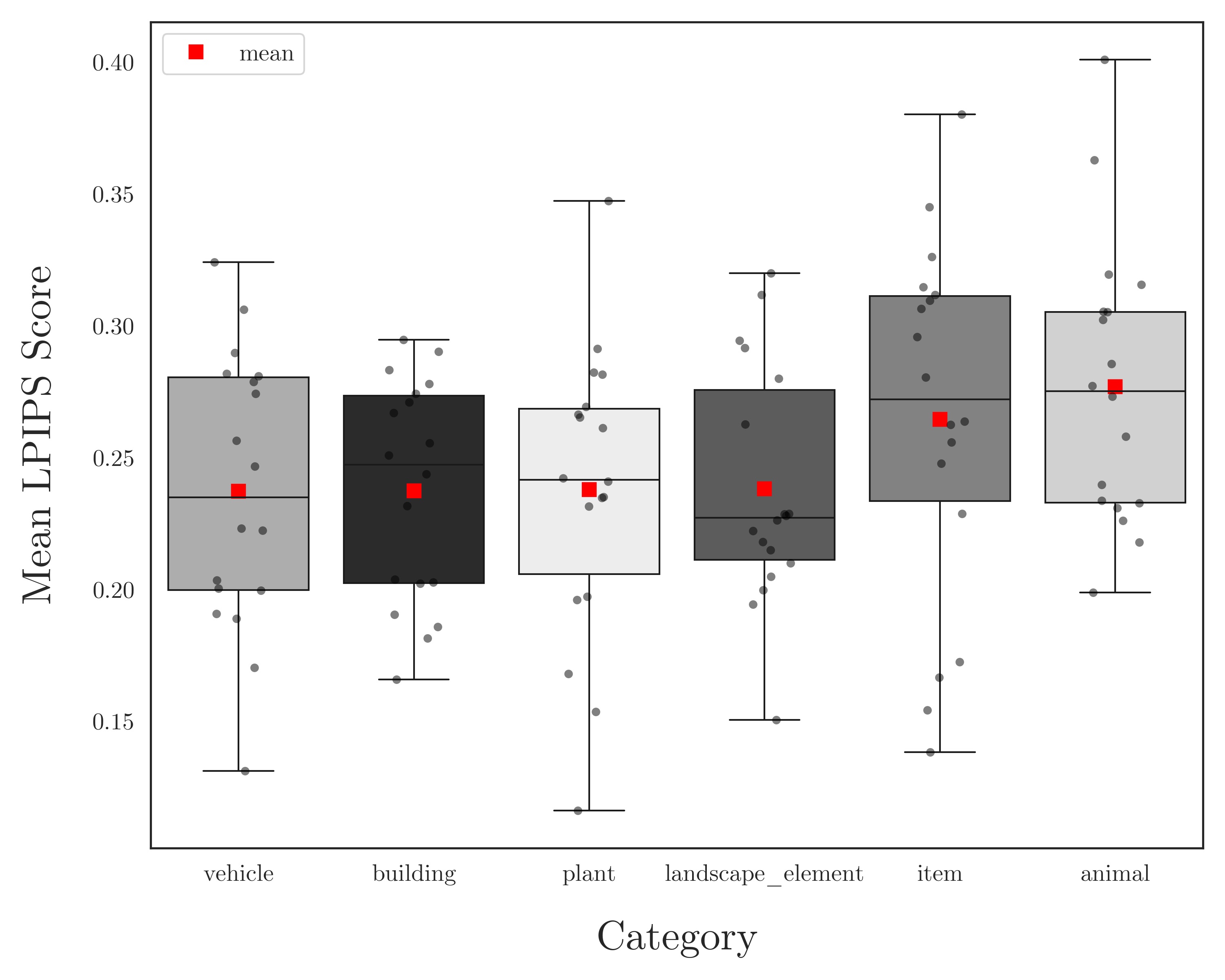}
        \label{fig:lipis_scores_first_and_interpol_category}
    \end{subfigure}
    
    \vspace{0.3cm}
    
    \begin{subfigure}[b]{\textwidth}
        \centering
        \caption{}
        \includegraphics[width=\linewidth]{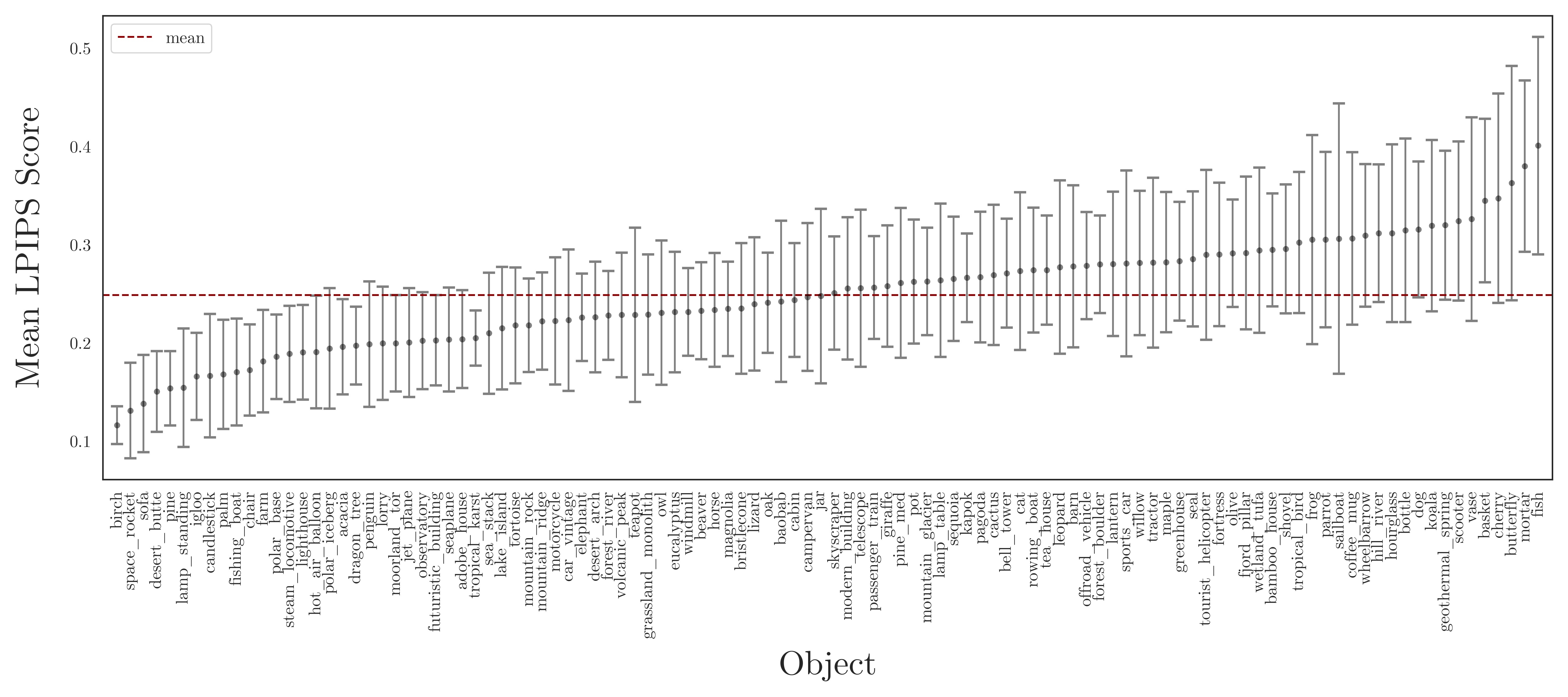}
        \label{fig:lipis_scores_first_and_interpol_obj}
    \end{subfigure}
    \caption{Mean LPIPS scores between \textit{anchor images} and each interpolated image, grouped by a) category and b) single objects.}
    \label{fig:lpips_interpols}
\end{figure}

Results are shown in Figure~\ref{fig:lpips_interpols}. Overall, the mean LPIPS score across all objects was $0.249$ ($SD = 0.055$). The object with the highest mean LPIPS score was fish ($M = 0.401$, $SD = 0.111$), and the one with the lowest was birch ($M = 0.116$, $SD = 0.019$). The object with the highest standard deviation in LPIPS scores was sailboat in the vehicle category ($M = 0.306$, $SD = 0.138$), whereas the object with the lowest standard deviation in LPIPS scores was birch in the plant category ($M = 0.116$, $SD = 0.019$). We then examined the mean LPIPS scores for each category. Animal had a mean LPIPS score ($M = 0.277$, $SD = 0.053$), followed by item ($M = 0.264$, $SD = 0.069$), landscape element ($M = 0.238$, $SD = 0.045$), plant ($M = 0.238$, $SD = 0.055$), building ($M = 0.237$, $SD = 0.042$), and vehicle ($M = 0.237$, $SD = 0.053$). A one-way ANOVA revealed no statistically significant difference in mean LPIPS scores across categories ($F(5, 24) = 1.919$, $p = 0.098$).

\begin{figure}[htbp]
    \centering
    \includegraphics[width=0.7\textwidth]{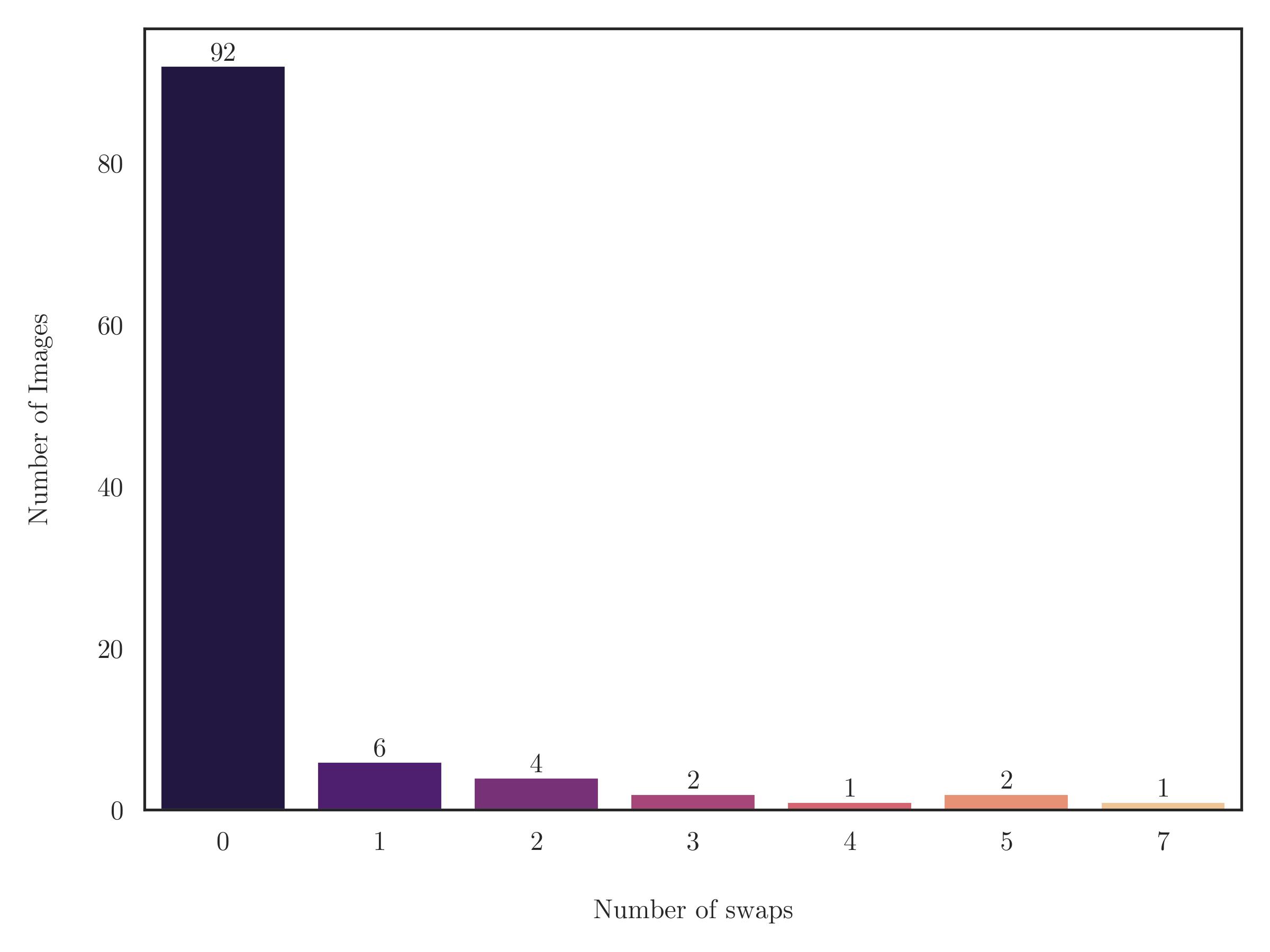}
    \caption{Swapped interpolations. Specifically, the following objects required swaps: \textit{butterfly}, \textit{horse}, \textit{koala}, and \textit{owl} each required $1$ swap ($10\%$). The mortar in the \textit{item} category required $1$ swap ($10\%$), and the \textit{table lamp} required $2$ swaps ($20\%$). The \textit{fjord pillar} in \textit{landscape elements} required $1$ swap ($10\%$). Among \textit{animals}, the \textit{cat}, \textit{leopard}, and \textit{parrot} each required $2$ swaps ($20\%$). For \textit{vehicles}, the \textit{passenger train} and \textit{seaplane} required $3$ swaps each ($30\%$), and the \textit{rowing boat} required $5$ swaps ($50\%$). The \textit{teapot} required $4$ swaps ($40\%$), and the \textit{telescope} required $5$ swaps ($50\%$) in the item category. The \textit{giraffe} had the highest number of swaps, with $7$ swaps ($70\%$).
    }
    \label{fig:swapped_interpols}
\end{figure}

We defined target similarity scores that were linearly spaced between the minimum and maximum LPIPS values observed for each object. For each target score, we selected the image whose similarity score was closest to the target score, ensuring no duplicates. We visually inspected the interpolated images to exclude those with artefacts and maintain a high level of quality and consistency. Given the vast number of images (21,600 in total), the visual inspection was performed only after selecting the interpolated images of interest. For each object, we assessed how many interpolations needed to be swapped with one of their neighbours. The vast majority of objects ($87.62\%$) required no swaps, while a total of $16$ objects ($14.81\%$) required one or more swaps. Among those, $8$ were from \textit{animal} ($44.4\%$ of the objects from that category), $4$ from \textit{item} ($22.2\%$), $3$ from \textit{vehicle} ($16.7\%$), and $1$ from \textit{landscape element} ($5.5\%$). See Figure~\ref{fig:swapped_interpols}.

\begin{figure}[htbp]
    \centering
    \includegraphics[width=0.7\textwidth]{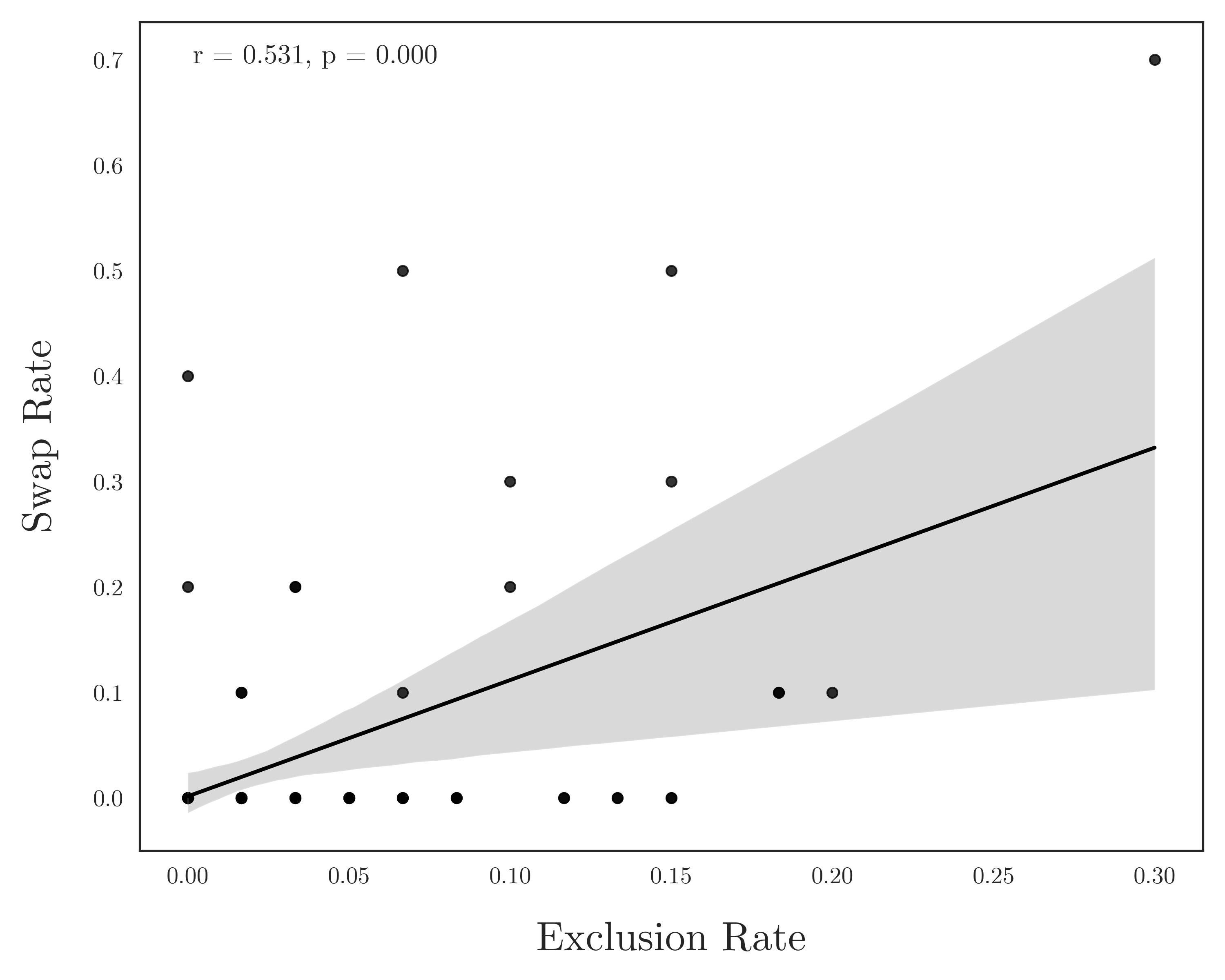}
    \caption{Correlation between exclusion and swap rates.}
    \label{fig:rate_correlation}
\end{figure}

We performed a Pearson correlation to assess the relationship between exclusion rates in anchor images and swap rates in interpolated images. The analysis revealed a significant positive correlation ($r = 0.531$, $p < 0.001$), indicating that objects with higher exclusion rates in exemplar images also tended to have higher swap rates in interpolated images, suggesting consistent issues with the generation of certain objects (Figure~\ref{fig:rate_correlation}). The most problematic was the \textit{animal} category, with an average exclusion rate of $0.068$ and an average swap rate of $0.094$. The least problematic categories were \textit{building} and \textit{plant}, both with no exclusions. Additionally, the \textit{building} category had no swaps, while the \textit{plant} category had a minimal swap rate ($0.016$).

\subsubsection{Physical characteristics of the stimulus set}\label{appendix:physical-properties}

\begin{figure}[htp]
    \centering
    \begin{subfigure}{\textwidth}
        \centering
         \caption{}
        \includegraphics[width=\textwidth]{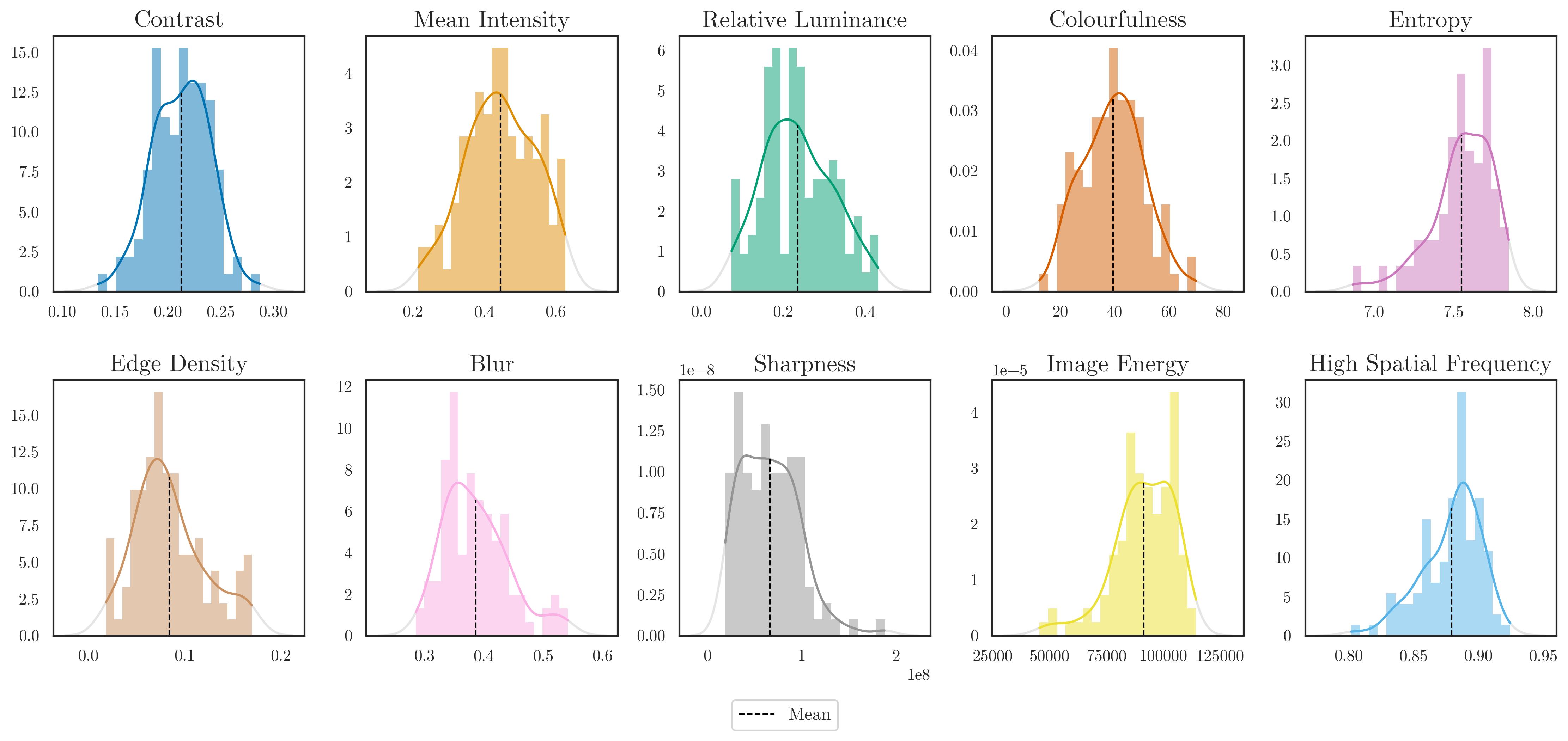}
        \label{fig:img_physical_characteristics}
    \end{subfigure}
    \hfill
    \begin{subfigure}{\textwidth}
        \centering
        \caption{}
        \includegraphics[width=\textwidth]{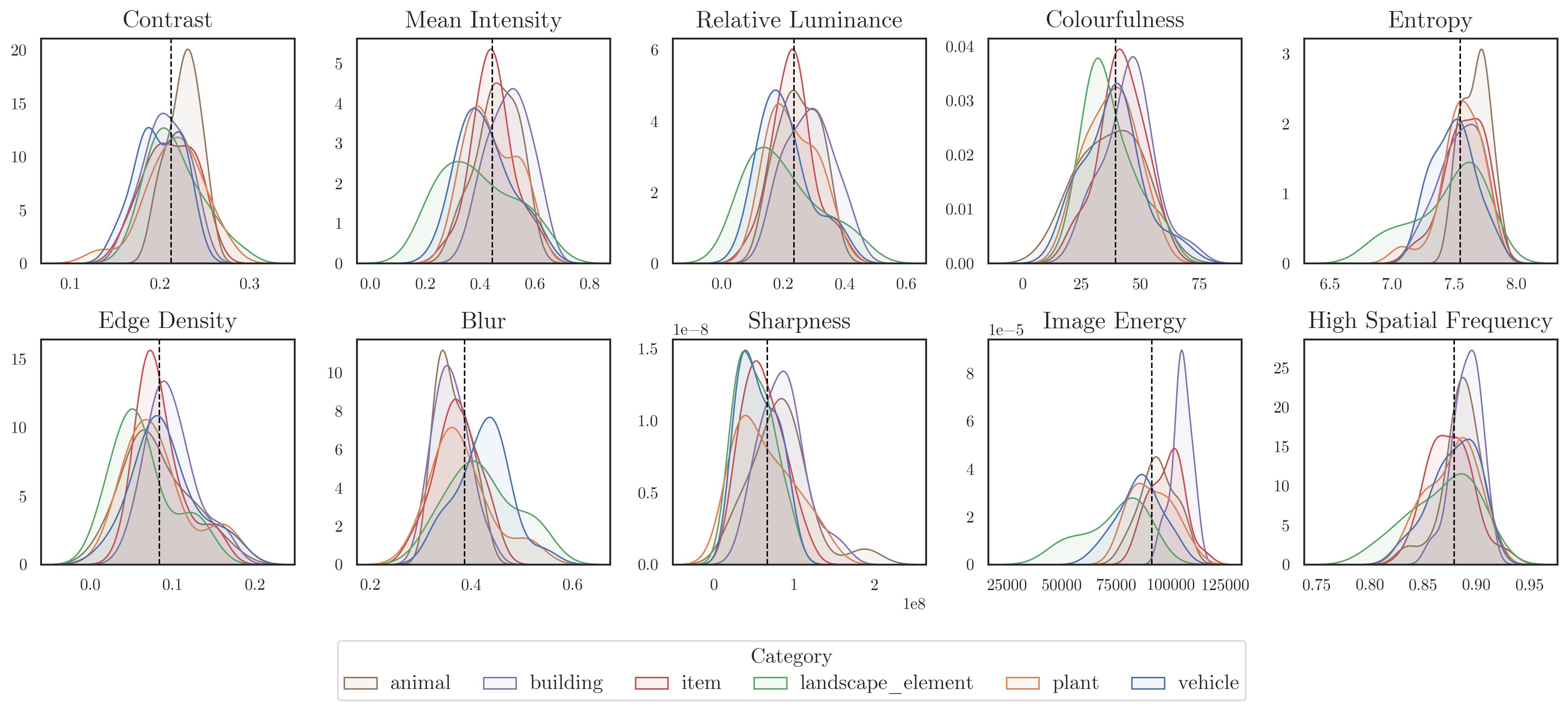}
        \label{fig:img_physical_characteristics_by_category}
    \end{subfigure}
    \caption{Physical characteristics of the \textit{anchor images} of the final stimulus set. a) The first set of histograms illustrates the overall distributions of various image features across the stimulus set. Each feature is accompanied by a fitted density curve and a dashed line indicating the mean value. b) The second set of plots provides a comparison of feature distributions across different image categories (animal, building, item, landscape element, plant, vehicle). Each plot shows the distribution of a feature for each category, with the overall mean indicated by a dashed line.}
    \label{fig:combined_physical_characteristics}
\end{figure}

Figure~\ref{fig:img_physical_characteristics_by_category} shows how image features vary by category. Contrast is higher in \textit{animal} and \textit{building}, with symmetrical distributions. Mean pixel intensity is elevated in \textit{building} and \textit{item}, both normally distributed. Relative luminance is uniform across categories, with a slight increase in \textit{animal}. Colourfulness peaks in \textit{plant} and \textit{landscape element}, with normal distributions. Entropy is consistent, but higher in \textit{item} and \textit{vehicle}, with slight left skew. Edge density is higher in \textit{animal} and \textit{item}, and right-skewed. Blur increases in \textit{building} and \textit{item}, with symmetrical distributions. Sharpness is higher in \textit{animal} and \textit{item}, and right-skewed. Image energy is elevated in \textit{building} and \textit{item}, with right skew. High spatial frequency content is also higher in \textit{building} and \textit{item}, with slight left skew.

\clearpage
\subsection{Similarity judgement task}
\subsubsection{Online task}

\begin{figure}[htp]
    \centering
    \begin{subfigure}[b]{0.48\textwidth}
        \centering
        \caption{}
        \includegraphics[width=\linewidth, height=5cm, keepaspectratio]{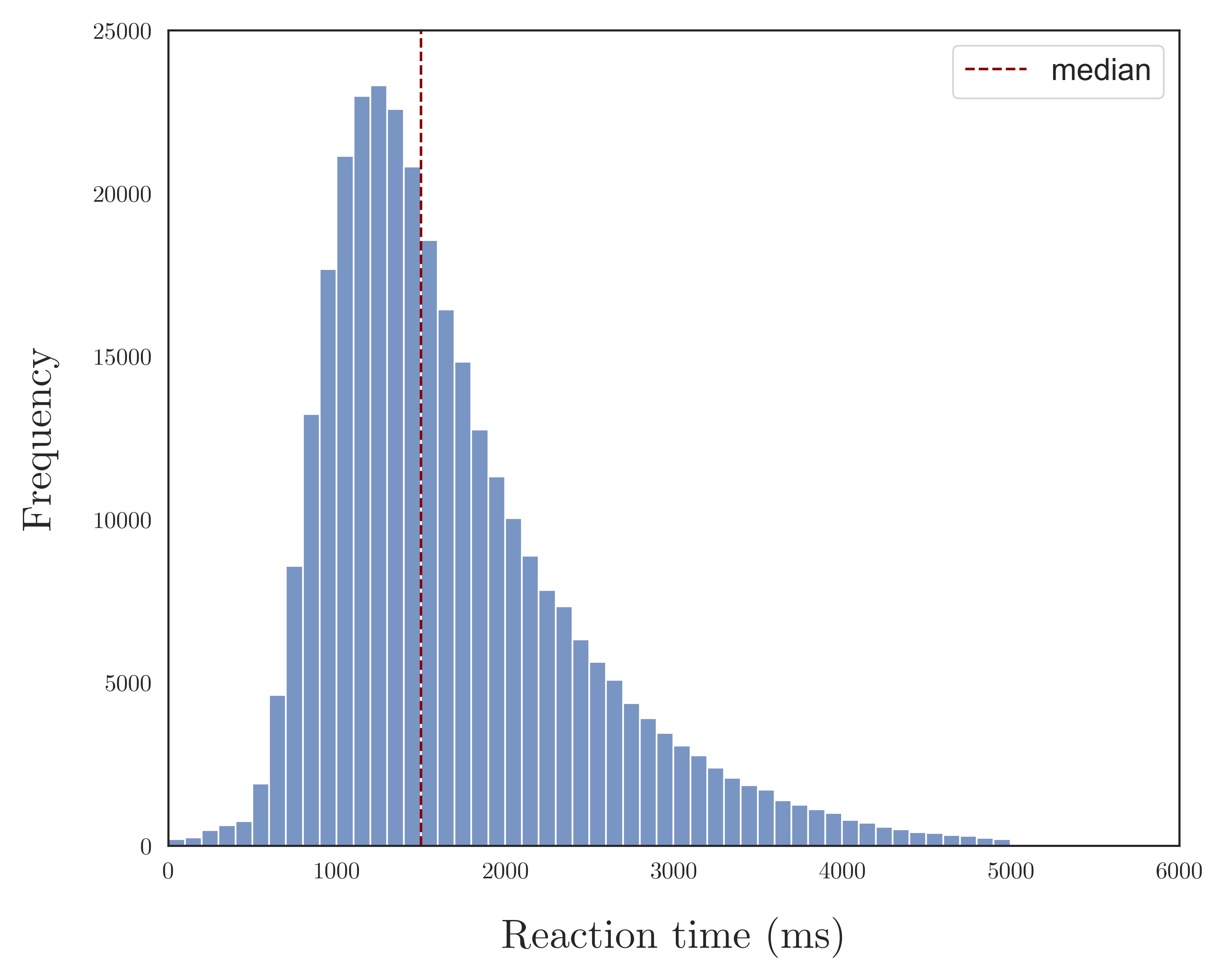}
        \label{fig:RTs}
    \end{subfigure}
    \hfill
    \begin{subfigure}[b]{0.48\textwidth}
        \centering
        \caption{}        
        \includegraphics[width=\linewidth, height=5cm, keepaspectratio]{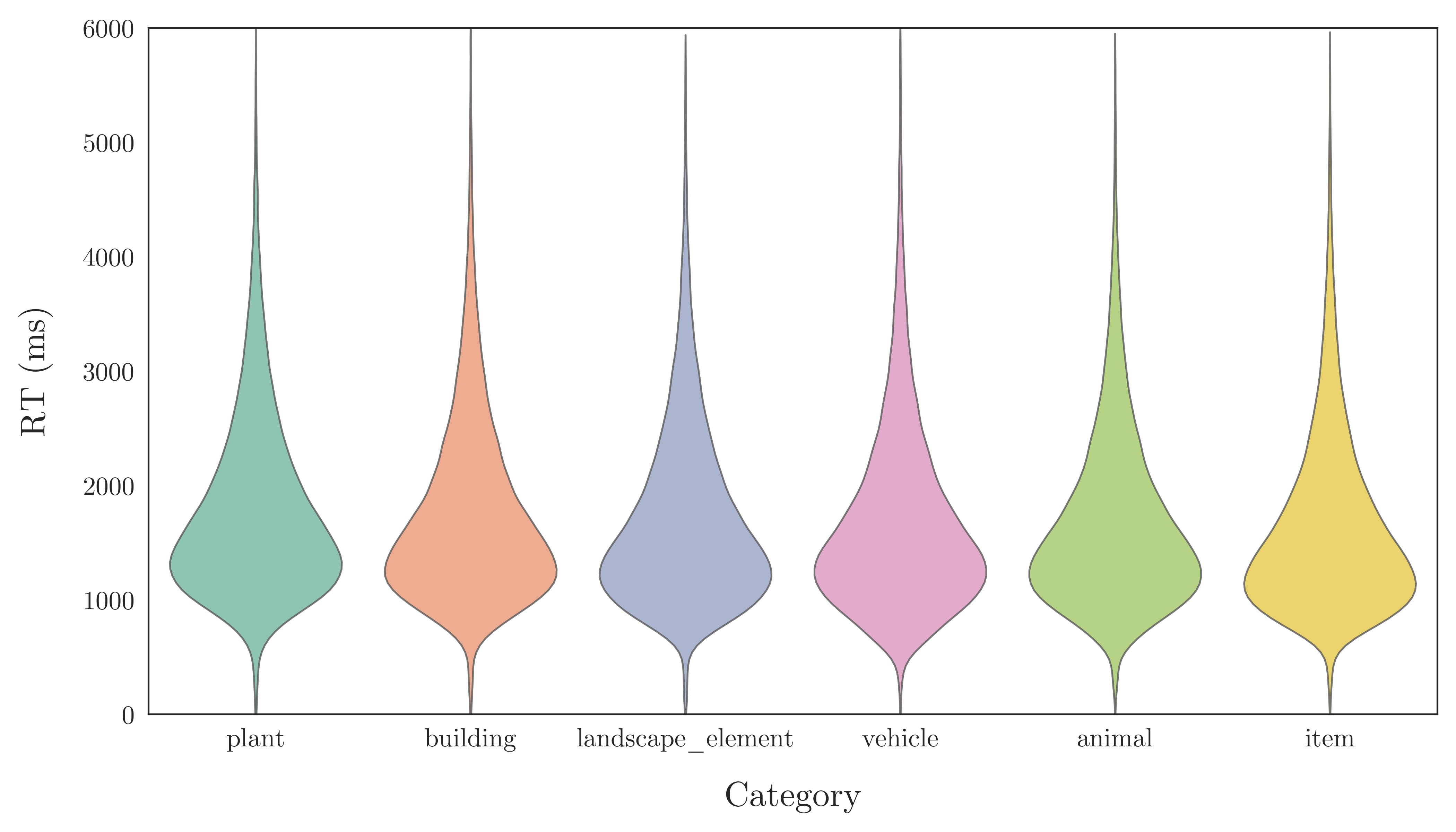}
        \label{fig:rt_distributions_by_category}
    \end{subfigure}
    
    \vspace{0.5cm} %
    
    \begin{subfigure}[b]{\textwidth}
        \centering
        \caption{}
        \includegraphics[width=\linewidth]{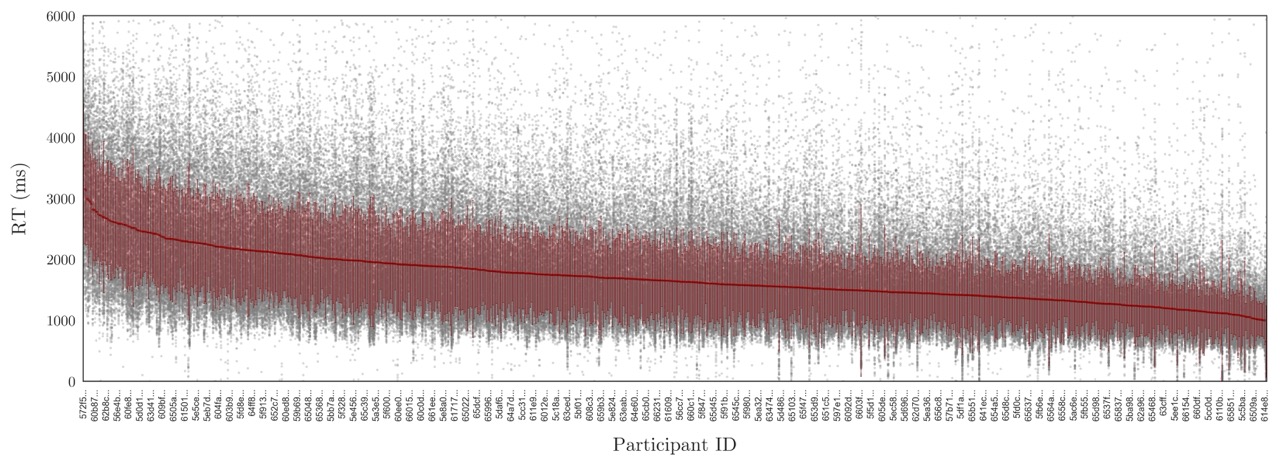}
        \label{fig:rt_final_sample_by_participant}
    \end{subfigure}

    \caption{Distributions aggregated over all participants' data: (a) Reaction time distribution, (b) Reaction time distributions for each category, and (c) Reaction time distributions for all participants in the final sample (\(RT < 6000 \, \text{ms}\)), sorted by participant ID in descending order of mean reaction time.}
    \label{fig:distributions}
\end{figure}

The median experiment duration was $24.11$ minutes and the median duration of the triplet comparison task (excluding instructions and training) was $20.5$ minutes. The median reaction time in the task was $1501.0$ ms (Figure~\ref{fig:RTs}). 

\subsubsection{Evaluation of the embeddings}\label{appendix:evaluation_embeddings}
\begin{figure}[H]
    \centering
    \begin{subfigure}[b]{0.48\textwidth}
        \centering
        \caption{} %
        \includegraphics[height=5cm]{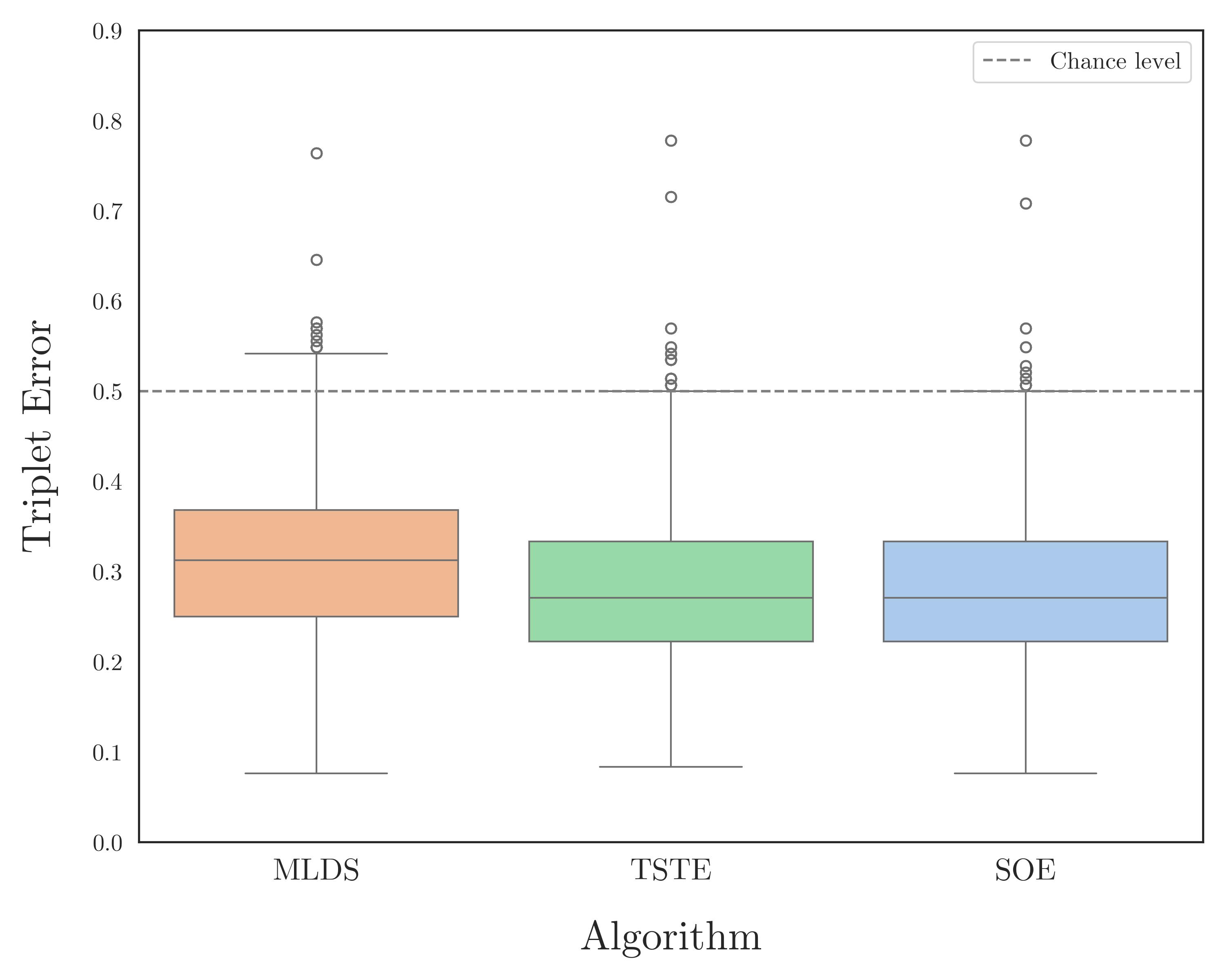}  
        \label{fig:model_performance_1}
    \end{subfigure}
    \hfill
    \begin{subfigure}[b]{0.48\textwidth}
        \centering
        \caption{}
        \includegraphics[height=5cm]{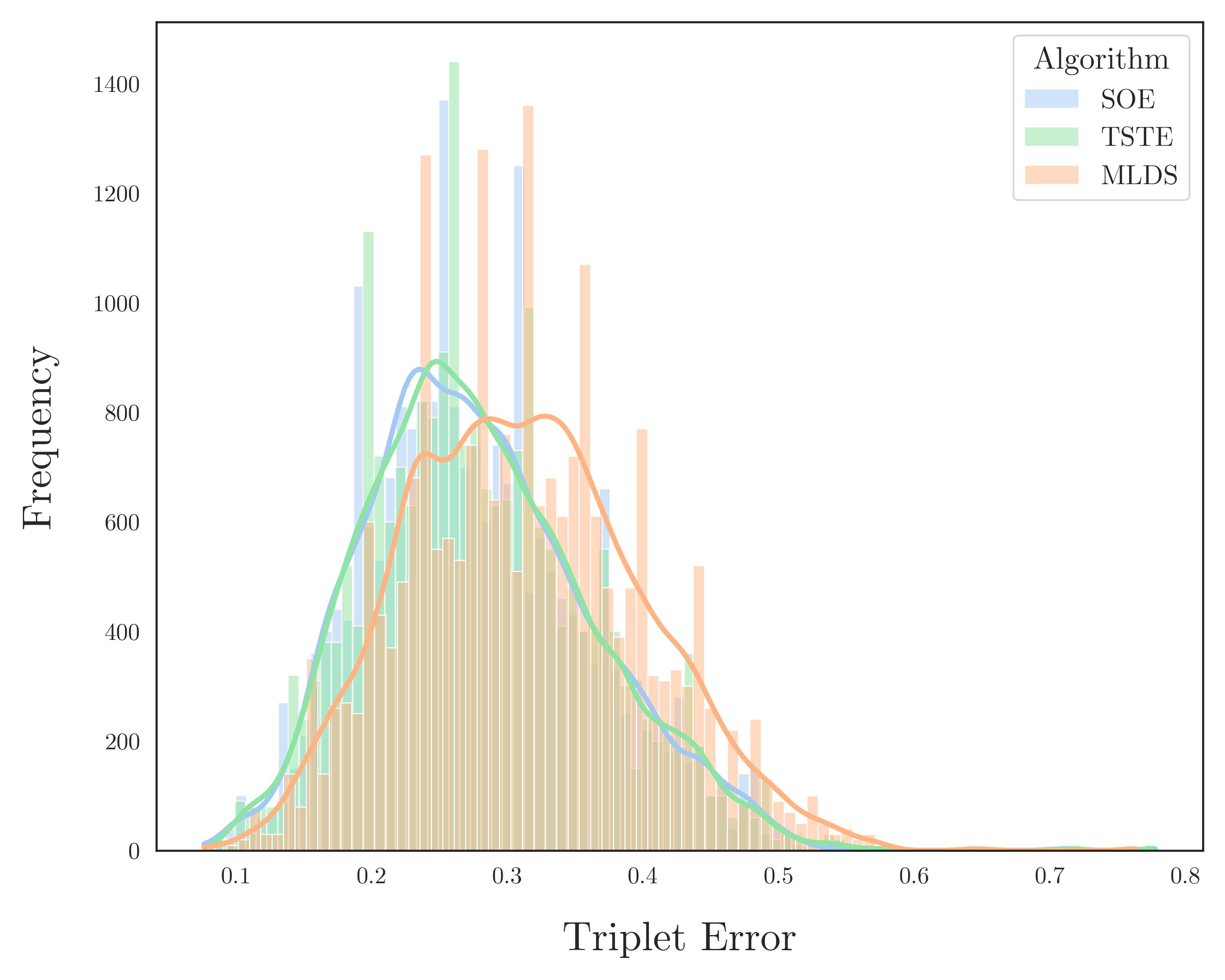}
        \label{fig:model_performance_distr_2}
    \end{subfigure}
    
    \vspace{0cm}
    
    \begin{subfigure}[b]{0.7\textwidth}
        \centering
        \caption{} 
        \includegraphics[height=5cm]{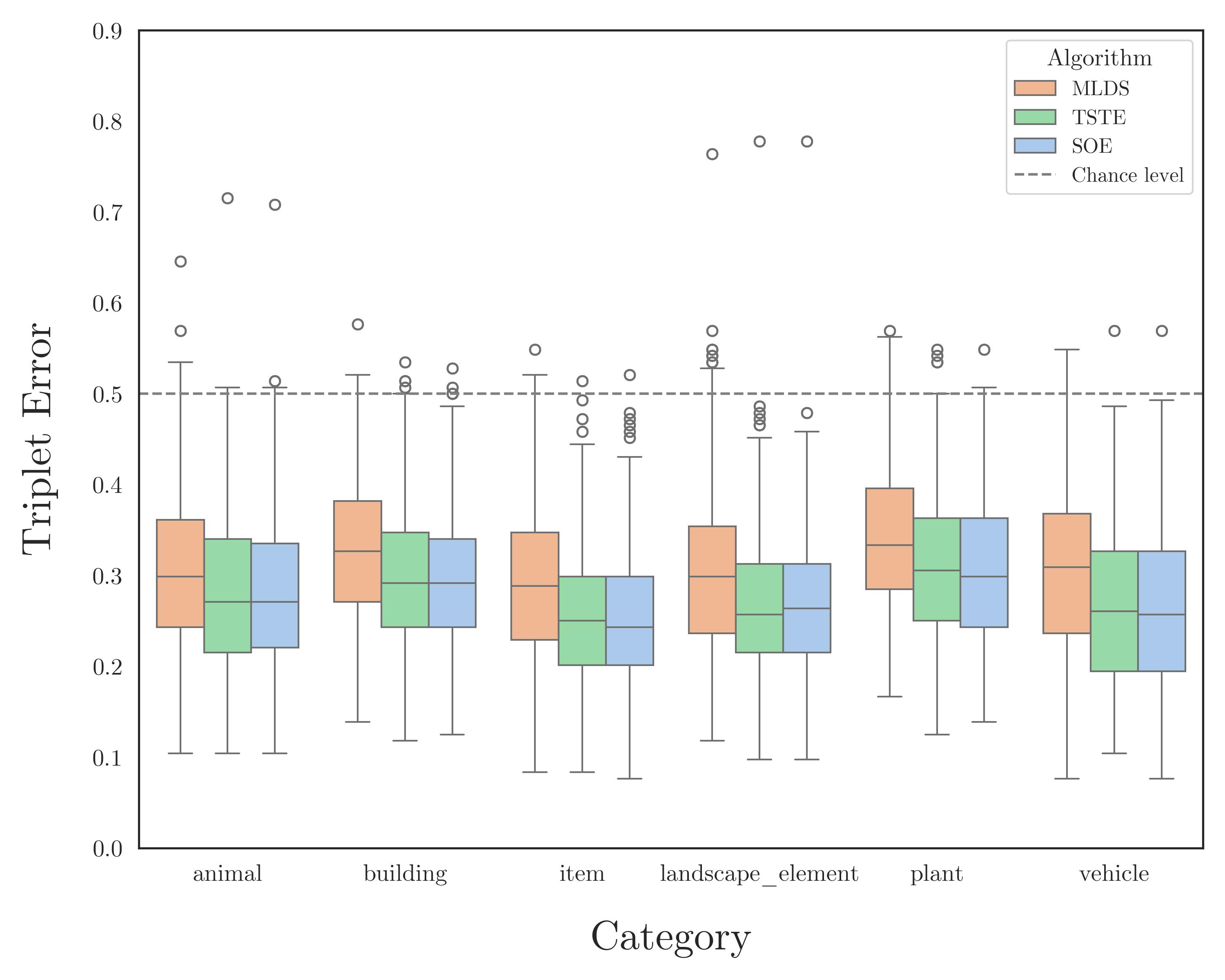}
        \label{fig:model_performance_categ_4}
    \end{subfigure}
    
    \vspace{0cm}
    
    \begin{subfigure}[b]{1\textwidth}
        \centering
        \caption{} 
        \includegraphics[width=\linewidth]{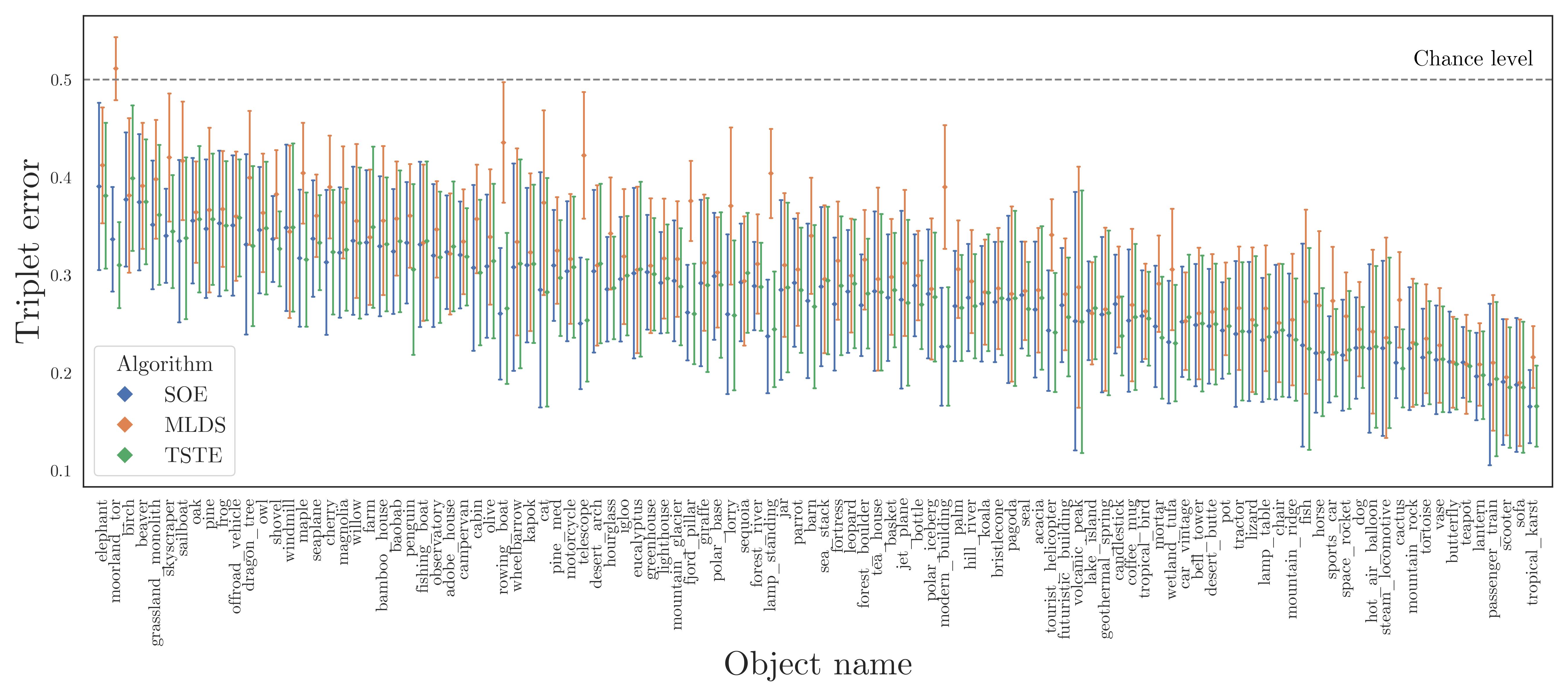}
        \label{fig:model_performance_obj_3}
    \end{subfigure}
    
    \caption{
        Comparative performance analysis of SOE, t-STE, and MLDS algorithms using triplet error. (a) Overall performance: mean triplet error across conditions. (b) Error distributions: variability per algorithm as histograms. (c) Category-specific performance: triplet error across object categories. (d) Object-specific performance: triplet error for individual objects.
    }

    \label{fig:model_performance} 
\end{figure}

The performance of the three embedding algorithms of interest — t-STE, SOE, and MLDS — was evaluated using the Leave-One-Participant-Out cross-validated triplet error (see Methods in Appendix~\ref{appendix:similarity_judgement_analysis}). In order to assess the overall performance of each model, we calculated the mean cross-validated error by method, averaging across all objects. The mean cross-validated triplet error rates were 0.313 for MLDS, 0.280 for t-STE, and 0.279 for SOE (Figure~\ref{fig:model_performance_1}). A one-way ANOVA revealed a significant effect of the embedding algorithm on the triplet error rate, \(F(2, 87) = 114.3, p < 0.0001\). Post-hoc comparisons using the Tukey HSD test indicated that both t-STE and SOE had significantly lower triplet error rates compared to MLDS (\(p < 0.05\)). Specifically, the mean difference between MLDS and SOE was -0.0335 (\(95\% \, \text{CI} = [-0.0395, -0.0276]\)), and between MLDS and t-STE was -0.0332 (\(95\% \, \text{CI} = [-0.0391, -0.0272]\)), both significant at \(p < 0.05\). There was no significant difference between t-STE and SOE (mean difference = 0.0004, \(95\% \, \text{CI} = [-0.0056, 0.0064]\), \(p = 0.9872\)).

We also assessed object-specific and category-specific performance by aggregating and averaging the cross-validated errors for each object (Figure~\ref{fig:model_performance_obj_3}) and category (Figure~\ref{fig:model_performance_categ_4}). Our analysis revealed that t-STE and SOE consistently performed well across different objects. In contrast, MLDS showed significant performance variability, with a notable error increase for certain objects. For instance, the triplet error for ``moorland tor'' was 0.48 with MLDS, compared to 0.304 for SOE and 0.305 for t-STE. The analysis of category-specific performance showed that t-STE and SOE consistently outperformed MLDS across all categories. Notably, t-STE and SOE had very similar performance, with negligible differences in triplet errors. 

\subsubsection{Stability of the embedding values}\label{appendix:stability_embedding_values}
\begin{figure}[H]
\includegraphics[width=\textwidth]{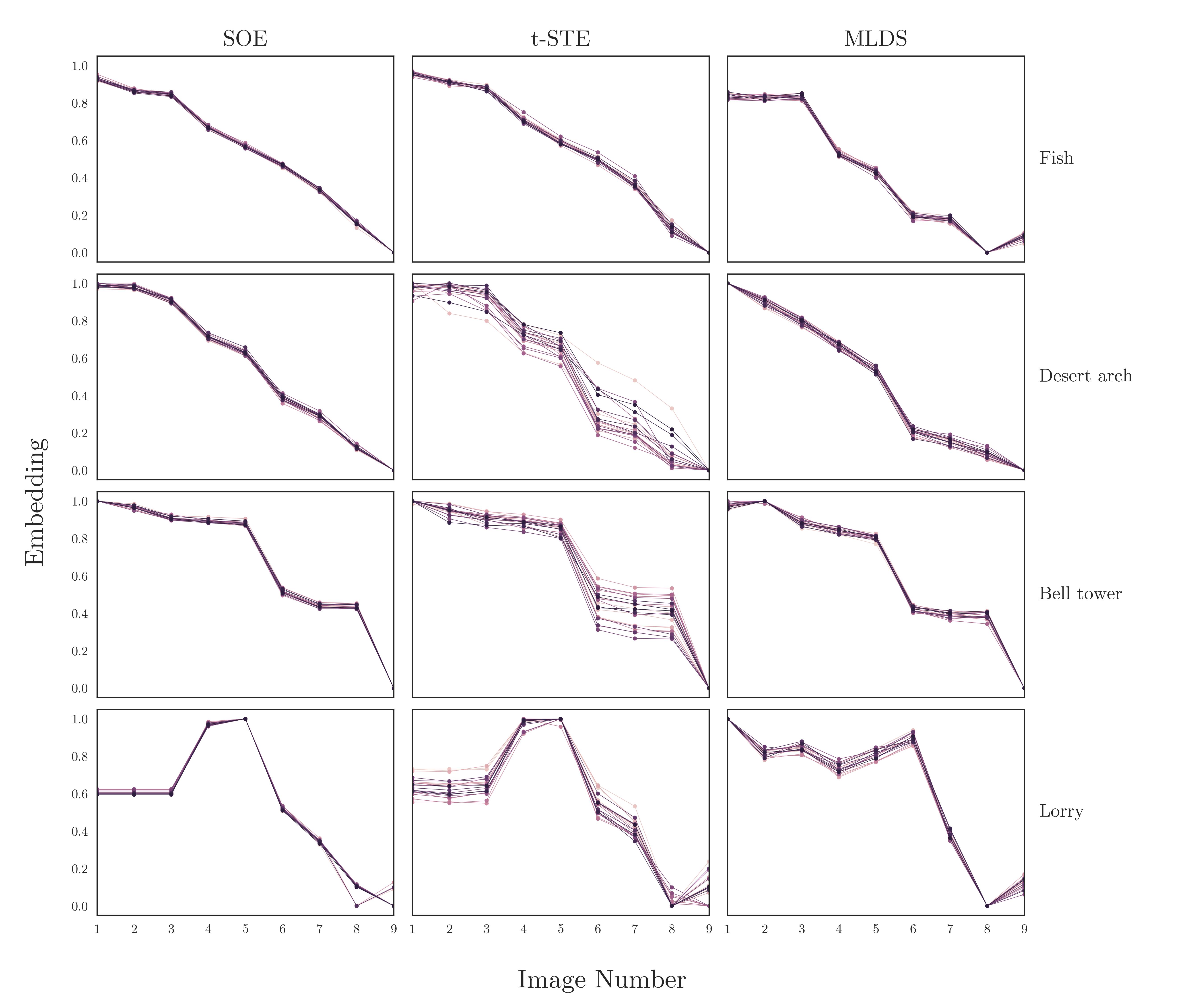}
\caption{Visualisation of the variability of embedding estimates across algorithms for a) \textit{desert arch}, b) \textit{fish}, c) \textit{bell tower}, d) \textit{lorry}. Each line represents embedding values estimated from one cross-validation fold, where one participant is left out. This illustrates the sensitivity of the estimates to individual participants. SOE yields more robust estimates compared to the other algorithms, though all perform well overall. The image order on the x-axis follows the model-based similarity metric (LPIPS), not the participants' reordered judgements. This explains why the lines are not monotonic, as the model ranking does not necessarily correspond to the participants' evaluations.}
\label{fig:embedding_stability}
\end{figure}

The cross-validated triplet error provides a performance estimate for each algorithm, indicating how well the estimated embeddings generalise to unseen triplets across participants. Since we were ultimately interested in using the participants’ judgements to order the images along a perceptual scale determined by the embedding values, we assessed their stability. Embedding estimates are only useful to this purpose if they are reasonably coherent across the sample and not overly sensitive to variations in the source data. We analysed the influence of data from individual participants on the final embedding values by comparing the object-specific embedding estimates across different folds for each algorithm. In general, we found that SOE provided more stable embedding estimates, making it more robust to variations in the source data. A few object-specific examples are provided in Figure~\ref{fig:embedding_stability}.

\subsubsection{Reordering the image variations}\label{appendix:reordering}
\begin{figure}[h]
\includegraphics[width=\textwidth]{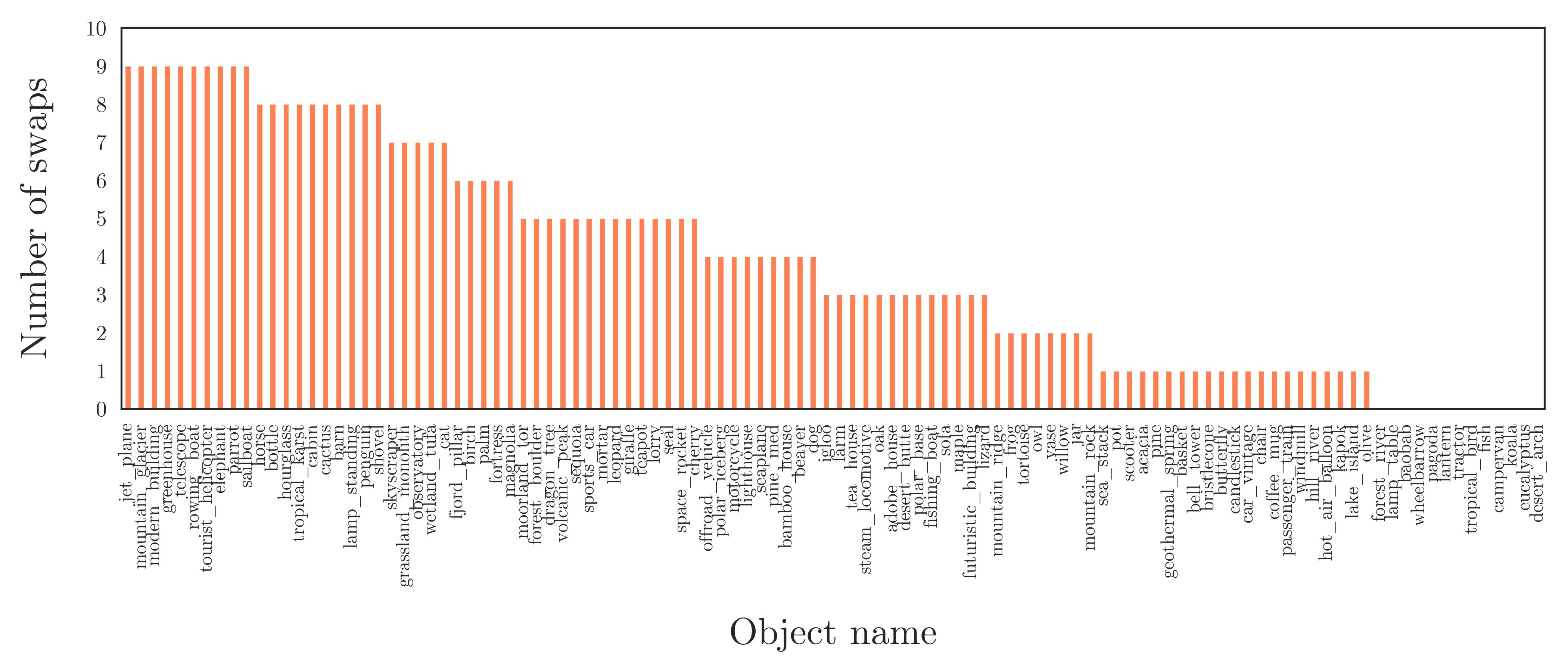}
\caption{Reordering count by object.}
\label{fig:reordering_count}
\end{figure}

Although we found no significant difference in performance between SOE and t-STE in terms of triplet error, SOE showed more consistent embedding values across iterations, reducing variability in the results (Figure~\ref{fig:embedding_stability}). Therefore, we decided to use the embedding values estimated by SOE to order the image variations on a perceptual scale for each object. The values were computed object-wise, using the aggregate data from all participants. In total, 95 objects (88\%) saw at least one position reordering, whereas 13 objects maintained the order provided by the LPIPS score (Figure~\ref{fig:reordering_count}). The median number of reorderings per object was 3. To quantify the degree of agreement between the ordinal positions derived from LPIPS scores and those from SOE embeddings, we calculated a Spearman’s rank correlation. The analysis revealed a positive correlation, $\rho(1078) = 0.73$, $p < 0.001$, indicating that the ordinal positions based on the LPIPS metric align with those based on the SOE embedding values. 
We provide some examples of the stimulus set after psychophysical validation in Figure~\ref{fig:final_stimulus_set} (main text). For a comparison between the LPIPS-based order and the order after the behavioural task.

\begin{figure}[htbp]
    \centering
    \begin{subfigure}[b]{0.40\textwidth}
        \caption{}
        \includegraphics[width=\textwidth]{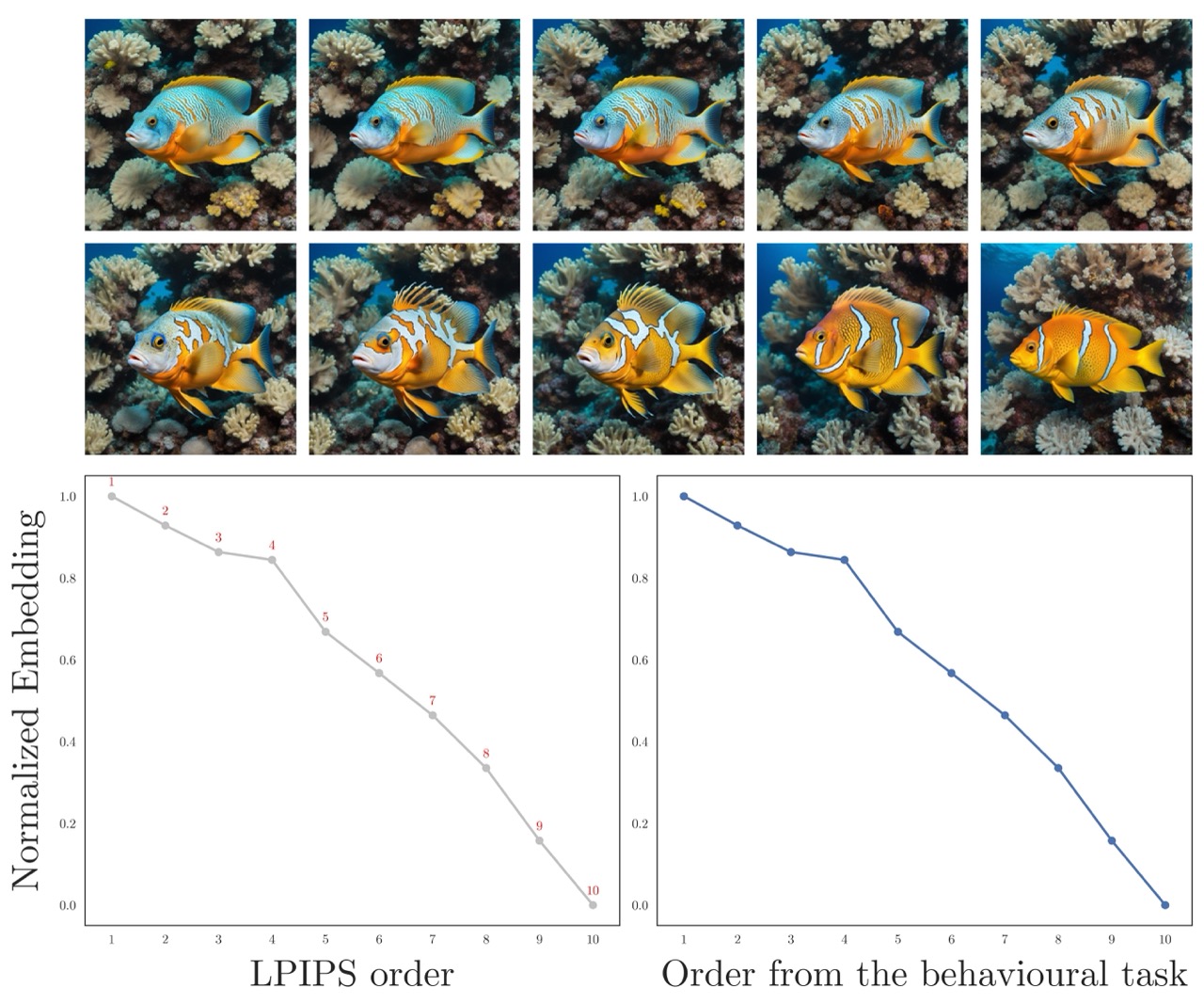}
        \label{fig:example_swaps_1}
    \end{subfigure}
    \hfill
    \begin{subfigure}[b]{0.40\textwidth}
        \caption{}
        \includegraphics[width=\textwidth]{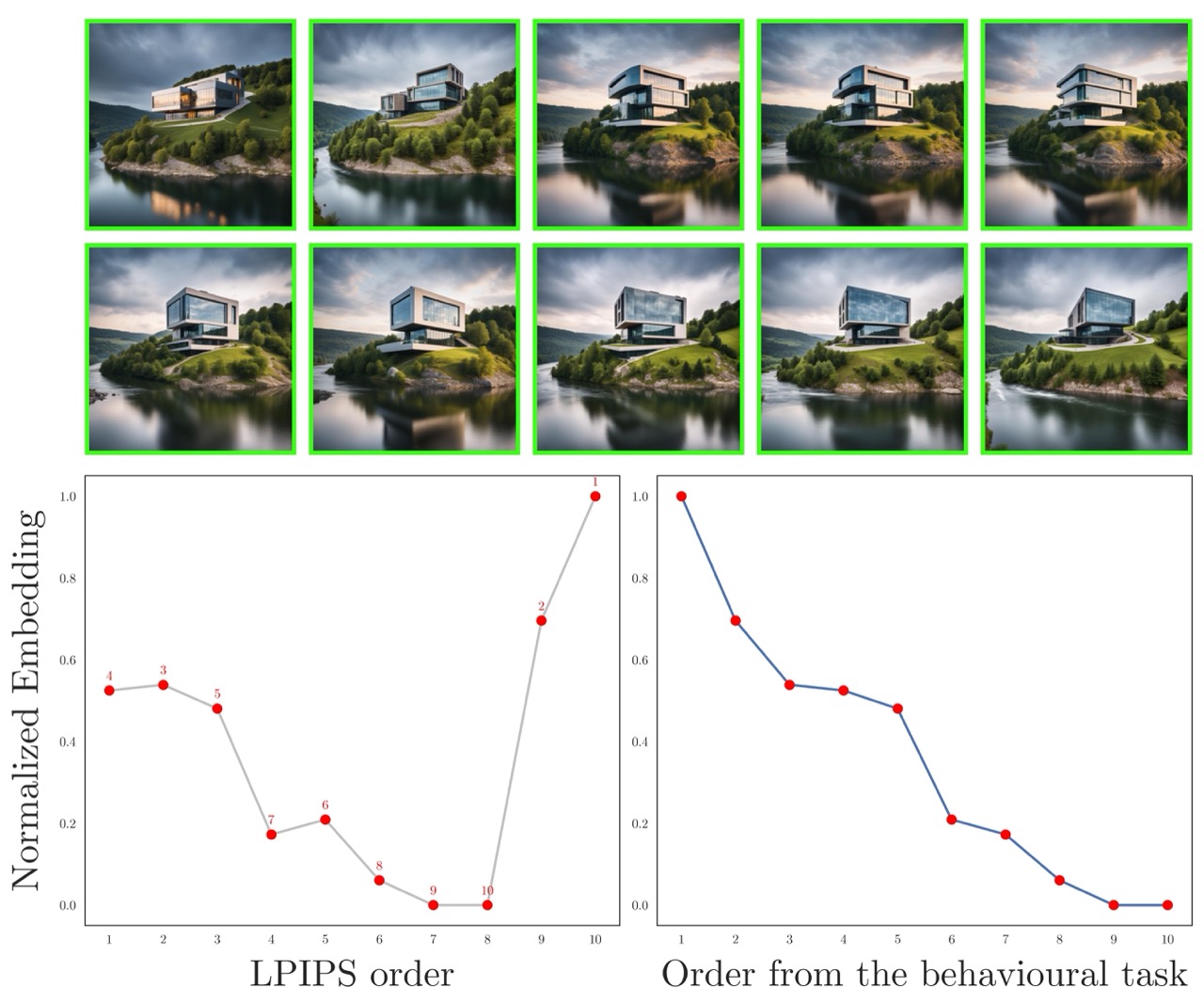}
        \label{fig:example_swaps_2}
    \end{subfigure}
    \hfill
    \begin{subfigure}[b]{0.40\textwidth}
        \caption{}
        \includegraphics[width=\textwidth]{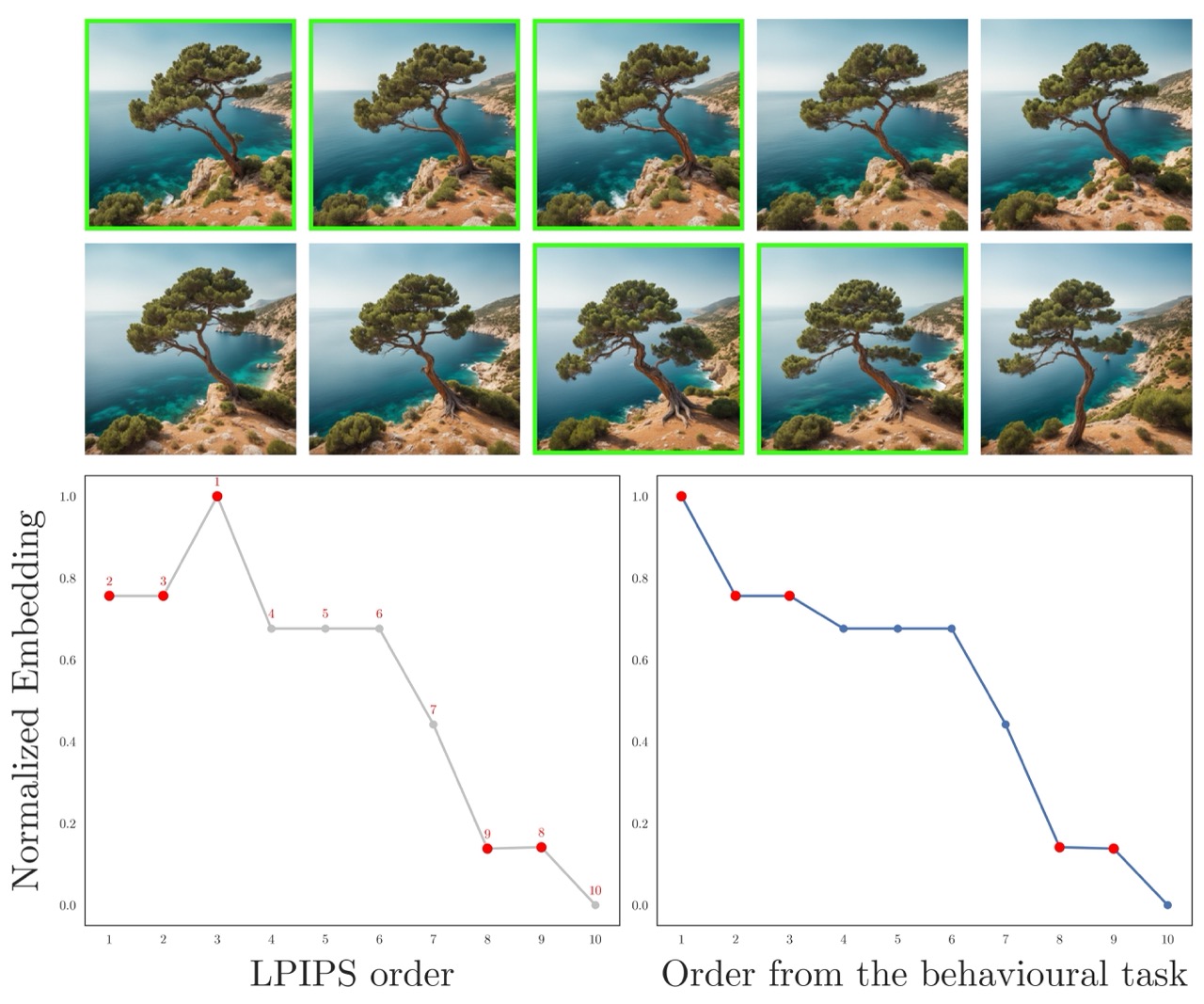}
        \label{fig:example_swaps_3}
    \end{subfigure}
    \hfill
        \begin{subfigure}[b]{0.40\textwidth}
        \caption{}
        \includegraphics[width=\textwidth]{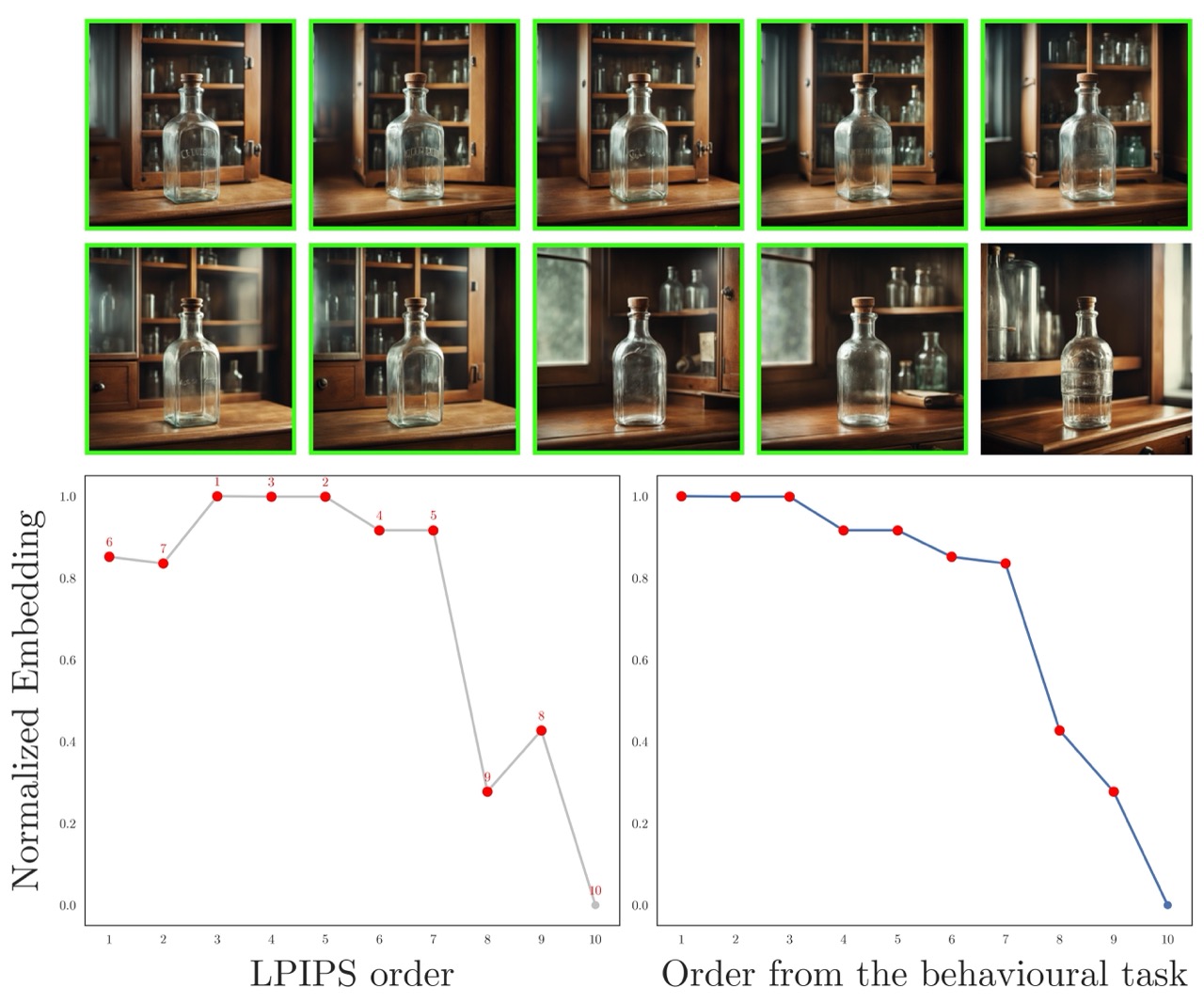}
        \label{fig:example_swaps_4}
    \end{subfigure}
    \hfill
    \begin{subfigure}[b]{0.40\textwidth}
        \caption{}
        \includegraphics[width=\textwidth]{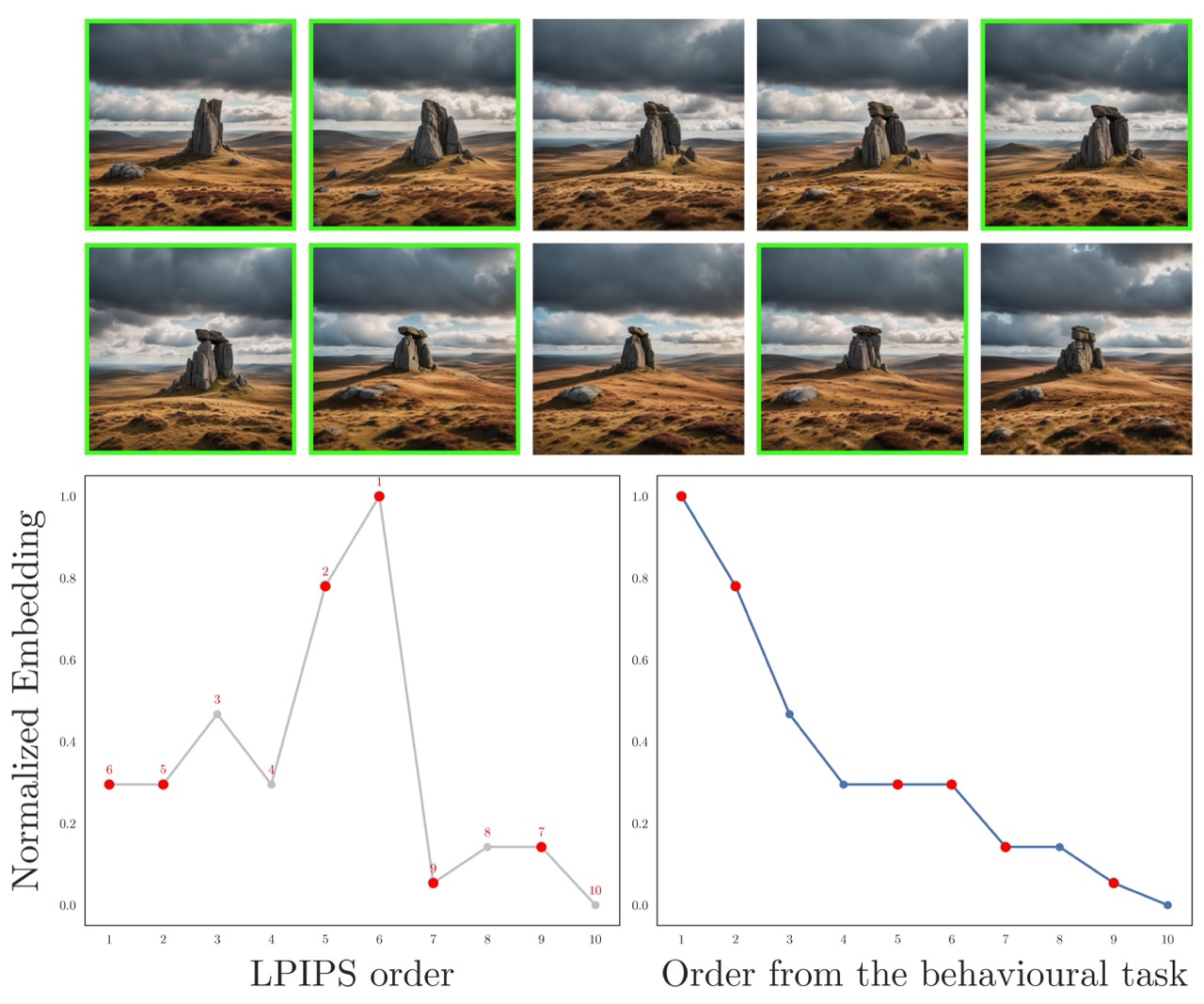}
        \label{fig:example_swaps_5}
    \end{subfigure}
    \hfill
    \begin{subfigure}[b]{0.40\textwidth}
        \caption{}
        \includegraphics[width=\textwidth]{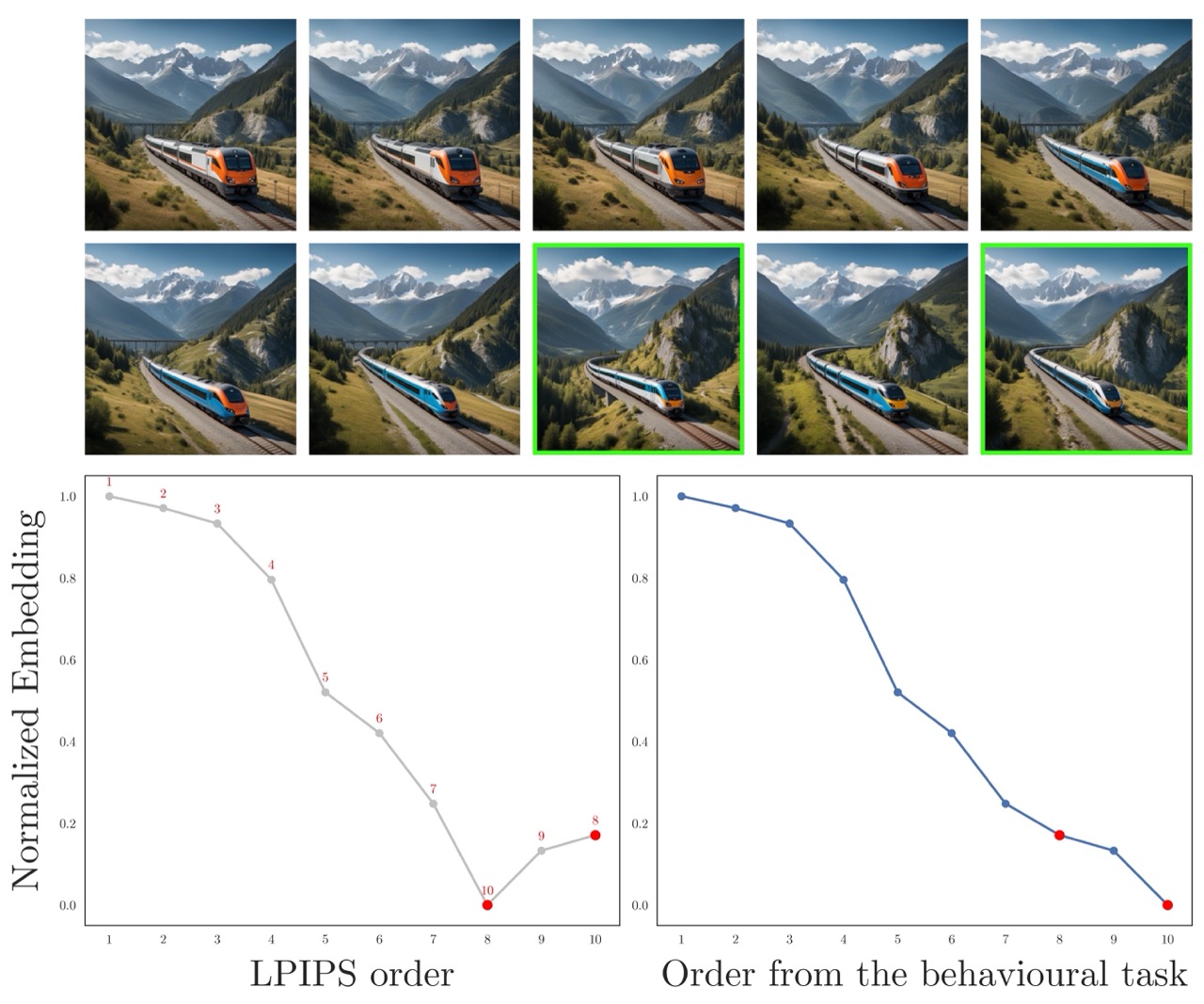}
        \label{fig:example_swaps_6}
    \end{subfigure}
    \caption{\footnotesize Example reorderings based on embedding values. Each panel displays images ordered by their embedding values. Images where the order was changed compared to the LPIPS model are highlighted in green. The left plot shows the original order, with red numbers indicating new positions post-reordering. The right plot displays the reordered objects sorted by embedding value. The x-axis represents object positions, while the y-axis shows embedding values. Red dots mark changed positions, highlighting adjustments in object order. Panels: a) \textit{fish}, b) \textit{modern building}, c) \textit{Pine (Mediterranean)}, d) \textit{bottle}, e) \textit{moorland tor}, f) \textit{passenger train}.}
    \label{fig:example_swaps}
\end{figure}

\subsubsection{Embedding analyses}\label{appendix:embedding_analyses}
MLDS, T-STE and SOE generally produced comparable results. This corroborates other studies relying on simulations that showed a substantial equivalence in performance for low-dimensional cases \parencite{vankadara_insights_2023}. One open issue is to what extent the embedding values reflect real differences in perceptual similarity. 

One limitation of using the embedding values to reorder the images on a perceptual scale is  that it is not very robust to small variations in embedding due to noise. This is due to the fact that some of the image variations are very similar to one another and have very similar embedding values. In these cases, the ordinal position of the stimuli is susceptible to  noise fluctuations. At the same time, it should be noted that not all reorderings are perceptually meaningful. When stimuli have similar embedding values, it means that participants did not perceive significant differences between them. They can therefore be considered a “perceptual cluster”, within which the order of the stimuli is relatively stochastic. 

We made the simplifying assumption that each set of image variations can be ranked using a single dimension that follows the LPIPS metric. An open question is what stimulus features within the objects drove the similarity judgements more strongly and future work should address that. However, we found that – in most cases – it was possible to order naturalistic stimuli along one perceptual dimension in good agreement with the LPIPS metric. Our results extend the existing literature showing that the triplet comparison method can be effectively used for naturalistic images, even for perceptually subtle variations.

\pagebreak

\subsection{Memory experiment}\label{appendix:memory_experiment_results}

\subsubsection{Bayesian model results}\label{appendix:bayesian_model_results}

,Baseline accuracy for Easy, non-repeated trials (intercept) was moderately high, with \(\hat{\beta_0}=1.52\) and a 95\% credible interval $\mathrm{CI}_{95}=[1.30, 1.73]$, corresponding to a baseline probability of success of approximately 78\%. Participants exhibited progressive learning gains, with a positive block effect, $\hat{\beta_1} = 0.15$, $\mathrm{CI}_{95}=[0.09, 0.21]$, indicating a 17\% increase in the odds of success per block, $\mathrm{OR}= 1.17$. Task difficulty significantly reduced performance, with Hard trials ($\hat{\beta_3}= -0.89$, $\mathrm{CI}_{95}=[-1.13, -0.64]$, $\mathrm{OR}= 0.41$) showing a 59\% decrease in odds, and Medium trials ($\hat{\beta_2}= -0.52$, $\mathrm{CI}_{95}=[-0.78, -0.27]$, $\mathrm{OR}= 0.59$) showing a 41\% reduction. While stimulus repetition alone showed no reliable main effect ($\hat{\beta_4}= -0.13$, $\mathrm{CI}_{95}=[-0.41, 0.16]$), its interaction with block progression demonstrated robust learning improvements (block $\times$ repeated: $\hat{\beta_7}= 0.13$, $\mathrm{CI}_{95}=[0.05, 0.22]$, $\mathrm{OR}= 1.14$ per block). Higher-order interactions involving difficulty levels showed no credible effects, with confidence intervals spanning zero. Individual differences were pronounced: baseline performance varied substantially across participants (random intercept $SD= 0.72$, $\mathrm{CI}_{95}=[0.62, 0.83]$), and learning rates exhibited moderate heterogeneity (random block slope $SD= 0.16$, $\mathrm{CI}_{95}=[0.13, 0.18]$). A negative correlation between baseline and learning rate ($r= -0.54$, $\mathrm{CI}_{95}=[-0.67, -0.38]$) suggested that participants with higher initial accuracy improved less over time. Bayesian hypothesis tests provided further corroboration of these findings. We summarise these tests using posterior probabilities ($\mathrm{pp}$), indicating the probability of the tested hypothesis given the observed data, and evidence ratios ($\mathrm{ER}$), quantifying how much more likely the data are under the tested hypothesis compared to the alternative. We observed decisive support for progressive learning across blocks ($H_1$: $\mathrm{ER}= \infty$, $\mathrm{pp}= 1.00$) and pronounced accuracy reductions under harder task conditions ($H_2$: Hard vs. Easy, $\mathrm{ER}= \infty$, $\mathrm{pp}=1.00$; $H_3$: Medium vs. Easy, $\mathrm{ER}=15,999$, $\mathrm{pp}=1.00$). While repeated stimuli alone showed no reliable benefit ($H_4$: $\mathrm{ER}=0.24$, $\mathrm{pp}=0.19$), their interaction with block progression strongly improved learning ($H_5$: $\mathrm{ER}=2,132$, $\mathrm{pp}=1.00$). For interactions involving difficulty, we found moderate evidence that Hard trials attenuated learning gains ($H_6$: $\mathrm{ER}=25.8$, $\mathrm{pp}=0.96$), though the effect size was small ($\hat{\beta_6}=-0.06$). In contrast, other interactions including Medium-difficulty effects ($H_7$: $\mathrm{ER}=4.48$, $\mathrm{pp}=0.82$) and three-way combinations ($H_8–H_{11}$: $\mathrm{ER} < 4.5$, $\mathrm{pp} \leq 0.79$) received negligible support, suggesting repetition effects remained stable across difficulty levels.

\begin{sidewaystable}[htbp]
\centering
\caption{Summary of posterior estimates from the Bayesian hierarchical model. The table reports posterior means (\textit{Estimate}), standard errors (\textit{Est.\ Error}), and 95\% credible intervals (\(\mathrm{CI}_{95}\)). Model diagnostics include the Gelman--Rubin convergence statistic (\(\hat{R}\)) and effective sample sizes (Bulk ESS, Tail ESS), indicating sufficient sampling efficiency.}
\begin{tabular}{lccccccc}
\toprule
& \multicolumn{3}{c}{Model Estimates} & \multicolumn{3}{c}{Diagnostics} \\
\cmidrule(lr){2-4}\cmidrule(lr){5-7}
Parameter & $\beta$ & Estimate & Est.\ Error & $\mathrm{CI}_{95}$ & $\hat{R}$ & Bulk ESS & Tail ESS \\
\midrule
Intercept & $\beta_0$ & 1.516 & 0.110 & [1.303, 1.734] & 1.000 & 9834 & 16524 \\
Block (Learning Effect) & $\beta_1$ & 0.152 & 0.029 & [0.094, 0.209] & 1.000 & 9181 & 15058 \\
Medium Difficulty (vs.\ Easy) & $\beta_2$ & -0.521 & 0.131 & [-0.779, -0.266] & 1.000 & 9998 & 17470 \\
Hard Difficulty (vs.\ Easy) & $\beta_3$ & -0.886 & 0.126 & [-1.133, -0.639] & 1.000 & 9808 & 15487 \\
Repeated Condition (vs.\ Non-Repeated) & $\beta_4$ & -0.126 & 0.144 & [-0.412, 0.157] & 1.000 & 8258 & 14999 \\
\addlinespace
Block $\times$ Medium Difficulty & $\beta_5$ & -0.032 & 0.035 & [-0.102, 0.038] & 1.000 & 9553 & 16399 \\
Block $\times$ Hard Difficulty & $\beta_6$ & -0.061 & 0.034 & [-0.128, 0.006] & 1.000 & 9325 & 16067 \\
Block $\times$ Repeated Condition & $\beta_7$ & 0.134 & 0.041 & [0.054, 0.216] & 1.000 & 8348 & 15042 \\
Medium Difficulty $\times$ Repeated Condition & $\beta_8$ & -0.070 & 0.189 & [-0.441, 0.301] & 1.000 & 9071 & 16165 \\
Hard Difficulty $\times$ Repeated Condition & $\beta_9$ & 0.028 & 0.182 & [-0.327, 0.387] & 1.000 & 8771 & 15203 \\
\addlinespace
Block $\times$ Medium Difficulty $\times$ Repeated Condition & $\beta_{10}$ & 0.042 & 0.054 & [-0.063, 0.148] & 1.000 & 9145 & 17181 \\
Block $\times$ Hard Difficulty $\times$ Repeated Condition & $\beta_{11}$ & -0.024 & 0.051 & [-0.125, 0.076] & 1.000 & 8634 & 14974 \\
\bottomrule
\multicolumn{7}{l}{\textit{Note.} $\mathrm{CI}_{95}$ = 95\% credible interval; ESS = effective sample size; $\mathrm{\hat{R}}$ = Gelman--Rubin convergence statistic.}
\end{tabular}
\label{tab:bayesian_model_results}
\end{sidewaystable}

\begin{figure}
    \centering
    \includegraphics[width=0.9\linewidth]{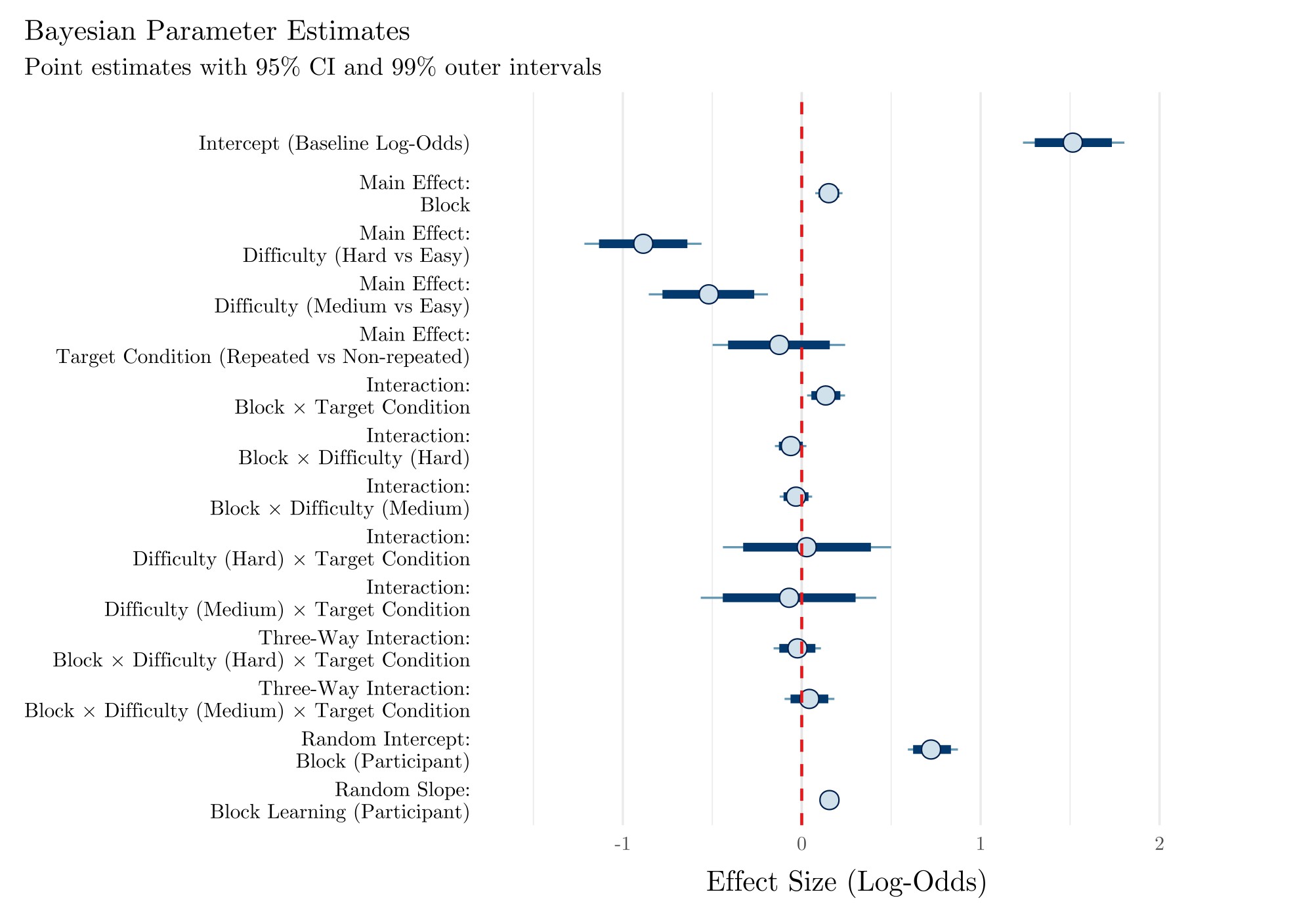}
    \caption{Posterior parameter estimates from hierarchical Bayesian logistic regression. Point estimates (posterior medians) with 95\% (thick lines) and 99\% (thin lines) credible intervals. All effects represent log-odds differences relative to baseline (Easy difficulty, Non-repeated targets, Block 1). The hierarchical structure accounts for participant-level variability in baseline performance (Random Intercept) and learning trajectories (Random Slope). Three-way interactions (top) quantify how repetition benefits depend on both block progression and difficulty. Uncertainty intervals derive from MCMC posterior draws across four chains. }
    \label{fig:parameter-estimates}
\end{figure}

\begin{figure}[htbp]
    \centering

    \begin{subfigure}[t]{0.8\textwidth}
        \includegraphics[width=\linewidth]{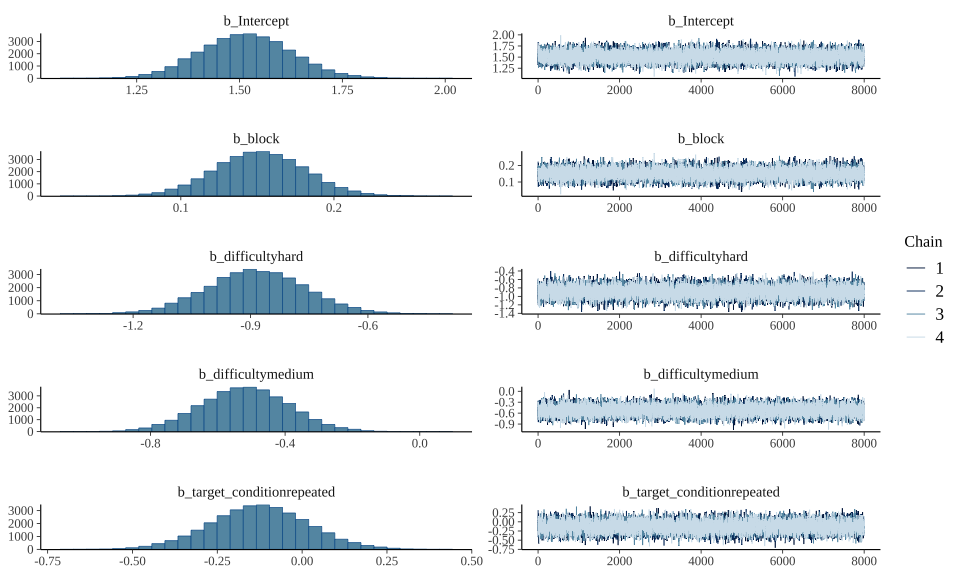}
    \end{subfigure}

    \begin{subfigure}[t]{0.8\textwidth}
        \includegraphics[width=\linewidth]{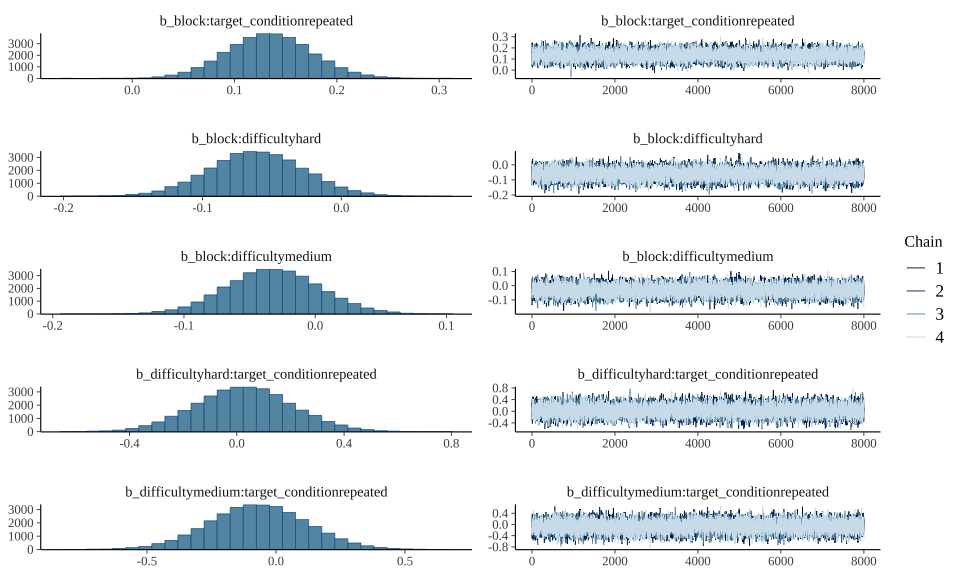}
    \end{subfigure}

    \begin{subfigure}[t]{0.8\textwidth}
        \includegraphics[width=\linewidth]{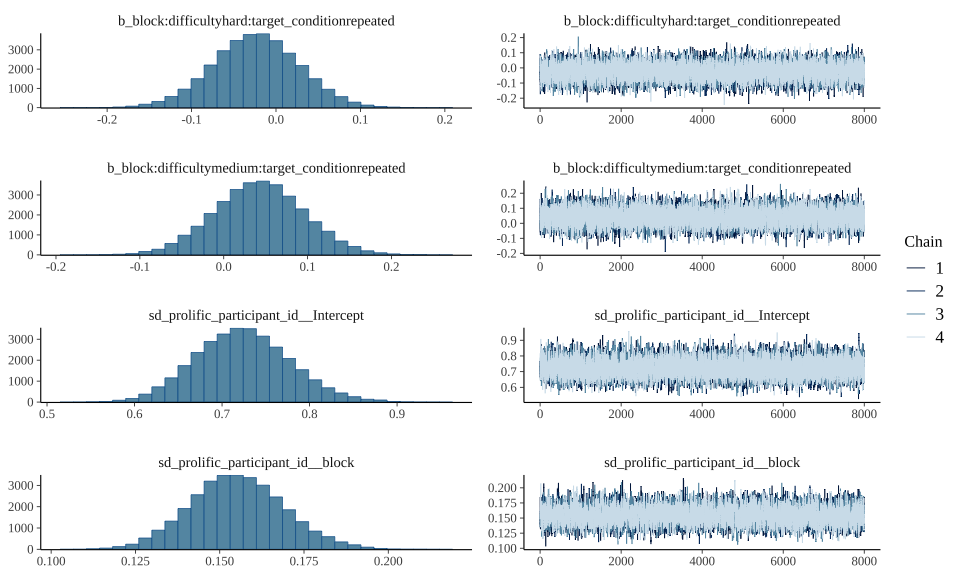}
    \end{subfigure}

    \caption{MCMC diagnostics plots.}
    \label{fig:diagnostic-panel}
\end{figure}

\end{document}